\documentclass[letterpaper,english,aip,pop,reprint]{revtex4-1}
\usepackage[T1]{fontenc}
\usepackage[utf8]{inputenc}
\usepackage{color}
\usepackage{babel}
\usepackage{units}
\usepackage{mathrsfs}
\usepackage{mathtools}
\usepackage{amsmath}
\usepackage{amssymb}
\usepackage{graphicx}
\usepackage[bookmarks=true,bookmarksnumbered=false,bookmarksopen=false,
 breaklinks=false,pdfborder={0 0 0},pdfborderstyle={},backref=false,colorlinks=true]
 {hyperref}
\hypersetup{pdftitle={Charging of dust grains by RKD plasmas},
 linkcolor=blue, citecolor=blue}

\makeatletter

\pdfpageheight\paperheight
\pdfpagewidth\paperwidth

\allowdisplaybreaks

\makeatother

\begin{document}
\global\long\def\sech{\mathop{\mathrm{sech}}\nolimits}%

\global\long\def\Uratio#1#2{\prescript{[#1]}{\vphantom{#1}}{\mathscr{U}}_{\vphantom{#2}}^{[#2]}}%

\global\long\def\Uratiol#1{\prescript{[#1]}{\vphantom{#1}}{\mathscr{U}}}%

\global\long\def\Uratior#1{\mathscr{U}^{[#1]}}%

\global\long\def\Uratiodl#1{\prescript{[#1]}{\vphantom{#1}}{\mathscr{U}}_{d}}%

\preprint{}
\title{Collisional charging of dust particles by suprathermal plasmas.\\
II - Regularized Kappa distributions}
\author{Rudi Gaelzer}
\email{rudi.gaelzer@ufrgs.br}

\author{Luiz F. Ziebell}
\email{luiz.ziebell@ufrgs.br}

\affiliation{Instituto de Física, Universidade Federal do Rio Grande do Sul, CP
15051, Porto Alegre, Rio Grande do Sul, 91501-970, Brazil}
\date{\today}
\begin{abstract}
We study the effects of the velocity distribution functions of the
plasma particles on the equilibrium charge of dust grains, acquired
through inelastic collisions of the particles with the grains. This
paper is the second in a series of two papers on the subject. Here,
we consider the charging process when the plasma particles are statistically
described by the recently proposed regularized Kappa distribution
functions, which allow for extreme suprathermal states, characterized
by extremely low values of the kappa index, previously forbidden to
the standard Kappa distributions, whose effects on dust charging were
studied in Paper I of this series. We analyse the effects that extreme
suprathermal states of the plasma particles have on dust charging
and verify conditions for the uncommon result of positive equilibrium
charge, employing two different models for the regularized Kappa distributions,
one with kinetic temperature dependent on the kappa index, and another
where the temperature is kappa-independent.
\end{abstract}

\maketitle

\section{Introduction}

Space and astrophysical plasmas are rarely composed by electrons and
ionized atoms alone, but usually contain a variable fraction of their
total mass in form of contaminants ranging from nanoparticles to heavier
bodies with various chemical compositions and shapes. These contaminants
in different plasma environments are usually referred to as ``dust'',
and can have different origins. In planetary systems, dust can be
produced by volcanism, collisions between asteroids, or be present
in planetary rings and comet tails. Dust is also present in the interstellar
medium and can comprise a significant part of the stellar winds emanating
from carbon-rich stars, as inferred from spectroscopic studies in
the infrared region. There are several reviews available in the literature
discussing the origin and composition of dust in space and laboratory
plasmas, of which Refs. \onlinecite{Goertz89,Northrop92,MendisRosenberg94/09,ShuklaMamun02,Tsytovich+04/10,Fortov+05/12,Mann+10/12,Mann+14/03,Spahn+19/02,Melzer+21/11,Sterken+22/12,GodenkoIzmodenov23/12}
are a representative sample. 

Recently, observations from the James Webb Space Telescope, processed
with a new spectroscopic technique, have given direct evidence of
carbon-enriched stellar wind emanating in form of dust shells from
the orbiting Wolf-Rayet binary WR 140.\citep{Lieb+25/01} The dust
shells are propagating outward with a measured velocity consistent
with the theoretical wind speed and the distance between the shells
correlates with the period of maximum dust production by the binary,
which is associated with the orbital period of the pair.

The dust contained in a space plasma environment acquires an electric
charge through a number of mechanisms, the most important of which
are: absorption of plasma particles due to inelastic collisions, photoionization,
secondary electron emission and field emission. Refs. \onlinecite{KimuraMann98/05,ShuklaMamun02,Fortov+05/12,Ignatov09/09,GalvaoZiebell12/09,Mann+14/03,Ibanez-Mejia+19/05,Kodanova+19/07,Masheyeva+22/11,Sterken+22/12,GodenkoIzmodenov23/12},
among others, give detailed account on the most important charging
mechanisms for space plasmas. After acquiring electric charge, the
dust grains start to interact with the plasma particles not only through
collisions, but also via electromagnetic fields. This complex system
is then referred to as ``dusty plasma''.

Although the dust grains in space and astrophysical plasmas are almost
always subjected to more than one of the charging processes mentioned
above, each with varying degree of importance on the final charge
due to circumstances such as density and temperature of the plasma,
magnetic field, ultraviolet flux intensity and chemical composition,
a detailed description of each process is a complex task that demands
expertise not only on plasma physics, but also on quantum physics
and materials research. Moreover, the final result can mask the detailed
processes that take place in a particular charging mechanism. Therefore,
in the present series of two papers we investigate solely the collisional
charging of dust grains immersed in Kappa plasmas, in order to stress
the particular physics involved in this process.

One important effect due to the presence of dust in a plasma is related
to the propagation of electrostatic or electromagnetic waves and their
interaction with the plasma particles and dust grains. The presence
of dust gives rise to entirely new wave modes, which are of extremely
low frequency, on the order of the dust plasma or dust cyclotron frequencies\@.\citep{DAngelo90/09,Amin96/09,ShuklaMamun02,BalukuHellberg15/08,Bhakta+17/02,Momot+18/07}
Moreover, the dust component also affects the usual wave modes propagating
in the plasma, particularly the Alfvén waves.\citep{deJuliSchneider98/09,Falceta-GoncalvesJatenco-Pereira02/09,deJuli+05/05,VidottoJatenco-Pereira06/03,deJuli+07/02,deJuli+07/10,Gaelzer+09/01,deJuli+09/03,Jatenco-Pereira+14/03,deToniGaelzer21/11,deToni+22/11,deToni+22/05,deToni+24/02}

On the other hand, space plasma environments are usually found in
a turbulent (quasi-) steady state that is far from the thermodynamic
equilibrium\@.\citep{Treumann+04/04,Goldstein+15/04,Howes15/05,Ferriere20/01,Sahraoui+20/12}
Of particular importance for the kinetic theory of space plasmas are
the velocity distribution functions (VDFs) that better describe the
statistical distribution of velocities of the plasma particles. In
recent years, it has become a consensus in the space plasma community
that the VDFs that describe the turbulent state of the plasma have
distinct nonthermal features, such as high-energy tails that follow
a power-law dependency, instead of a Gaussian decay, typical of thermal
plasmas. These distributions are frequently modelled by the so-called
\emph{Kappa} velocity distribution functions and the recent compilations
\onlinecite{Livadiotis17} and \onlinecite{LazarFichtner21} contain
a comprehensive discussion on the properties of Kappa plasmas.

Hence, the propagation and interaction dynamics of waves in dusty
Kappa plasmas can be an important research topic regarding particle
and energy transport in dust-rich environments, such as planetary
rings, magnetospheres of gas giants, star nebulae and the stellar
winds of carbon-type stars. Notwithstanding the potential importance,
relatively few contributions can be found in the literature investigating
the propagation and absorption/amplification of electrostatic/electromagnetic
waves in dusty Kappa plasmas.\citep{Gaelzer+10/09,Rubab+10/10,Deeba+11/07,Galvao+11/12,Galvao+12/12,ShahmansouriTribeche12/11,dosSantos+16/01,dosSantos+17/01,Ziebell+17/10}

In the first paper (hereafter called Paper I),\citep{ZiebellGaelzer24/11-arxiv}
we investigated the case where the plasma particles are described
by standard Kappa distributions, both isotropic and anisotropic. In
the case of anisotropic distributions, we have considered the bi-Kappa
and product-bi-Kappa models. In the present paper (Paper II), we will
consider plasma particles described by \emph{regularized} Kappa distributions,
which are categorized below. As far as we are aware, this is the first
time that such systematic study is reported in the literature.

The plan of the paper is as follows. In section \ref{sec:DRKAP1:Charge-equilibrium-eq}
the charge-equilibrium equation is briefly derived from the orbital
motion limited (OML) theory. In section \ref{sec:DRKAP1:RKD} the
regularized Kappa velocity distribution function is introduced and
some of its properties are discussed. In section \ref{sec:DRKAP1:Charging-RKD-models}
the charge-equilibrium equation is solved for the two different models
of the regularized distribution considered in this Paper. The results
are presented and analyzed in section \ref{sec:DRKAP1:Numerical_solutions},
and in section \ref{sec:DRKAP1:Final_remarks} our conclusions are
drawn.

\section{The charge equilibrium equation\label{sec:DRKAP1:Charge-equilibrium-eq}}

We consider a spherical dust grain with radius $a$ immersed in a
thermal plasma, which acquires a variable charge $q\left(t\right)$
uniformly distributed over its surface. As mentioned in the Introduction,
in this series of two Papers we consider only the collisional charging
mechanism; hence, the dust charge equation is simply $dq/dt=I_{0}\left(q\right)$,
where $I_{0}\left(q\right)$ is the net current on the surface of
the grain when the charge is $q$\@.

Inside the heliosphere, the dust is composed by different materials
and has a distribution of sizes that usually follows a power law.\citep{Mann+11/11,MisraMishra13/07,Sterken+22/12}
Moreover, interplanetary and interstellar dust can also be composed
by irregularly shaped aggregates which could accumulate a large charge-to-mass
ratio due to its porous nature.\citep{Ma+13/02} In the present Paper,
we will assume for simplicity that all grains are spherical with the
same radius. Hence, the charging current $I_{0}\left(q\right)$ is
given, in the nonrelativistic limit, by the integration in velocity
space\citep{deJuliSchneider98/09}
\[
I_{0}\left(q\right)=\sum_{\beta}q_{\beta}\int d^{3}v\,v\sigma_{\beta}\left(q,v\right)f_{\beta0}\left(\boldsymbol{v}\right),
\]
where the summation is performed over the plasma species $\left(\beta=e,i,\dots\right)$,
$\sigma_{\beta}\left(q,p\right)$ is the cross section for dust charging,
obtained from the OML theory,\citep{Allen92/05} given by 
\[
\sigma_{\beta}\left(q,v\right)=\pi a^{2}\left(1-\frac{2qq_{\beta}}{m_{\beta}av^{2}}\right)H\left(1-\frac{2qq_{\beta}}{m_{\beta}av^{2}}\right),
\]
where $H\left(x\right)$ is the Heaviside function and where $f_{\beta0}\left(\boldsymbol{v}\right)$,
the velocity distribution function for the $\beta$-species, is normalized
to its number density $n_{\beta0}$ as 
\[
\int d^{3}v\,f_{\beta0}\left(\boldsymbol{v}\right)=n_{\beta0}.
\]
Moreover, $q_{\beta}$ and $m_{\beta}$ are, respectively, the electric
charge and the mass of the $\beta$-species particle. The dust grains
are assumed to be all spherical with radius $a$.

The OML theory assumes that the plasma particles move in the vicinity
of an immovable grain with a dynamic charge $q$ along collisionless
orbits that satisfy the conservation of mechanical energy and angular
momentum, similar to Rutherford scattering.\citep{Allen92/05,Mann+11/11,Melzer19}
Usually, the plasma is assumed to be in a local (quasi-) stationary
state in such a way that the static VDF $f_{\beta0}\left(\boldsymbol{v}\right)$
can be employed to describe the collisional charging throughout the
dynamical evolution of the charge. Moreover, the effect of the ambient
magnetic field is neglected.

Although it is possible to include both the effects of the magnetic
field and of the plasma fluctuations in the collisional charging process,\citep{SalimullahSandbergShukla03/08}
we will employ the simpler OML approximation. Ref. \onlinecite{ChangSpariosu93/01}
developed a generalization of the OML theory including magnetic field
and concluded that the traditional approach is valid for a magnetized
plasma as long as the dust radius is much smaller than the electron
Larmor radius. For the physical parameters considered in this work,
this condition is always satisfied. More recent particle simulations\citep{Kodanova+19/07,Davari+23/01}
have corroborated this conclusion and shown that although the total
charging time can diminish with the field intensity, the equilibrium
charge varies with the magnetic field around an average value, depending
on the plasma and dust conditions.

The equilibrium condition for the dust-plasma system is obtained by
assuming vanishing charging current over the dust particles. The charging
time (the time for the grain to reach the equilibrium charge) for
interplanetary and interstellar dust can vary from a few seconds to
months depending on the grain composition and radius and its position
inside the heliosphere.\citep{Mann+11/11,Ma+13/02,MisraMishra13/07,Sterken+22/12,GodenkoIzmodenov23/12}
In this work, we will assume that the dust grain is fully charged.
We denote by $q_{d0}=-Z_{d}e$ the dust charge at the equilibrium,
where $Z_{d}$ is the dust charge number and $e$ is the elementary
charge. In the vast majority of cases, $Z_{d}>0$, meaning that the
dust is negatively charged due to collisions with plasma particles.
However, we will show below that under certain circumstances, the
dust can become positively charged (\emph{i.e.}, $Z_{d}<0$) when
immersed in a suprathermal plasma.

The average collision frequency for collisional charging at the equilibrium
is then given by\citep{deJuliSchneider98/09}
\begin{equation}
\nu_{\beta d}=\frac{n_{d0}}{n_{\beta0}}\int d^{3}v\,v\sigma_{\beta}\left(q_{d0},v\right)f_{\beta0}\left(\boldsymbol{v}\right),\label{eq:DRKAP1:Collision_frequency-general}
\end{equation}
where $n_{d0}$ is the number density of dust grains.\textcolor{blue}{{}
}Hence, the condition of zero current on the surface of the dust particle,
$I_{0}\left(q_{d0}=-Z_{d}e\right)=0$, can be written as 
\begin{gather*}
\sum_{\beta}n_{\beta0}q_{\beta}\nu_{\beta d}=0.
\end{gather*}
We must also take into account the charge neutrality condition 
\[
\sum_{\beta}n_{\beta0}q_{\beta}+n_{d0}q_{d0}=0.
\]

Considering now an electron-ion plasma with dust, where the ion charge
is $q_{i}=Ze$ $\left(Z=1,2,\dots\right)$, the charge neutrality
can be written as $Zn_{i0}-n_{e0}-Z_{d}n_{d0}=0$\@. In this case,
the zero-current condition results 
\begin{equation}
Z\nu_{id}\left(Z_{d}\right)-\left(Z-Z_{d}\frac{n_{d0}}{n_{i0}}\right)\nu_{ed}\left(Z_{d}\right)=0.\label{eq:DRKAP1:Charge-equilibrium_condition}
\end{equation}
The solution of (\ref{eq:DRKAP1:Charge-equilibrium_condition}) provides
the dust charge number $Z_{d}$ at equilibrium.

\section{The $\kappa$-regularized velocity distribution function\label{sec:DRKAP1:RKD}}

Space and astrophysical plasmas are frequently observed (or inferred)
to be in a (quasi-) stationary turbulent state that is far from the
thermal equilibrium. When the plasma is in the equilibrium state,
the velocity distribution functions of the particle components are
the well-known Maxwell-Boltzmann distributions (the Maxwellian)\@. 

On the other hand, direct observations performed by space probes in
planetary magnetospheres and in the solar wind have shown that when
the plasma is in the stationary turbulent state, the measured VDFs
are endowed with distinct non-Maxwellian features, particularly with
high-energy tails that follow a power-law dependence on velocity,
instead of the Gaussian decay of a Maxwellian. Due to this particular
nonthermal feature, remarked since the late 1960s in the plasma of
Earth's magnetosphere, the VDFs of space plasmas have been since then
modelled by \emph{Lorentzian} or, more recently, by the \emph{Kappa}
distribution functions\@.\citep{Livadiotis17,LazarFichtner21} Plasmas
in this particular nonthermal equilibrium state are often referred
to in the literature as \emph{superthermal} or \emph{suprathermal}
plasmas. In this work, we will employ the latter denomination throughout.

The first model of a Kappa VDF to appear in a journal publication
was introduced by Ref. \onlinecite{Vasyliunas68/05}\@. Over the
years, the mathematical forms of Kappa VDFs have acquired several
different versions (called here \emph{models}) and the physically
correct version is still a matter of controversy in the literature\@.\citep{Hellberg+09/09,Hau+09/09,Livadiotis17,Scherer+17/12,Fichtner+18/11,LazarFichtner21}
Most of the different models proposed in the literature can be written
in a generalized form introduced by Refs. \onlinecite{GaelzerZiebell14/12,GaelzerZiebell16/02}
and which corresponds to the isotropic version of the distributions
employed in Paper I, 
\begin{equation}
f_{w,\alpha}\left(v\right)=\frac{n_{0}}{\pi^{3/2}w^{3}}\frac{\kappa^{-3/2}\Gamma\left(\sigma\right)}{\Gamma\left(\sigma-\nicefrac{3}{2}\right)}\left(1+\frac{v^{2}}{\kappa w^{2}}\right)^{-\sigma},\label{eq:DRKAP1:SKD}
\end{equation}
where $\Gamma\left(z\right)$ is the Gamma function,\citealp{AskeyRoy-Full-NIST10}
$\sigma=\kappa+\alpha$, with $\kappa$ being the kappa index that
gives the VDF its name and $\alpha>0$ is a free parameter (usually
integer)\@. The quantity $w=w\left(\kappa\right)$ has the dimension
of velocity\@. Choosing different values and mathematical forms for
$\alpha$ and $w\left(\kappa\right)$, different models found in the
literature can be reproduced by (\ref{eq:DRKAP1:SKD}), which will
be called here the \emph{standard Kappa distribution} (SKD).

Since the initial proposal, the Kappa distribution (\ref{eq:DRKAP1:SKD}),
with particular values for the parameters $\alpha$ and $w$, and
its anisotropic generalization, the bi-kappa distribution function,\citep{Gaelzer+16/06}
have been consistently employed to fit observed VDFs and other related
physical quantities in various space ambients, such as the solar wind.\citep{Maksimovic+05/09,Stverak+08/03,Stverak+09/05,WilsonIII+19/07,WilsonIII+22/11}
The SKD was also employed to estimate the evolution of the proton
temperature and the spectral index of energy distribution of energetic
neutral atoms (ENAs) in the inner heliosheath from sky maps observed
by the Interstellar Boundary Explorer spacecraft.\citep{Livadiotis+22/10}

Notwithstanding the frequent use of the SKD model to interpret observed
space and astrophysical data, some inconsistencies have been noticed,
both in the observational and theoretical fronts.\textcolor{blue}{{}
}One important restriction of the SKD is that it can only be defined
for $\sigma>\nicefrac{3}{2}$ and higher-order moments of (\ref{eq:DRKAP1:SKD})
impose even more restrictive conditions. For instance, the second
moment, which measures the dispersion of particle velocities around
the mean, and hence is related to the concept of the temperature of
the plasma species, only exists for $\sigma>\nicefrac{5}{2}$\@.
For values of $\sigma$ below these constraints the VDF moments diverge.
Such drawbacks of the SKD have created difficulties for certain applications
of the model, such as an hydrodynamic formulation of a Kappa plasma
or small-Larmor-radius expansions to study the dynamics of kinetic
Alfvén waves.\citep{GaelzerZiebell14/12,LazarFichtner21}

One observational inconsistency is clearly shown by the energy spectral
index of the ENAs in the heliosheath. Using the SKD distribution (with
$\alpha=1$) to fit the observed energy distribution, the spectral
index agrees with the measured value of the kappa index only for $\kappa>\nicefrac{3}{2}$\@.
For smaller values the values disagree because the mean kinetic energy
derived from the SKD is not defined for $\kappa<\nicefrac{3}{2}$.\citep{Livadiotis+22/10}

In order to correct these deficiencies and find a divergence-free
formulation for a suprathermal plasma, a recent model for a nonequilibrium
stationary VDF that is valid for any $\kappa>0$ and contains all
the moments of the distribution, the $\kappa$-\emph{regularized}
velocity distribution function (RKD), was proposed initially by Ref.
\onlinecite{Scherer+17/12} and was later cast in a more generalized
form by Ref. \onlinecite{Scherer+20/07}\@. Here we will employ the
notation introduced by Ref. \onlinecite{Gaelzer+24/07} for the generalized
isotropic RKD, given by 
\[
\begin{aligned}f_{\kappa,\theta}^{(\eta,\zeta,\mu)}\left(v\right) & =n_{0}N_{\kappa,\theta}^{(\eta,\zeta,\mu)}\left(1+\frac{v^{2}}{\eta\theta^{2}}\right)^{-\zeta}e^{-\mu v^{2}/\theta^{2}},\end{aligned}
\]
where 
\begin{align*}
\frac{1}{N_{\kappa,\theta}^{(\eta,\zeta,\mu)}} & =\left(\pi\eta\right)^{3/2}\theta^{3}U\left(\frac{3}{2},\frac{5}{2}-\zeta,\mu\eta\right)\\
 & =\left(\frac{\pi}{\mu}\right)^{3/2}\left(\mu\eta\right)^{\zeta}\theta^{3}U\left(\zeta,\zeta-\frac{1}{2},\mu\eta\right)
\end{align*}
is the normalization constant and $U\left(a,b,z\right)$ is one of
the solutions of Kummer's equation\citealp{Daalhuis-Full-NIST10a}
(also known as the Tricomi function).

The distribution $f_{\kappa,\theta}^{(\eta,\zeta,\mu)}\left(v\right)$
contains the free dimensionless parameters $\eta=\eta\left(\kappa\right)$,
$\zeta=\zeta\left(\kappa\right)$, and $\mu=\mu\left(\kappa\right)$,
which must be such that the RKD exists for $\kappa>0$, and also $\theta=\theta\left(\kappa\right)$,
which must be related to the second moment of the distribution via
some form of ``thermal'' velocity and have the dimension of velocity. 

Notice that all free parameters can be functions of the kappa index,
with the proviso that the usual limit $\kappa\to\infty$ must necessarily
recover the Maxwellian. Other limits may also reproduce other model
VDFs employed in the literature. Finally, the quantity $\mu\left(\kappa\right)$
is called the \emph{regularization parameter} and it is assumed that
$0\leqslant\mu\ll1$, in such a way that the RKD reduces to the SKD
in the limit $\mu=0$ and is close to the latter in the low- and medium-energy
ranges. It is only in the very-high-energy range that the effect of
the Gaussian kicks in and regularizes the distribution. The value
of $\mu$ can also be judiciously chosen so as to minimize the effects
of nonphysical super-relativistic particles in a nonrelativistic formulation
or numerical simulation.

Since the original proposition, the RKD has been applied to reevaluations
of transport coefficients,\citep{Husidic+22/03} solar wind observations,\citep{Scherer+22/07,Pierrard+23/09}
Harris current sheet,\citep{Hau+23/10} and dusty plasmas.\citep{Liu24/02}

It is also important to remark here that the RKD is not merely a mathematical
model. It was shown that the $\kappa$-regularized distribution restores
the property of additivity of the Boltzmann-Gibbs entropy formula
for a continuous system\citep{Fichtner+18/11} and that the RKD is
a self-consistent solution for the quasi-equilibrium state between
Langmuir fluctuations and suprathermal electrons.\citep{Yoon+18/12,Yoon21}

With the previous observations in mind, we will adopt hereafter the
expressions 
\begin{align*}
\eta & =\kappa, & \zeta & =\sigma, & \sigma & =\kappa+\alpha, & \mu & =\mu_{0}\sech\left(\frac{\kappa}{\kappa_{0}}\right),
\end{align*}
where $\left\{ \alpha,\mu_{0},\kappa_{0}\right\} $ are new free parameters
$\left(\mu_{0}\ll1\right)$, and write $f_{\kappa,\theta}^{(\kappa,\sigma,\mu)}\left(v\right)\equiv f_{\kappa,\theta}^{(\alpha)}\left(v\right)$,
where 
\begin{equation}
f_{\kappa,\theta}^{(\alpha)}\left(v\right)=n_{0}N_{\kappa,\theta}^{(\kappa,\sigma,\mu)}\left(1+\frac{v^{2}}{\kappa\theta^{2}}\right)^{-\sigma}e^{-\mu v^{2}/\theta^{2}},\label{eq:DRKAP1:RKD-general}
\end{equation}
and where 
\begin{align*}
\frac{1}{N_{\kappa,\theta}^{(\kappa,\sigma,\mu)}} & =\left(\pi\kappa\right)^{3/2}\theta^{3}U\left(\frac{3}{2},\frac{5}{2}-\sigma,\mu\kappa\right)\\
 & =\left(\frac{\pi}{\mu}\right)^{3/2}\left(\mu\kappa\right)^{\sigma}\theta^{3}U\left(\sigma,\sigma-\frac{1}{2},\mu\kappa\right).
\end{align*}

The particular form adopted for the RKD in (\ref{eq:DRKAP1:RKD-general})
is still sufficiently general to encompass several different models.
For instance, setting the regularization parameter $\mu_{0}=0$ and
calling $\theta\left(\kappa\right)=w\left(\kappa\right)$, the RKD
reduces to (\ref{eq:DRKAP1:SKD})\@. Here, the normalization constant
$N_{\kappa,\theta}^{(\kappa,\sigma,\mu)}$ reduces to the expression
in (\ref{eq:DRKAP1:SKD}) by applying the limiting cases for the Tricomi
function discussed in Appendix \ref{sec:DRKAP1:Limiting_cases}.

On the other hand, in the limit $\kappa\to\infty$ the RKD transitions
smoothly to the Maxwellian, 
\[
\lim_{\kappa\to\infty}f_{\kappa,\theta}^{(\alpha)}\left(v\right)=f_{\mathrm{M}}\left(v\right)=n_{0}\frac{e^{-v^{2}/v_{T}^{2}}}{\pi^{3/2}v_{T}^{3}}.
\]
In order to obtain this limit, one must first notice that $\kappa\mu\left(\kappa\right)\xrightarrow{\kappa\to\infty}0$,
employ the Stirling formula for the Gamma functions\citealp{AskeyRoy-Full-NIST10}
and the exponential limit $\left(1+y^{2}/\kappa\right)^{-\kappa}\xrightarrow{\kappa\to\infty}e^{-y^{2}}$\@.
Moreover, the quantity $\theta\left(\kappa\right)$ must be such that
\[
\lim_{\kappa\to\infty}\theta\left(\kappa\right)=\lim_{\kappa\to\infty}w\left(\kappa\right)=v_{T}=\sqrt{\frac{2T}{m}},
\]
where $v_{T}$ is the thermal velocity and the temperature is evaluated
in energy units.

The explicit expressions for the parameter $\theta\left(\kappa\right)$,
or $w\left(\kappa\right)$, will be given by two different models,
the same models considered by Paper I and which will be discussed
below. These models are concerned with the measure of the second moment
of the RKD and by the definition of the ``temperature'' of a suprathermal
plasma.

The kinetic temperature obtained from a generic VDF is 
\[
T_{K}=\frac{1}{3}m\left\langle v^{2}\right\rangle ,
\]
where 
\[
\left\langle v^{2}\right\rangle =\frac{1}{n_{0}}\int d^{3}v\,v^{2}f_{\alpha0}\left(\boldsymbol{v}\right)
\]
is the second moment of the distribution.

Evaluating $T_{K}$ from the SKD (\ref{eq:DRKAP1:SKD}), we obtain
\begin{equation}
T_{K}=\frac{1}{2}m\frac{\kappa w^{2}}{\kappa+\alpha-\nicefrac{5}{2}},\label{eq:DRKAP1:Kinetic_temperature-SKD}
\end{equation}
which shows that a physically relevant expression for the kinetic
temperature can only be obtained for $\kappa+\alpha>\nicefrac{5}{2}$.

For one of the most employed models, the kinetic temperature is defined
in such a way that it is the same measure of the second moment, regardless
of the particular value of the $\kappa$ index. The idea is that the
temperature is a measure of the thermality of the stationary system
and it must be the same regardless whether the system is in thermal
equilibrium or in a nonequilibrium steady state. In this way, independent
measures of the temperature, either as the second moment of the VDF
or as an intensive parameter obtained from the entropy of a nonequilibrium
system are the same and can be considered as a ``physical'' temperature.\citep{LivadiotisMcComas09/11,Livadiotis17}

For this model, it has been shown\citep{LivadiotisMcComas09/11,Livadiotis17}
that using the SKD (\ref{eq:DRKAP1:SKD}) with $\alpha=1$ and $w^{2}\left(\kappa\right)=\left(\kappa-\nicefrac{3}{2}\right)v_{T}^{2}/\kappa$,
there results that $T=T_{K}$\@. Moreover, the Kappa distribution
in this model maximizes the Tsallis entropy $S_{\kappa}$ (in terms
of the $\kappa$ index) within the constraints of a canonical ensemble
and, consequently, can be evaluated also as $T^{-1}=\partial S_{\kappa}/\partial E$,
where $E$ is the internal energy of the system. More recently, the
concept of ``entropy deficiency'' has been proposed, based on the
additivity formula satisfied by the Tsallis entropy, $S_{A+B}=S_{A}+S_{B}-\kappa^{-1}S_{A}S_{B}$,
where $A$ and $B$ are two independent statistical systems.\citep{LivadiotisMcComas21/12,LivadiotisMcComas22/11}
Using this concept, it was shown that this form of the SKD is the
only one that can satisfy Tsallis's additivity formula.\citep{LivadiotisMcComas24/05}

This model is completely self-consistent with Tsallis's entropic principle
of nonadditive, nonequilibrium statistical mechanics and, for this
reason, has been extensively employed in the literature.\citep{Livadiotis17}

For the other model considered in this work, the parameter $w$ in
(\ref{eq:DRKAP1:SKD}) is a constant, given by $w^{2}=2T_{\kappa}/m$,
where $T_{\kappa}$ is some measure of the second moment.\citep{Vasyliunas68/05,Leubner02/11}
In this case, the kinetic temperature (\ref{eq:DRKAP1:Kinetic_temperature-SKD})
becomes kappa-dependent\@. This form of the SKD is sometimes called
``Olbertian'' and has also been extensively used (usually with $\alpha=1$)
in the literature,\citep{Lazar+17/06,LazarFichtner21,Eyelade+21/04,Kirpichev+21/10,Zheng+24/12}
since clear correlations of the measured temperature (and related
parameters) of the plasma species with the measured kappa index were
found in the solar wind,\citep{Lazar+17/06} the Earth's magnetosphere,\citep{Eyelade+21/04,Kirpichev+21/10}
and in the inner heliosheath.\citep{Livadiotis+22/10}

One of the objectives of this Paper II is to analyse the differences
on the charging of the dust grains using both interpretations of the
kinetic temperature, but now for the regularized Kappa distribution.
Accordingly,\textcolor{blue}{{} }evaluating now $\left\langle v^{2}\right\rangle $
with the RKD $f_{\kappa,\theta}^{(\alpha)}\left(v\right)$ given by
(\ref{eq:DRKAP1:RKD-general}), we arrive at the integral
\[
\left\langle v^{2}\right\rangle =4\pi N_{\kappa,\theta}^{(\kappa,\sigma,\mu)}\int_{0}^{\infty}dv\,v^{4}\left(1+\frac{v^{2}}{\kappa\theta^{2}}\right)^{-\sigma}e^{-\mu v^{2}/\theta^{2}}.
\]
Upon defining the new integration variable $t=v^{2}/\kappa\theta^{2}$,
we can identify the integral identity (\ref{eq:DRKAP1:Tricomi-Integ_rep-1})
and then obtain 
\[
\left\langle v^{2}\right\rangle =\frac{3}{2}\mu^{-1}\frac{U\left(\sigma,\sigma-\nicefrac{3}{2},\mu\kappa\right)}{U\left(\sigma,\sigma-\nicefrac{1}{2},\mu\kappa\right)}\theta^{2},
\]
where we have also used the Kummer identity (\ref{eq:DRKAP1:Kummer_transformation}).

\begin{subequations}
\label{eq:DRKAP1:U_scr}

A more compact notation can be achieved if we introduce the symbols
\begin{align}
\Uratio mn\left(\eta,\zeta,\mu\right) & =\frac{\left(\mu\eta\right)^{-m/2}U\left(\zeta,\zeta-\frac{1+m}{2},\mu\eta\right)}{\left(\mu\eta\right)^{-n/2}U\left(\zeta,\zeta-\frac{1+n}{2},\mu\eta\right)},\label{eq:DRKAP1:U_scr^mn-1}\\
\Uratiol m\left(\eta,\zeta,\mu\right) & =\Uratio m0\left(\eta,\zeta,\mu\right),\nonumber \\
\Uratior n\left(\eta,\zeta,\mu\right) & =\Uratio 0n\left(\eta,\zeta,\mu\right),\nonumber 
\end{align}
which were originally defined by Ref. \onlinecite{Scherer+20/07}\@.
In this work, we also introduce the symbol
\begin{equation}
\Uratiodl m\left(\eta,\zeta,\mu,\chi\right)=\frac{U\left(\zeta,\zeta-\frac{1+m}{2},\mu\eta\left(1+\chi/\eta\right)\right)}{\left(\mu\eta\right)^{m/2}U\left(\zeta,\zeta-\frac{1}{2},\mu\eta\right)}e^{-\mu\chi},\label{eq:DRKAP1:Ud_scr^m}
\end{equation}
which is the dusty plasma counterpart to $\Uratiol m\left(\eta,\zeta,\mu\right)$\@.
Obviously, $\Uratiodl m\left(\eta,\zeta,\mu,0\right)=\Uratiol m\left(\eta,\zeta,\mu\right)$\@. 
\end{subequations}

Therefore, the kinetic temperature for a RKD plasma can be written
as 
\begin{equation}
T_{K}=\frac{1}{2}m\kappa\Uratiol 2\left(\kappa,\kappa+\alpha,\mu\right)\theta^{2}.\label{eq:DRKAP1:Kinetic_temperature-RKD}
\end{equation}

The expression for the kinetic temperature given by (\ref{eq:DRKAP1:Kinetic_temperature-RKD})
is important for the following discussion concerning the models adopted
in this work.

\section{The charging of dust grains with different RKD models\label{sec:DRKAP1:Charging-RKD-models}}

The dust charge equilibrium equation (\ref{eq:DRKAP1:Charge-equilibrium_condition})
will be solved for two different models of the $\kappa$-regularized
velocity distribution function. We will first obtain a general expression
for the collision frequencies (\ref{eq:DRKAP1:Collision_frequency-general})
of the dust grains in a electron-ion plasma whose velocity distribution
functions are given by the generalized form (\ref{eq:DRKAP1:RKD-general}).

\subsection{The general form of the collision frequencies}

As the results will show, in extreme cases of suprathermality the
dust grains can acquire a positive net charge, although the charge
will be negative for the vast majority of the cases. Because of this
possibility, we have to solve the integrations in (\ref{eq:DRKAP1:Collision_frequency-general})
taking into account both possible charge signs. Inserting then the
distribution (\ref{eq:DRKAP1:RKD-general}) into (\ref{eq:DRKAP1:Collision_frequency-general}),
we have the collision frequency of the dust grains with the $\beta$-th
plasma species $\left(\beta=e,i\right)$ given by 
\begin{multline*}
\nu_{\beta d}=4\pi^{2}a^{2}n_{d0}N_{\kappa_{\beta},\theta_{\beta}}^{\left(\kappa_{\beta},\sigma_{\beta},\mu_{\beta}\right)}\int_{0}^{\infty}dv\,v\left(v^{2}-V_{\beta}^{2}\right)\\
\times H\left(v^{2}-V_{\beta}^{2}\right)\left(1+\frac{v^{2}}{\kappa_{\beta}\theta_{\beta}^{2}}\right)^{-\sigma_{\beta}}e^{-\mu_{\beta}v^{2}/\theta_{\beta}^{2}},
\end{multline*}
where we have defined the constant 
\[
V_{\beta}^{2}=\frac{2q_{d0}q_{\beta}}{m_{\beta}a},
\]
which would correspond to the square of the velocity acquired at infinity
by a $\beta$-species particle initially at rest at the surface of
a dust grain with charge $q_{d0}$ and radius $a$, if $q_{d0}q_{\beta}>0$\@.
Here, $V_{\beta}^{2}$ is just convenient definition for a real constant
that can have either positive or negative signs. 

Now we have to consider the 2 possible signs of $V_{\beta}^{2}$ separately.

\paragraph{$\boldsymbol{V_{\beta}^{2}<0}$\@.}

In this case, we write $\mathcal{V}_{\beta}^{2}=\left|V_{\beta}^{2}\right|$
and obtain
\[
\nu_{\beta d}=4\pi^{2}a^{2}n_{d0}N_{\kappa_{\beta},\theta_{\beta}}^{\left(\kappa_{\beta},\sigma_{\beta},\mu_{\beta}\right)}\left(I^{(1)}+\mathcal{V}_{\beta}^{2}I^{(0)}\right),
\]
where
\[
I^{(\ell)}=\int_{0}^{\infty}dv\,v^{2\ell+1}\left(1+\frac{v^{2}}{\kappa_{\beta}\theta_{\beta}^{2}}\right)^{-\sigma_{\beta}}e^{-\mu_{\beta}v^{2}/\theta_{\beta}^{2}}.
\]
Changing the integration variable to $v^{2}=\kappa_{\beta}\theta_{\beta}^{2}t$,
we employ again identities (\ref{eq:DRKAP1:Tricomi-Integ_rep-1})
and (\ref{eq:DRKAP1:Kummer_transformation}) and obtain 
\begin{multline*}
\nu_{\beta d}^{\mathrm{(RKD)}}=2\pi^{2}a^{2}n_{d0}\kappa_{\beta}^{\sigma_{\beta}}\theta_{\beta}^{2}\mu_{\beta}^{\sigma_{\beta}-1}N_{\kappa_{\beta},\theta_{\beta}}^{\left(\kappa_{\beta},\sigma_{\beta},\mu_{\beta}\right)}\\
\times\left[\theta_{\beta}^{2}\mu_{\beta}^{-1}U\left(\sigma_{\beta},\sigma_{\beta}-1,\mu_{\beta}\kappa_{\beta}\right)+\mathcal{V}_{\beta}^{2}U\left(\sigma_{\beta},\sigma_{\beta},\mu_{\beta}\kappa_{\beta}\right)\right].
\end{multline*}

\paragraph{$\boldsymbol{V_{\beta}^{2}\geqslant0}$\@.}

In this case, starting with 
\begin{multline*}
\nu_{\beta d}=4\pi^{2}a^{2}n_{d0}N_{\kappa_{\beta},\theta_{\beta}}^{\left(\kappa_{\beta},\sigma_{\beta},\mu_{\beta}\right)}\\
\times\int_{V_{\beta}}^{\infty}dv\,v\left(v^{2}-V_{\beta}^{2}\right)\left(1+\frac{v^{2}}{\kappa_{\beta}\theta_{\beta}^{2}}\right)^{-\sigma_{\beta}}e^{-\mu_{\beta}v^{2}/\theta_{\beta}^{2}},
\end{multline*}
we define the new integration variable
\[
v^{2}=\kappa_{\beta}\theta_{\beta}^{2}\left(\psi_{\beta}t+E_{\beta}\right),
\]
where 
\begin{align*}
E_{\beta} & =\frac{V_{\beta}^{2}}{\kappa_{\beta}\theta_{\beta}^{2}}, & \psi_{\beta} & =1+E_{\beta},
\end{align*}
and then we obtain
\begin{multline*}
\nu_{\beta d}^{\mathrm{(RKD)}}=2\pi^{2}a^{2}n_{d0}N_{\kappa_{\beta},\theta_{\beta}}^{\left(\kappa_{\beta},\sigma_{\beta},\mu_{\beta}\right)}\left(\kappa_{\beta}\theta_{\beta}^{2}\right)^{2}\left(\mu_{\beta}\kappa_{\beta}\right)^{\sigma_{\beta}-2}\\
\times e^{-\mu_{\beta}V_{\beta}^{2}/\theta_{\beta}^{2}}U\left(\sigma_{\beta},\sigma_{\beta}-1,\mu_{\beta}\kappa_{\beta}\left(1+\frac{V_{\beta}^{2}}{\kappa_{\beta}\theta_{\beta}^{2}}\right)\right).
\end{multline*}

\begin{subequations}
\label{eq:DRKAP1:Collision_freqs-general}
\begin{widetext}
The collision frequencies for a species/population whose VDF is a
RKD can be cast in a single expression from the results above as 
\begin{equation}
\nu_{\beta d}^{\left(\kappa_{\beta},\alpha_{\beta},\mu_{\beta}\right)}=2\sqrt{\pi}a^{2}n_{d0}\kappa_{\beta}^{1/2}\theta_{\beta}\times\begin{cases}
{\displaystyle \Uratiol 1\left(\kappa_{\beta},\sigma_{\beta},\mu_{\beta}\right)+\Uratiol{-1}\left(\kappa_{\beta},\sigma_{\beta},\mu_{\beta}\right)\frac{\left|V_{\beta}^{2}\right|}{\kappa_{\beta}\theta_{\beta}^{2}}}, & q_{d0}q_{\beta}<0\\
{\displaystyle \Uratiodl 1\left(\kappa_{\beta},\sigma_{\beta},\mu_{\beta},\frac{V_{\beta}^{2}}{\theta_{\beta}^{2}}\right)}, & q_{d0}q_{\beta}\geqslant0,
\end{cases}\label{eq:DRKAP1:CFG-RKD}
\end{equation}
in terms of the symbols defined in (\ref{eq:DRKAP1:U_scr}a, b).

In the limit $\mu_{\beta0}\to0$, employing again the limiting properties
shown in the Appendix, we have, for $\sigma_{\beta}>\nicefrac{5}{2}$
and $\theta_{\beta}=w_{\beta}$, the collision frequencies for a SKD
plasma, 
\begin{equation}
\nu_{\beta d}^{\left(\kappa_{\beta},\alpha_{\beta}\right)}=2\sqrt{\pi}a^{2}n_{d0}w_{\beta}\frac{\kappa_{\beta}^{1/2}\Gamma\left(\sigma_{\beta}-2\right)}{\Gamma\left(\sigma_{\beta}-\nicefrac{3}{2}\right)}\times\begin{cases}
{\displaystyle 1+\frac{\sigma_{\beta}-2}{\kappa_{\beta}}\frac{\left|V_{\beta}^{2}\right|}{w_{\beta}^{2}},} & q_{d0}q_{\beta}<0\\
{\displaystyle \left(1+\frac{1}{\kappa_{\beta}}\frac{V_{\beta}^{2}}{w_{\beta}^{2}}\right)^{-\left(\sigma_{\beta}-2\right)},} & q_{d0}q_{\beta}\geqslant0,
\end{cases}\label{eq:DRKAP1:CFG-SKD}
\end{equation}
which correspond to the isotropic limit of the models considered in
Paper I.

\end{widetext}

Finally, in the limit $\kappa_{\beta}\to\infty$, we have $\mu_{\beta}\kappa_{\beta}\to0$,
$w_{\beta}=v_{T\beta}$ and then we obtain the collision frequencies
of dust with Maxwellian particles, 
\begin{equation}
\nu_{\beta d}^{\mathrm{(M)}}=2\sqrt{\pi}a^{2}n_{d0}v_{T\beta}\times\begin{cases}
{\displaystyle 1+\frac{\left|V_{\beta}^{2}\right|}{v_{T\beta}^{2}},} & q_{d0}q_{\beta}<0\\
e^{-V_{\beta}^{2}/v_{T\beta}^{2}}, & q_{d0}q_{\beta}\geqslant0.
\end{cases}\label{eq:DRKAP1:CFG-Maxwellian}
\end{equation}

\end{subequations}

We can now propose the following relations and nondimensional quantities
to be used by both models,
\begin{gather*}
\begin{aligned}\tilde{a} & =\frac{a}{\lambda_{i}}, & \epsilon & =\frac{n_{d0}}{n_{i0}}, & \tau_{e} & =\frac{T_{e}}{T_{i}}, & \chi_{e} & =\frac{Z_{d}e^{2}}{aT_{e}}=\frac{Z_{d}}{\tau_{e}\tilde{a}},\end{aligned}
\\
\begin{aligned}v_{Ti}^{2} & =v_{A}^{2}\beta_{i} & v_{Te}^{2} & =\frac{m_{i}}{m_{e}}\tau_{e}v_{A}^{2}\beta_{i}, & \gamma & =\frac{\lambda_{i}^{2}n_{i0}v_{A}}{\Omega_{i}},\end{aligned}
\end{gather*}
where 
\begin{align*}
\lambda_{i} & =\frac{e^{2}}{T_{i}}, & \beta_{i} & =\frac{8\pi n_{i0}T_{i}}{B_{0}^{2}},\\
v_{A} & =\frac{B_{0}}{\sqrt{4\pi n_{i0}m_{i}}}, & \Omega_{i} & =\frac{ZeB_{0}}{m_{i}c},
\end{align*}
are, respectively, the classical distance of minimum approach for
ions, the beta parameter for ions, the Alfvén speed and the ion (angular)
cyclotron frequency.

Now the general form for the collision frequency $\nu_{\beta d}$
will be first written in terms of a normalized frequency $\tilde{\nu}_{\beta d}$,
as $\nu_{\beta d}=\Omega_{i}\tilde{\nu}_{\beta d}$, which in turn
will be written in terms of a collision frequency factor $\hat{\nu}_{\beta d}$,
where 
\begin{align*}
\tilde{\nu}_{\beta d} & =C\hat{\nu}_{\beta d}, & C & =2\sqrt{\pi}\gamma\epsilon\tilde{a}^{2}\beta_{i}^{1/2}.
\end{align*}

In this way, the equilibrium charge equation (\ref{eq:DRKAP1:Charge-equilibrium_condition})
can be written as an equation for the normalized potential $\chi_{e}$
in terms of the frequency factors as 
\begin{equation}
\left(Z-\tilde{a}\tau_{e}\epsilon\chi_{e}\right)\hat{\nu}_{ed}\left(\chi_{e}\right)-Z\hat{\nu}_{id}\left(\chi_{e}\right)=0,\label{eq:DRKAP1:Charge-equilibrium_condition-normalized}
\end{equation}
where expressions for the collision frequencies from different VDF
models can be introduced.

\subsection{Model 1\@. Kappa-dependent temperature\label{subsec:DRKAP1:Model-1-Kappa-dependent}}

The first model to be considered for the RKD assumes that the parameter
$\theta$ for the $\beta$-species is a constant, given by $\theta_{\beta}^{2}=v_{T\beta}^{2}=2T_{\beta}/m_{\beta}$,
where $T_{\beta}$ is a measure of the second moment of the VDF\@.
With this model, the kinetic temperature given by (\ref{eq:DRKAP1:Kinetic_temperature-RKD})
is not $\kappa$-invariant. In fact, 
\[
T_{K}\left(\kappa\right)=\kappa\Uratiol 2\left(\kappa,\sigma,\mu\right)T.
\]

\textcolor{red}{}

\begin{widetext}
\begin{subequations}
\label{eq:DRKAP1:Collision_freqs-electron-ion-RKD-gen-M2}

In this model, the collision factors for a RKD plasma, from (\ref{eq:DRKAP1:CFG-RKD}),
become 
\begin{align}
\hat{\nu}_{id}^{\left(\kappa_{i},\alpha_{i},\mu_{i}\right)} & =\kappa_{i}^{1/2}\times\begin{cases}
{\displaystyle \Uratiol 1\left(\kappa_{i},\sigma_{i},\mu_{i}\right)+Z\frac{\tau_{e}\chi_{e}}{\kappa_{i}}\Uratiol{-1}\left(\kappa_{i},\sigma_{i},\mu_{i}\right)}, & \chi_{e}\geqslant0\\
\Uratiodl 1\left(\kappa_{i},\sigma_{i},\mu_{i},-Z\tau_{e}\chi_{e}\right), & \chi_{e}<0
\end{cases}\\
\hat{\nu}_{ed}^{\left(\kappa_{e},\alpha_{e},\mu_{e}\right)} & =\sqrt{\frac{m_{i}}{m_{e}}\tau_{e}\kappa_{e}}\times\begin{cases}
\Uratiodl 1\left(\kappa_{e},\sigma_{e},\mu_{e},\chi_{e}\right), & \chi_{e}\geqslant0\\
{\displaystyle \Uratiol 1\left(\kappa_{e},\sigma_{e},\mu_{e}\right)-\frac{\chi_{e}}{\kappa_{e}}\Uratiol{-1}\left(\kappa_{e},\sigma_{e},\mu_{e}\right)}, & \chi_{e}<0.
\end{cases}
\end{align}
\end{subequations}
\end{widetext}

\begin{subequations}
\label{eq:DRKAP1:Collision_freqs-electron-ion-M}

We will also need the collision factors for a Maxwellian plasma, which
will be given from (\ref{eq:DRKAP1:CFG-Maxwellian}) as
\begin{align}
\hat{\nu}_{id}^{\mathrm{(M)}} & =\begin{cases}
{\displaystyle 1+Z\tau_{e}\chi_{e},} & \chi_{e}\geqslant0\\
e^{Z\tau_{e}\chi_{e}}, & \chi_{e}<0
\end{cases}\\
\hat{\nu}_{ed}^{\mathrm{(M)}} & =\sqrt{\frac{m_{i}}{m_{e}}\tau_{e}}\times\begin{cases}
e^{-\chi_{e}}, & \chi_{e}\geqslant0\\
{\displaystyle 1-\chi_{e},} & \chi_{e}<0.
\end{cases}
\end{align}
\end{subequations}

In the next section, expressions (\ref{eq:DRKAP1:Collision_freqs-electron-ion-RKD-gen-M2}a,
b) or (\ref{eq:DRKAP1:Collision_freqs-electron-ion-M}a, b), depending
on the suprathermality state of either species, will be introduced
into the charge equation (\ref{eq:DRKAP1:Charge-equilibrium_condition-normalized})
in order to obtain the equilibrium dust charge for this model.

\subsection{Model 2\@. Kappa-invariant temperature\label{subsec:DRKAP1:Model-2-Kappa-independent}}

In this model, the kinetic temperature will be evaluated in such a
way that is kappa-invariant. If we want the kinetic temperature independent
of $\kappa$ (and also of $\alpha$ and $\mu$), we must have from
(\ref{eq:DRKAP1:Kinetic_temperature-RKD}), 
\[
T_{K}=\frac{1}{2}m\kappa\Uratiol 2\left(\kappa,\kappa+\alpha,\mu\right)\theta^{2}=\frac{1}{2}mv_{T}^{2}=T,
\]
where $T$ is the physical temperature. Therefore, we have the $\kappa$-dependent
parameter 
\[
\theta^{2}\left(\kappa\right)=\kappa^{-1}\Uratior 2\left(\kappa,\kappa+\alpha,\mu\right)v_{T}^{2},
\]
according to the symmetry property (\ref{eq:DRKAP1:U_src-permutation})\@.
The quantity $\theta\left(\kappa\right)$ measures the second moment
of the RKD in model 2, and on the limit $\mu_{0}\to0$ it reduces
to the SKD value 
\[
\theta^{2}\left(\kappa\right)=w^{2}\left(\kappa\right)=\frac{\kappa+\alpha-\nicefrac{5}{2}}{\kappa}v_{T}^{2},
\]
which imposes the constraint $\kappa+\alpha>\nicefrac{5}{2}$ on the
possible kappa values.

\textcolor{red}{}

On the other hand, for a plasma species described by the RKD, any
$\kappa>0$ is possible. However, a $\kappa$-invariant temperature
is only meaningful for integer $\alpha=0,1$, because although $\left\langle v^{2}\right\rangle $
exists for any $\alpha$ and $\kappa$, for a finite $\theta$ we
have: 
\begin{align*}
\alpha=0: &  & \lim_{\kappa\to0}\left\langle v^{2}\right\rangle = & \frac{3}{2}\mu_{0}^{-1}\theta^{2}\\
\alpha=1: &  & \lim_{\kappa\to0}\left\langle v^{2}\right\rangle = & \frac{1}{2}\mu_{0}^{-1}\theta^{2}\\
\alpha\geqslant2: &  & \lim_{\kappa\to0}\left\langle v^{2}\right\rangle = & 0.
\end{align*}
Hence, for $\alpha\geqslant2$ the second moment is zero and the measure
of the physical temperature become meaningless.
\begin{widetext}
\begin{subequations}
\label{eq:DRKAP1:Collision_freqs-electron-ion-RKD-gen-M1}

Now, for this model, the collision factors (\ref{eq:DRKAP1:CFG-RKD})
become
\begin{align}
\hat{\nu}_{id}^{\left(\kappa_{i},\alpha_{i},\mu_{i}\right)} & =\sqrt{\phi_{i}}\times\begin{cases}
{\displaystyle \Uratiol 1\left(\kappa_{i},\sigma_{i},\mu_{i}\right)+Z\frac{\tau_{e}\chi_{e}}{\phi_{i}}\Uratiol{-1}\left(\kappa_{i},\sigma_{i},\mu_{i}\right)}, & \chi_{e}\geqslant0\\
{\displaystyle \Uratiodl 1\left(\kappa_{i},\sigma_{i},\mu_{i},-\frac{Z\kappa_{i}\tau_{e}}{\phi_{i}}\chi_{e}\right)}, & \chi_{e}<0
\end{cases}\\
\hat{\nu}_{ed}^{\left(\kappa_{e},\alpha_{e},\mu_{e}\right)} & =\sqrt{\frac{m_{i}}{m_{e}}\tau_{e}\phi_{e}}\times\begin{cases}
{\displaystyle \Uratiodl 1\left(\kappa_{e},\sigma_{e},\mu_{e},\frac{\kappa_{e}}{\phi_{e}}\chi_{e}\right)}, & \chi_{e}\geqslant0\\
{\displaystyle \Uratiol 1\left(\kappa_{e},\sigma_{e},\mu_{e}\right)-\frac{\chi_{e}}{\phi_{e}}\Uratiol{-1}\left(\kappa_{e},\sigma_{e},\mu_{e}\right)}, & \chi_{e}<0,
\end{cases}
\end{align}
where we have also defined the quantity $\phi_{\beta}=\Uratior 2\left(\kappa_{\beta},\sigma_{\beta},\mu_{\beta}\right)$,
for brevity.

\end{subequations}
\end{widetext}

For this model we can also consider the case where one of the species
is Maxwellian, in which case we will employ the factors (\ref{eq:DRKAP1:Collision_freqs-electron-ion-M}a,
b) again.

Some solutions of the charge equation (\ref{eq:DRKAP1:Charge-equilibrium_condition-normalized}),
obtained from both models are discussed in the next section.

\section{Numerical solutions\label{sec:DRKAP1:Numerical_solutions}}

Here we will discuss some solutions of the charge equilibrium equation
(\ref{eq:DRKAP1:Charge-equilibrium_condition-normalized})\@. Solutions
from each of the models introduced in the previous section will be
presented separately.

In order to reduce the quantity of free indices in the analysis, we
will restrict ourselves to the particular case $\alpha_{e}=\alpha_{i}=Z=1$,
with $m_{i}/m_{e}=1836.152673426$\@. Moreover, we will take $\kappa_{0}=50$,
which will render the parameter $\mu\left(\kappa\right)\approx\mu_{0}$
for most of the considered cases, and we will also keep the same physical
parameters $a=\unit[10^{-4}]{cm}$, $n_{i0}=\unit[10]{cm}^{-3}$,
$\epsilon=10^{-5}$, $\beta_{i}=2$, $v_{A}/c=10^{-4}$, and $\tau_{e}=1$
used by Paper I, because we are interested in exploring the effect
of suprathermality on the charging process, and these parameters are
independent of the suprathermal state of the plasma species. Nevertheless,
it is important to point out that these values are typical of stellar
winds from carbon-rich stars.\citealp{Tsytovich+04/10}

Plots of the solutions of the charge equilibrium equation (\ref{eq:DRKAP1:Charge-equilibrium_condition-normalized})
will be presented for the following cases:
\begin{description}
\item [{Case~1}] Ions: Maxwellian $(\hat{\nu}_{id}=\hat{\nu}_{id}^{(\mathrm{M})})$\@.
Electrons: RKD $(\hat{\nu}_{ed}=\hat{\nu}_{ed}^{\left(\kappa_{e},1,\mu_{e}\right)})$.
\item [{Case~2}] Ions: RKD $(\hat{\nu}_{id}=\hat{\nu}_{id}^{\left(\kappa_{i},1,\mu_{i}\right)})$\@.
Electrons: Maxwellian $(\hat{\nu}_{ed}=\hat{\nu}_{ed}^{(\mathrm{M})})$.
\item [{Case~3}] Ions: RKD $(\hat{\nu}_{id}=\hat{\nu}_{id}^{\left(\kappa_{i},1,\mu_{i}\right)})$
(varying $\kappa_{i}$)\@. Electrons: RKD $(\hat{\nu}_{ed}=\hat{\nu}_{ed}^{\left(\kappa_{e},1,\mu_{e}\right)})$
(fixed $\kappa_{e}$).
\item [{Case~4}] Ions: RKD $(\hat{\nu}_{id}=\hat{\nu}_{id}^{\left(\kappa_{i},1,\mu_{i}\right)})$
(varying $\kappa_{i}$)\@. Electrons: RKD $(\hat{\nu}_{ed}=\hat{\nu}_{ed}^{\left(\kappa_{e},1,\mu_{e}\right)})$
(varying $\kappa_{e}$).
\end{description}

Since the results from the RKD asymptotically approach the results
from the SKD as $\kappa$ grows (because $\mu\left(\kappa\right)\to0$),
most of the results will be restricted to the small-kappa range. The
results obtained for large $\kappa$ are essentially the same as those
obtained by the isotropic versions of the SKDs considered in Paper
I\@. In this way, we emphasize the particular behavior of dust charging
by extreme suprathermal plasmas.

\subsection{Model 1}

\begin{figure}
\includegraphics[width=1\columnwidth]{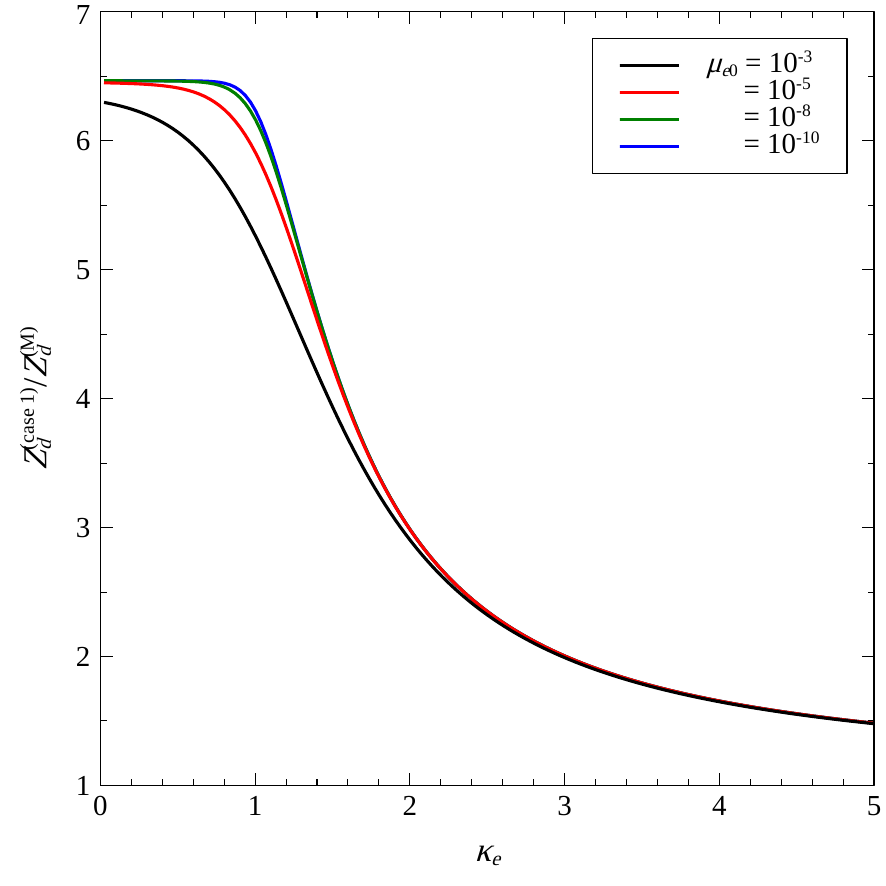}

\caption{Plots of the charge number ratios $Z_{d}^{\mathrm{(case}1)}/Z_{d}^{\mathrm{(M)}}$
for RKD electrons and Maxwellian ions (case 1) as function of $\kappa_{e}$.\label{fig:DRKAP1:Charge_M1-c1}}
\end{figure}

Results from the model 1, introduced in section \ref{subsec:DRKAP1:Model-1-Kappa-dependent},
are shown here. Figure \ref{fig:DRKAP1:Charge_M1-c1} is an example
of solutions of the charge equilibrium equation (\ref{eq:DRKAP1:Charge-equilibrium_condition-normalized})\@.
This plot was obtained considering case 1: Maxwellian ions and RKD
electrons, for different values of the regularization parameter $\mu_{e0}$,
showing the ratio of the dust charge number $Z_{d}^{\mathrm{(case}1)}$
over the charge number of a fully Maxwellian plasma $Z_{d}^{\mathrm{(M)}}$,
which is $Z_{d}^{\mathrm{(M)}}=15466.77$ for the parameters used. 

The results are consistent with the results obtained in Paper I; as
$\kappa_{e}$ diminishes, the ratio $Z_{d}^{\mathrm{(case}1)}/Z_{d}^{\mathrm{(M)}}$
grows, due to the excess of suprathermal electrons. However, differently
from the SKD results, a RKD plasma supports states with $\kappa_{e}<\nicefrac{3}{2}$\@.
When $\kappa_{e}\lesssim\nicefrac{3}{2}$ the ratio starts to saturate.
This happens because of the cut-off imposed by the regularization
parameter $\mu_{0}$ that imposes that high-energy particles are essentially
Gaussian, although they are still present in greater number than in
a Maxwellian plasma. As $\mu_{e0}$ diminishes, the ratio becomes
essentially invariant on the regularization parameter, because the
total number of electrons is always the same.

On the other hand, as $\kappa_{e}\to\infty$, the electronic VDF tends
to the Maxwellian and, consequently, the charge number ratio approaches
the unity.

The behavior shown by Fig. \ref{fig:DRKAP1:Charge_M1-c1} is in accordance
with previous studies about dust charging in suprathermal plasmas
when the electrons are suprathermal and the ions are Maxwellian (see,
\emph{e.g.}, Ref. \onlinecite{Gaelzer+10/09} or Paper I)\@. The
dust charge is enhanced due to the excess of suprathermal electrons,
and the charge is always negative.

A different picture emerges when the ions are more suprathermal than
the electrons. A recent paper (Ref. \onlinecite{Liu24/02}) has put
forth the possibility that in this case the dust charge can become
positive (\emph{i.e.}, $Z_{d}<0$)\@. The results we have obtained
corroborate this conclusion, but with much more stringent conditions
than those obtained by Ref. \onlinecite{Liu24/02}.

\begin{figure}
\includegraphics[width=1\columnwidth]{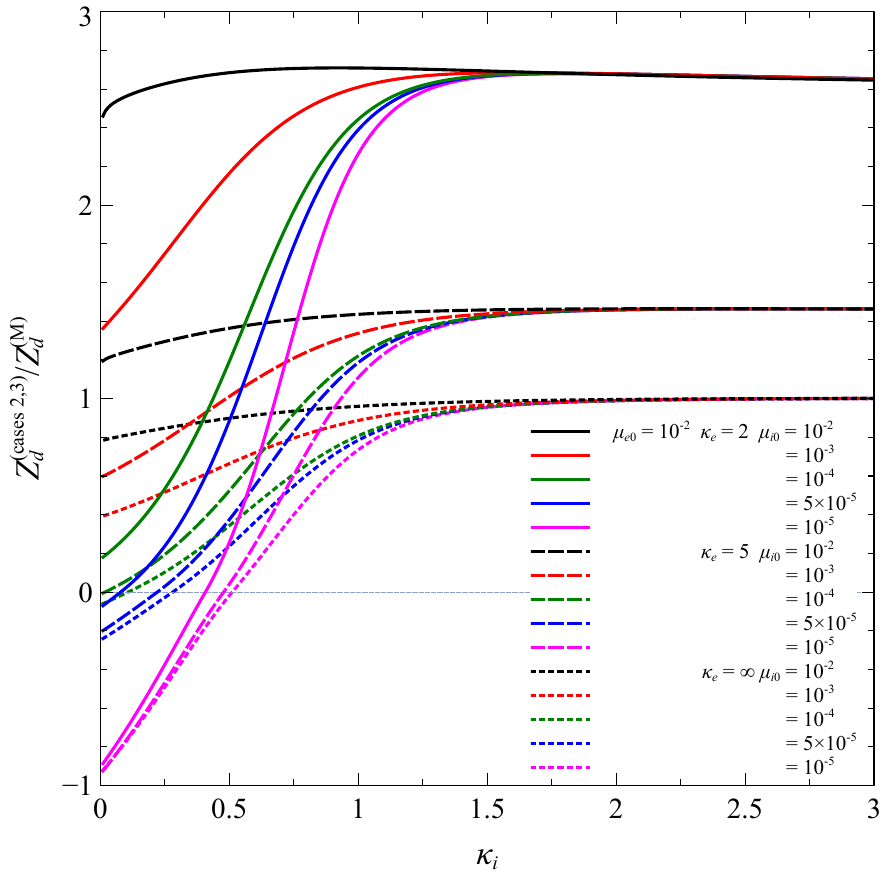}

\caption{Plots of the charge number ratios $Z_{d}^{\text{(cases 2,3)}}/Z_{d}^{\mathrm{(M)}}$
for RKD ions and RKD or Maxwellian electrons (cases 2 and 3) as functions
of $\kappa_{i}$, for $\mu_{i0}=10^{-2}$ (black), $10^{-3}$ (red),
$10^{-4}$ (green), $5\times10^{-5}$ (blue), and $10^{-5}$ (magenta)\@.
Continuous lines: constant $\mu_{e0}=10^{-2}$ and $\kappa_{e}=2$\@.
Dashed lines: same $\mu_{e0}$ and $\kappa_{e}=5$\@. Dotted lines:
$\kappa_{e}\to\infty$ (Maxwellian).\label{fig:DRKAP1:Charge_M1-c2_3}}
\end{figure}

A comprehensive study on the dust charging process with a electron-ion
RKD plasma would in principle demand a complete analysis of the four-dimensional
parameter space $\left\{ \mu_{0e},\kappa_{e},\mu_{0i},\kappa_{i}\right\} $\@.
However, a few choice cases are sufficient for us to reach the relevant
conclusions. Figure \ref{fig:DRKAP1:Charge_M1-c2_3} shows the charge
number ratios for several values of the RKD parameters, corresponding
to cases 2 and 3\@. Continuous lines are obtained for fixed $\mu_{e0}=10^{-2}$
and $\kappa_{e}=2$; dashed lines for the same $\mu_{e0}$ and $\kappa_{e}=5$;
and dotted lines for $\kappa_{e}\to\infty$ (Maxwellian)\@. The color
codes correspond to $\mu_{i0}=10^{-2}$ (black lines), $10^{-3}$
(red), $10^{-4}$ (green), $5\times10^{-5}$ (blue), and $10^{-5}$
(magenta)\@. All results are displayed for the ratios $Z_{d}^{\text{(cases 2,3)}}/Z_{d}^{\mathrm{(M)}}$
as functions of $\kappa_{i}$. 

The first striking behavior displayed in Fig. \ref{fig:DRKAP1:Charge_M1-c2_3}
is that the charge ratio is roughly constant for $\kappa_{i}\gtrsim\nicefrac{3}{2}$
and for any $\mu_{i0}$, irrespective to whether the electrons are
thermal $\left(\kappa_{e}\to\infty\right)$ or suprathermal $\left(\kappa_{e}=2\right)$\@.
Moreover, the overall behavior is similar to case 1: the dust charge
is negative $\left(Z_{d}>0\right)$, with larger charge numbers when
the electrons are more suprathermal.

However, as the ion kappa index approaches the regularized region
$\left(\kappa_{i}\lesssim1\right)$, the behavior starts to change.
The charge ratios split for different values of $\mu_{i0}$ and generally
diminish as $\kappa_{i}\to0$, in a opposite trend to that displayed
in case 1\@. When the electrons are very suprathermal ($\mu_{e0}=10^{-2}$
and $\kappa_{e}=2$), the ratio can still be positive even for $\kappa_{i}=0$,
although with $Z_{d}^{\text{(cases 2,3)}}<Z_{d}^{\mathrm{(M)}}$ in
some extreme cases. In even more extreme cases, one observes the continuous
curves crossing the null-charge boundary for very small values of
$\mu_{0i}$ and $\kappa_{i}$. This trend is more pronounced when
the electrons are moderately suprathermal $\left(\kappa_{e}=5\right)$
and even more pronounced when the electrons are thermal $\left(\kappa_{e}\to\infty\right)$\@.
In all considered cases, the dust charge ends up positive when $\mu_{i0}=10^{-5}$
and $\kappa_{i}\approx0$, with the upper-level case $Z_{d}^{\text{(cases 2,3)}}\approx\,-Z_{d}^{\mathrm{(M)}}$.

Therefore, the dust charge can indeed become positive when the ions
are highly suprathermal. In order to better elucidate this possibility,
we will examine the null-charge condition from Eq. (\ref{eq:DRKAP1:Charge-equilibrium_condition-normalized})\@.
Setting $\chi_{e}=0$ and introducing the collision factors (\ref{eq:DRKAP1:Collision_freqs-electron-ion-RKD-gen-M2}a,
b), the null-charge condition for Model 1 is obtained for a critical
$\kappa_{i}=\kappa_{ic}$, given by 
\begin{multline}
\sqrt{\kappa_{ic}}\Uratiol 1\left(\kappa_{ic},\kappa_{ic}+1,\mu_{i}\right)\\
=\sqrt{\frac{m_{i}}{m_{e}}\tau_{e}\kappa_{e}}\Uratiol 1\left(\kappa_{e},\kappa_{e}+1,\mu_{e}\right).\label{eq:DRKAP1:KIC-2-1}
\end{multline}

\textcolor{red}{}

\begin{figure}
\begin{centering}
\includegraphics[width=0.49\textwidth]{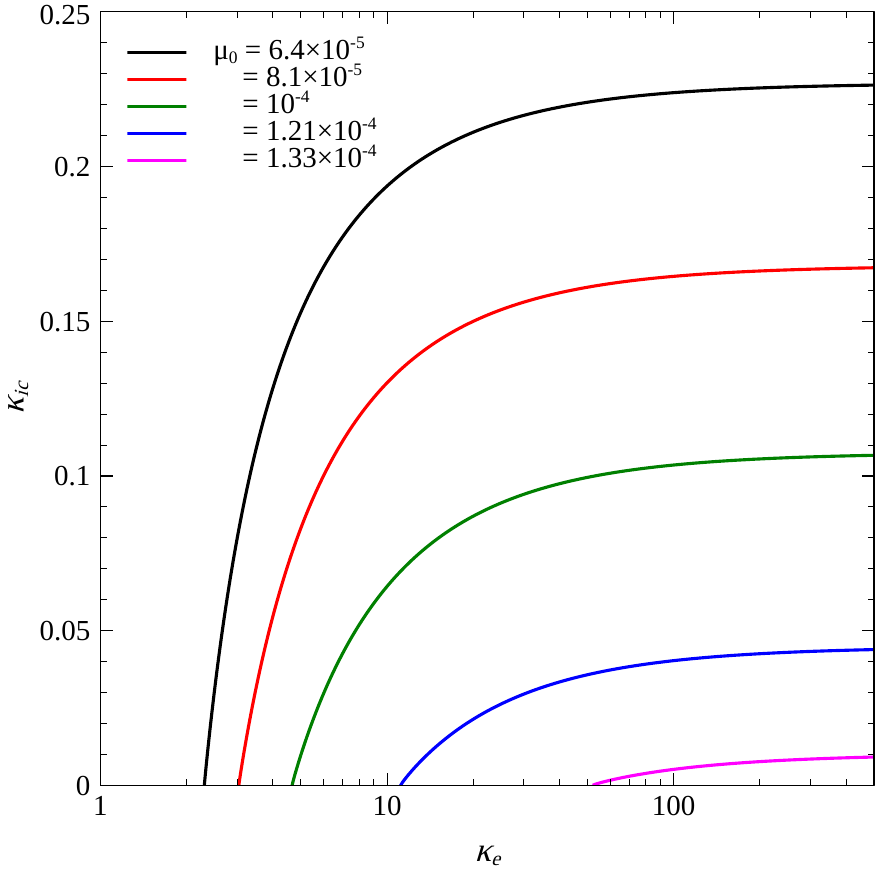}
\par\end{centering}
\caption{Plots of $\kappa_{ic}$, the solution of (\ref{eq:DRKAP1:KIC-2-1}),
as a function of $\kappa_{e}$, for several values of $\mu_{0e}=\mu_{0i}=\mu_{0}$.\label{fig:DRKAP1:kappa_ic-M2}}
\end{figure}

Figure \ref{fig:DRKAP1:kappa_ic-M2} shows some values of $\kappa_{ic}$
as a function of $\kappa_{e}$ for several values of the regularization
parameter $\mu_{e0}=\mu_{i0}=\mu_{0}$\@. For any set of RKD parameters,
the dust will only acquire a positive charge for $\kappa_{i}<\kappa_{ic}$
and, even so, when $\kappa_{e}$ is greater than a critical value
$\kappa_{ec}\gg\kappa_{ic}$\@. Moreover, the value of $\kappa_{ec}$
rapidly increases more than a order of magnitude with a relatively
smaller increase of $\mu_{0}$.

Fig. \ref{fig:DRKAP1:kappa_ic-M2} considers the range of electron
kappa-indices $1\leqslant\kappa_{e}\leqslant500$ and in all considered
cases we obtained $\kappa_{ic}\ll1$\@. Moreover, the results show
that for any $\mu_{0}$, the value of $\kappa_{ic}$ saturates even
when the electrons thermalize $\left(\kappa_{e}\to\infty\right)$\@.
The value of the regularization parameter is also crucial, favoring
the occurrence of positive dust when $\mu_{0}$ is very small, with
this possibility disappearing when $\mu_{0}\gtrsim1.33\times10^{-4}$\@.
This happens because a very small $\mu_{0}$ allows for the presence
of a substantial number of extremely suprathermal ions for a RKD with
$\kappa_{i}\ll1$.

From these results, one concludes that although positively-charged
dust grains are predicted when immersed in a RKD plasma, this can
happen only when there is an extreme difference between the suprathermal
states of ions and electrons, with the ions necessarily being much
more suprathermal than the electrons; the ionic kappa index must be
smaller than $\kappa_{e}$ by at least one order of magnitude. These
results contrast from those obtained by Ref. \onlinecite{Liu24/02},
which predicted a milder discrepancy in the suprathermal states, when
employing the same RKD model.

Such large disparity between the $\kappa_{i}$ and $\kappa_{e}$ indices,
necessary for null-charge, is unlikely to be detected in space and
astrophysical systems. Observations performed in the solar wind\citep{Stverak+09/05,Livadiotis+18/02,Cuesta+24/10}
measured kappa indices $\kappa_{i}\sim\kappa_{e}\lesssim5$, that
are roughly of the same order of magnitude for electrons and protons,
for a large range of solar radii.

One must also point out that the condition $\kappa_{i}\ll\kappa_{e}$
(with $\mu_{i0}=\mu_{e0}$) for $Z_{d}<0$ was obtained using model
1\@. In the next section we will present the results from model 2,
which are substantially different.

\begin{figure*}
\begin{minipage}[t]{0.49\textwidth}%
\includegraphics[width=1\columnwidth]{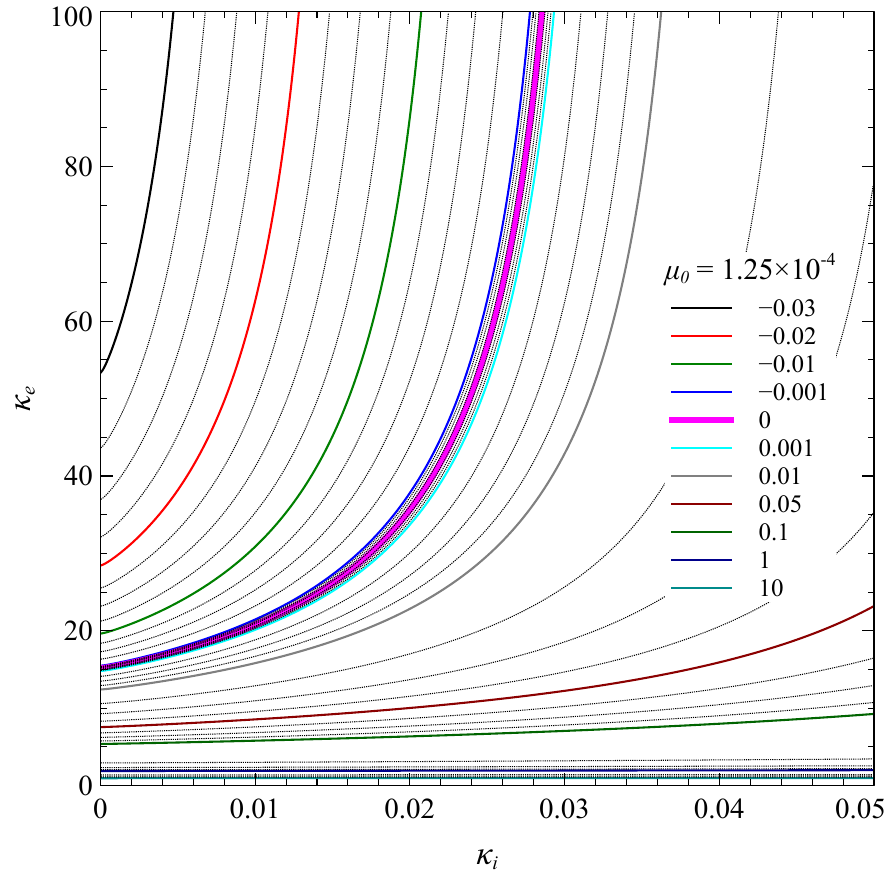}%
\end{minipage}\hfill{}%
\begin{minipage}[t]{0.49\textwidth}%
\includegraphics[width=1\columnwidth]{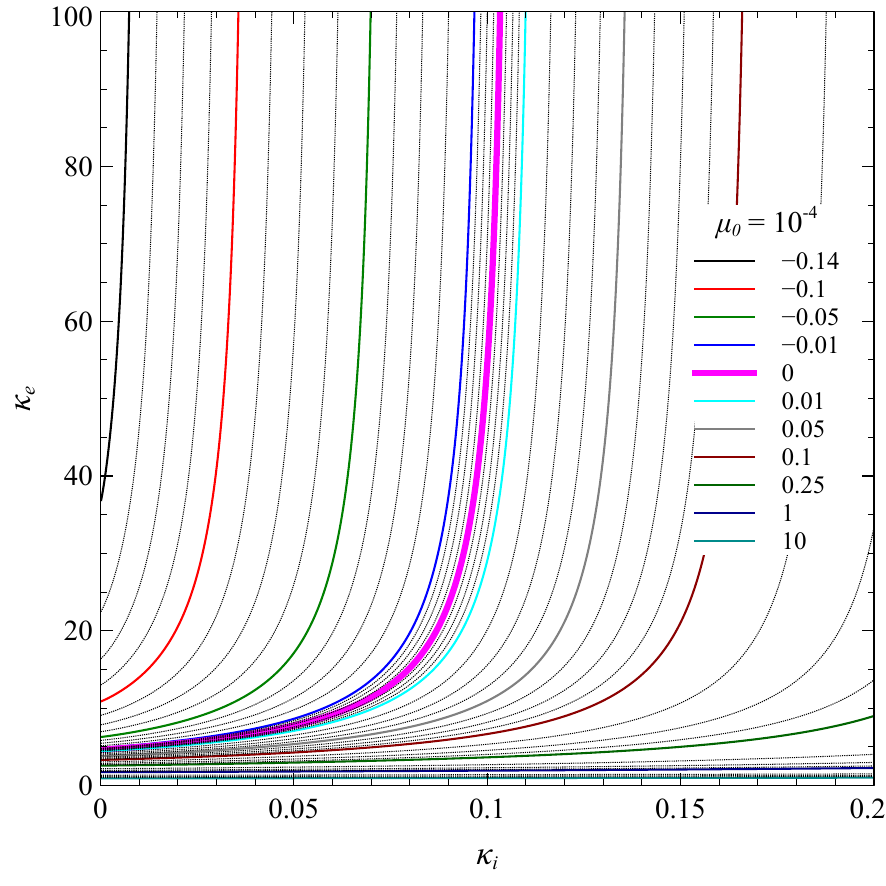}%
\end{minipage}

\begin{minipage}[t]{0.49\textwidth}%
\includegraphics[width=1\columnwidth]{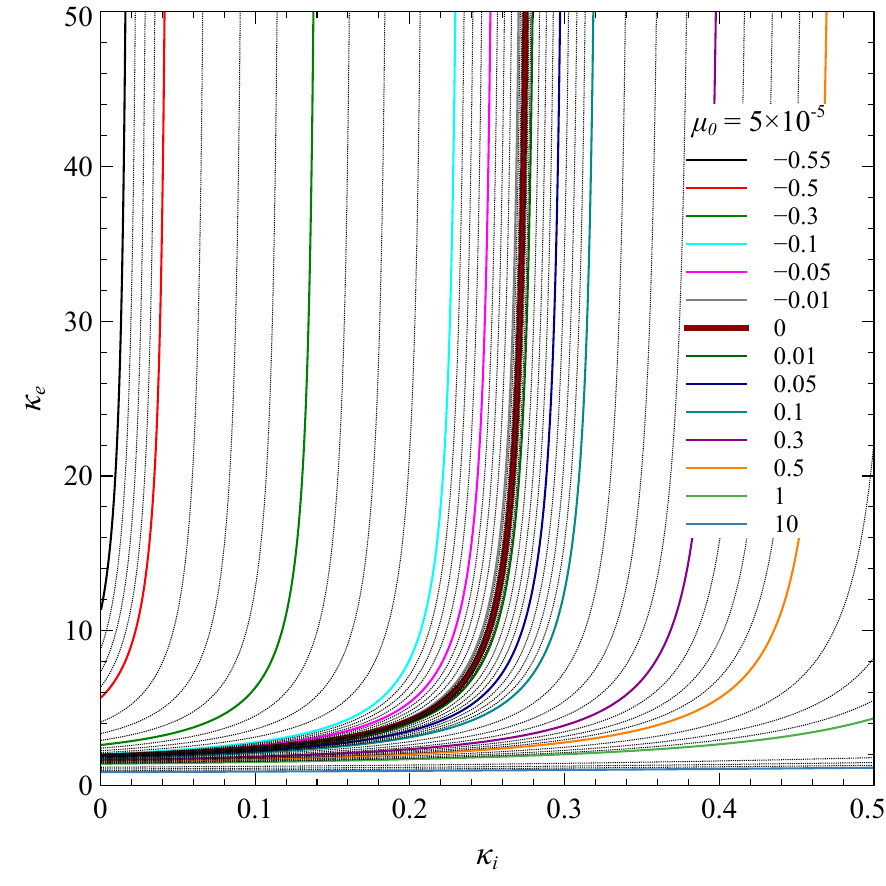}%
\end{minipage}\hfill{}%
\begin{minipage}[t]{0.49\textwidth}%
\includegraphics[width=1\columnwidth]{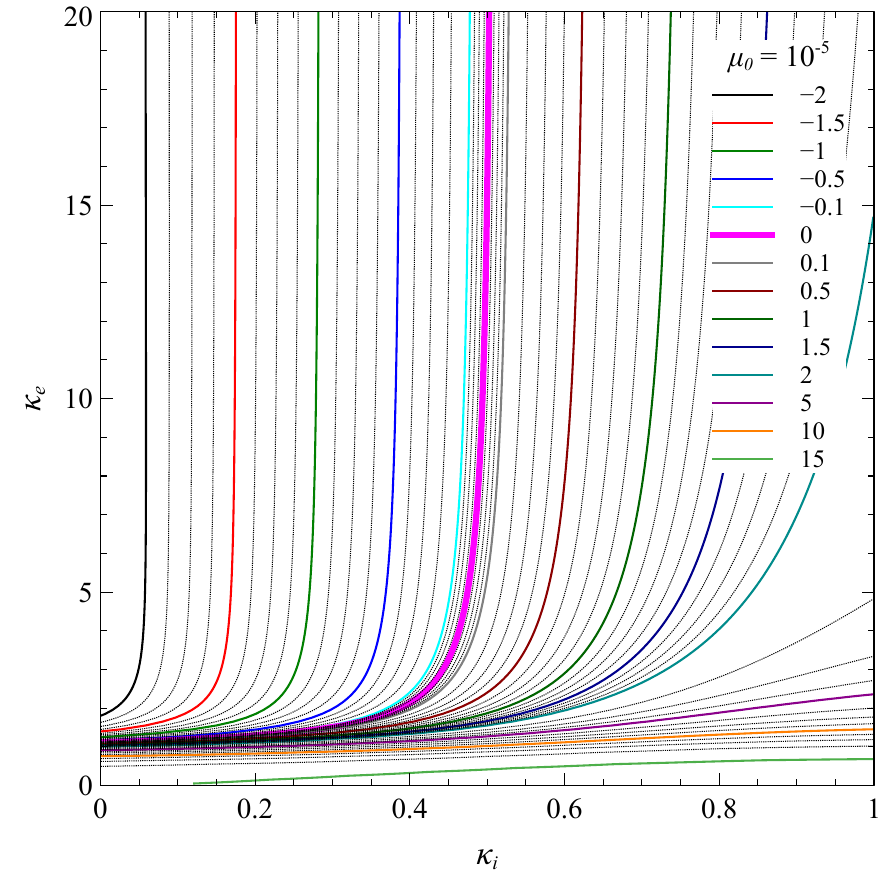}%
\end{minipage}

\caption{Contour plots of the normalized potential $\chi_{e}=Z_{d}/\tau_{e}\tilde{a}$
as function of $\kappa_{e}\times\kappa_{i}$ for several values of
$\mu_{e0}=\mu_{i0}=\mu_{0}$.\label{fig:DRKAP1:Chie_M2}}
\end{figure*}

As final results from model 1, figure \ref{fig:DRKAP1:Chie_M2} shows
contour plots of the normalized potential $\chi_{e}$, the solution
of (\ref{eq:DRKAP1:Charge-equilibrium_condition-normalized}), as
a function of $\kappa_{e}\times\kappa_{i}$ for several values of
the regularization parameter $\mu_{i0}=\mu_{e0}=\mu_{0}$ (this corresponds
to case 4: varying $\kappa_{i}$ and $\kappa_{e}$)\@. As a reference,
the potential for a fully Maxwellian plasma is $\chi_{e}^{\mathrm{(M)}}=2.374$.

In all considered situations, each panel of Fig. \ref{fig:DRKAP1:Chie_M2}
presents regions with either $\chi_{e}<0$ or $\chi_{e}>0$, with
the null-charge condition emphasized with a thicker contour line.
In all panels, the region on the left of the $\chi_{e}=0$ line corresponds
to positively-charged dust grains, whereas the region on the right
denotes negative charges.

The plots on Fig. \ref{fig:DRKAP1:Chie_M2} corroborate the conclusions
drawn from the analysis of the behavior of $\kappa_{ic}$\@. For
a given value of $\mu_{0}$, positive charge is only possible for
$\kappa_{i}<\kappa_{ic}$, when $\kappa_{e}>\kappa_{ec}$, and in
all considered situations, $\kappa_{ec}\gg\kappa_{ic}$ by at least
one order of magnitude.

Finally, the largest negative potential (corresponding to positive
charge) observed is relatively mild, rarely of the order (in absolute
value) of the Maxwellian $\chi_{e}^{\mathrm{(M)}}$\@. On the other
hand, all plots show that when $\kappa_{e}<\kappa_{ec}$, the potential
grows very fast and becomes much larger than $\chi_{e}^{\mathrm{(M)}}$
as $\kappa_{e}\to0$\@. At this point, the potential becomes largely
independent on the value of $\kappa_{i}$.

Summarizing the results from model 1, the dust charge is still negative
in the majority of the 4-dimensional RKD parameter space, with $Z_{d}^{\mathrm{(RKD)}}\gg Z_{d}^{\mathrm{(M)}}$
whenever the electrons are on the same level of suprathermality than
the ions, or higher $\left(\kappa_{e}\lesssim\kappa_{i}\right)$\@.
The occurrence of positively-charged dust grains by the collisions
with suprathermal plasma particles is only possible when the ions
are on an extreme level of suprathermality compared to the electrons
$\left(\kappa_{i}\ll\kappa_{e}\right)$\@. We believe that the possibility
of occurrence of such extreme differences in naturally-occurring environments
is highly unlikely and, hence, the collisional charging mechanism
alone should mostly still render the dust grains negatively charged,
even when the plasma environment is suprathermal.

\subsection{Model 2}

Model 2, which was discussed in section \ref{subsec:DRKAP1:Model-2-Kappa-independent},
is a model in which the kinetic temperature is $\kappa$-invariant
and thus serve as a measure of a physical temperature. Some results
of the equilibrium electric charge that a dust grain acquires due
to collisions of suprathermal plasma particles described by model
2 are presented here. The results will be presented again with the
combinations of distributions comprised by the cases 1 – 4 described
at the beginning of the section.

\begin{figure*}
\begin{minipage}[t]{0.49\textwidth}%
\includegraphics[width=1\columnwidth]{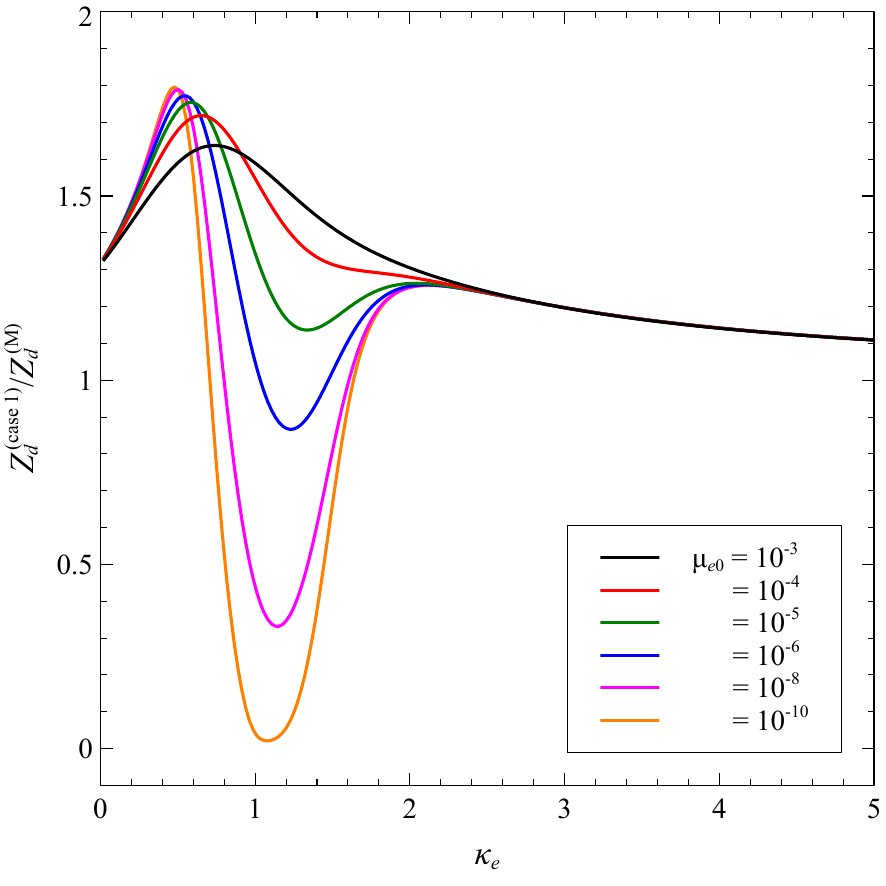}%
\end{minipage}\hfill{}%
\begin{minipage}[t]{0.49\textwidth}%
\includegraphics[width=1\columnwidth]{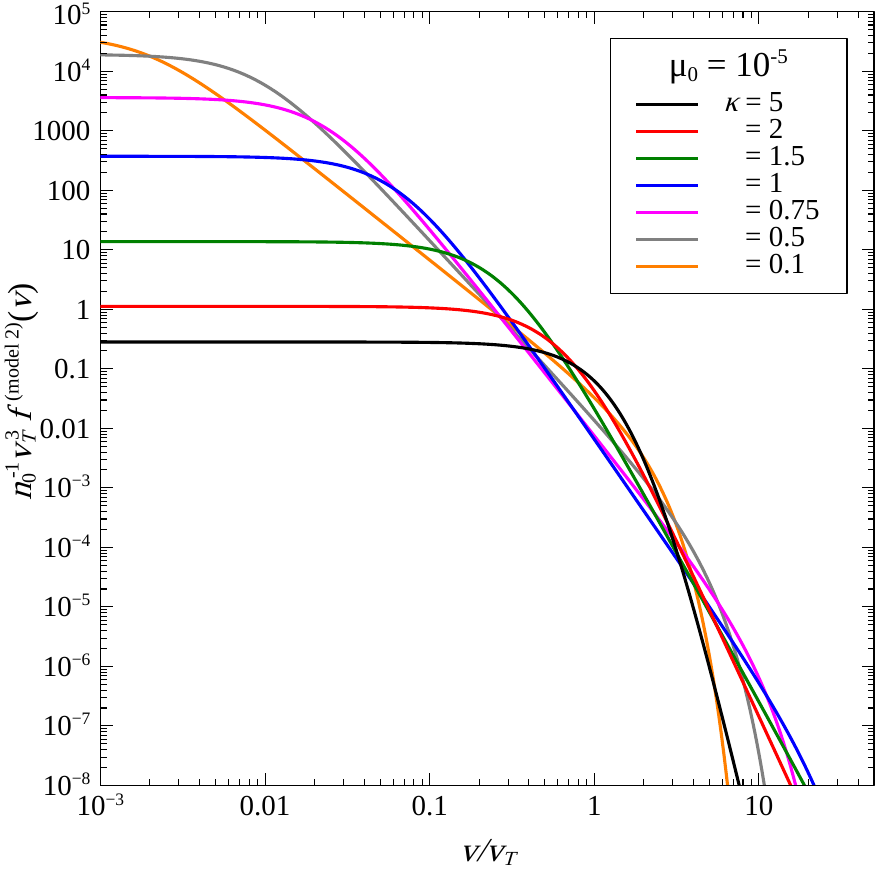}%
\end{minipage}

\caption{(Left panel) Plots of $Z_{d}^{\text{(case 1)}}/Z_{d}^{\mathrm{(M)}}$
as solutions of eq. (\ref{eq:DRKAP1:Charge-equilibrium_condition-normalized})
for model 2, case 1 as functions of $\kappa_{e}$\@. (Right panel)
log-log plots of the normalized RKD for fixed $\mu_{0}$ and several
values of $\kappa$.\label{fig:DRKAP1:Equilibrium_charge-M1-c1}}
\end{figure*}

Figure \ref{fig:DRKAP1:Equilibrium_charge-M1-c1} shows plots of the
dust charge number for cases 1 and 2\@. The charge numbers were obtained
from the solutions of (\ref{eq:DRKAP1:Charge-equilibrium_condition-normalized}),
employing the collision factors (\ref{eq:DRKAP1:Collision_freqs-electron-ion-RKD-gen-M1}a,
b) and (\ref{eq:DRKAP1:Collision_freqs-electron-ion-M}a, b).

The left panel of Fig. \ref{fig:DRKAP1:Equilibrium_charge-M1-c1}
shows the ratio $Z_{d}^{\text{(case 1)}}/Z_{d}^{\mathrm{(M)}}$ as
a function of $\kappa_{e}$ for several values of $\mu_{e0}$\@.
The results obtained from model 2 differ significantly from model
1, which were shown by Fig. \ref{fig:DRKAP1:Charge_M1-c1}\@. Now,
instead of a steady increase of the charge ratio with diminishing
$\kappa_{e}$, we observe initially a slight increase until $\kappa_{e}\approx2$
that is independent on $\mu_{e0}$\@. Then, for $\kappa_{e}\lesssim2$
the ratio curves split for different regularization parameters. For
the larger value $\left(\mu_{e0}=10^{-3}\right)$, the ratio continues
to grow until $\kappa_{e}\approx0.75$ and then it drops to the limiting
value of $Z_{d}^{\text{(case 1)}}\approx1.32Z_{d}^{\mathrm{(M)}}$
at $\kappa_{e}=0$\@. However, the results for $\mu_{e0}\leqslant10^{-5}$
present a more complex behavior. All these plots display a region
of diminishing charge ratio, with minima located in the region $1<\kappa_{e}<1.4$
that are smaller for smaller $\mu_{e0}$, and then the ratio eventually
starts to grow and diminish again as $\kappa_{e}$ keeps on reducing,
until eventually converging all to the same limiting value at $\kappa_{e}=0$\@.

The behavior of the dust charge displayed by the left panel of Fig.
\ref{fig:DRKAP1:Equilibrium_charge-M1-c1} hints at the possibility
of positively-charged grains when the regularization parameter for
the electronic VDF is sufficiently small, occurring at low (but not
too low) values of the kappa index. Notice that this could occur even
when the ions are thermal, which is the case here. Moreover, although
$\kappa_{e}\sim1$ is not too low, it is close to limit of validity
of standard Kappa distributions, which is where the difference between
RKDs and SKDs is the greatest and is the reason why this behavior
was not observed before. It is the regularization action carried out
by the parameter $\mu_{0}$ that allows the existence of plasmas with
a higher state of suprathermality.

The stark difference between the behavior observed with model 2 in
comparison with model 1 is due to the fact that in order to maintain
the kinetic temperature invariant when varying $\kappa$, both the
core and the tail of the VDF must be modified. The right panel of
Fig. \ref{fig:DRKAP1:Equilibrium_charge-M1-c1} shows log-log plots
of the model-2 VDF versus speed for fixed $\mu_{e0}=10^{-5}$ and
several values of $\kappa$\@. Looking separately the core and the
tail portions of the VDF, one notices that the core region steadily
grows with diminishing $\kappa_{e}$, but remains relatively constant
(in the log scale) up to $v\lesssim v_{T}$\@. However, for very
low kappa-index $\left(\kappa_{e}=0.1\right)$ the probability level
quickly reduces with speed. Looking now at the tail, one notices that
for $\kappa_{e}>1$ (at values permissible for SKDs), the population
level of the tail increases with speed, at least within the mid-energy
range displayed $\left(v/v_{T}<50\right)$\@. However for higher
energy, the action of the regularization parameter kicks in earlier.
This is noticeable in the plots for $\kappa_{e}<1$, where one already
notices a reduction in the population level at the tail that is more
pronounced for smaller $\kappa_{e}$\@. Hence, a model-2 RKD keeps
the kinetic temperature $\kappa$-invariant by ``removing'' particles
from the tail and ``injecting'' them in the core. Therefore, the
results of the left panel of Fig. \ref{fig:DRKAP1:Equilibrium_charge-M1-c1}
can be explained by the fact that for small $\kappa_{e}$ there are
in fact less high-energy electrons that can successfully collide and
become attached to the grain, thereby reducing the net negative charge.

\begin{figure}
\includegraphics[width=1\columnwidth]{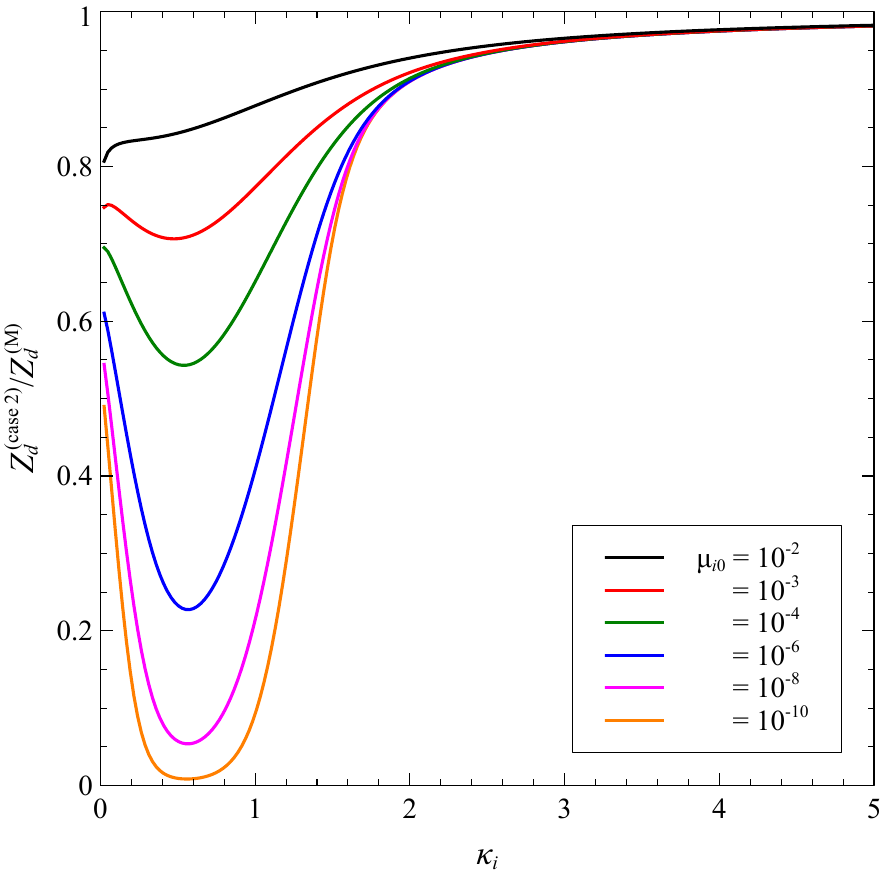}

\caption{Plots of $Z_{d}^{\mathrm{(RKD)}}/Z_{d}^{\mathrm{(M)}}$ as solutions
of eq. (\ref{eq:DRKAP1:Charge-equilibrium_condition-normalized})
for model 2, case 2 as functions of $\kappa_{i}$.\label{fig:DRKAP1:Equilibrium_charge-M1-c2}}
\end{figure}

Figure \ref{fig:DRKAP1:Equilibrium_charge-M1-c2} displays some results
for case 2\@. Once again, the solutions are displayed as charge number
ratio versus $\kappa_{i}$ and the results can be compared with Fig.
\ref{fig:DRKAP1:Charge_M1-c2_3}\@. One observes again the splitting
of the curves with $\mu_{i0}$ for $\kappa_{i}\lesssim2$ and the
possibility of charge inversion for very small values of the regularization
parameter. The depleted portions at $\kappa_{i}\lesssim1$ can be
understood in terms of the previous interpretation and by the fact
that an excess of slower ions increases the collisional cross-section
contained in the integration of the collision frequency (\ref{eq:DRKAP1:Collision_frequency-general}).

Investigation of the null-charge condition from model 2 also leads
to very different conclusions, as compared with model 1\@. Now, the
condition $\chi_{e}=0$ is met when the set $\left\{ \mu_{e0},\kappa_{e},\mu_{i0},\kappa_{i}\right\} $
satisfies
\begin{multline}
\sqrt{\phi_{i}}\Uratiol 1\left(\kappa_{i},\kappa_{i}+1,\mu_{i}\right)\\
=\sqrt{\frac{m_{i}}{m_{e}}\tau_{e}\phi_{e}}\Uratiol 1\left(\kappa_{e},\kappa_{e}+1,\mu_{e}\right),\label{eq:DRKAP1:Kappa_i-critic-Model-1}
\end{multline}
with $\phi_{\beta}=\Uratior 2\left(\kappa_{\beta},\kappa_{\beta}+1,\mu_{\beta}\right).$

\begin{figure}
\includegraphics[width=1\columnwidth]{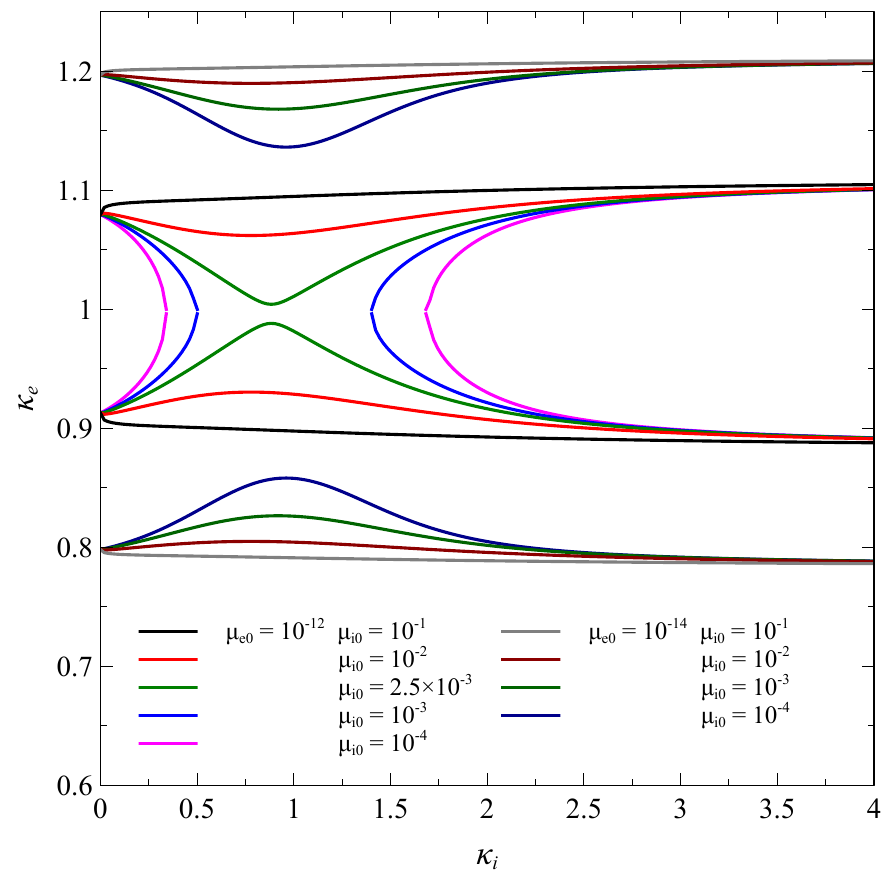}

\caption{Plots of $\kappa_{e}$ and $\kappa_{i}$ values as solutions of the
null-charge condition (\ref{eq:DRKAP1:Kappa_i-critic-Model-1}), for
different values of the set $\left\{ \mu_{e0},\kappa_{e},\mu_{i0},\kappa_{i}\right\} $.\label{fig:DRKAP1:kappa_eic-M1}}
\end{figure}

Figure \ref{fig:DRKAP1:kappa_eic-M1} shows some solutions of (\ref{eq:DRKAP1:Kappa_i-critic-Model-1})\@.
One obvious difference from Fig. \ref{fig:DRKAP1:kappa_ic-M2} is
that the values of the kappa indices are within the same order of
magnitude, whereas the regularization parameters are quite different.

Starting with the black curves, corresponding to a very small $\mu_{e0}=10^{-12}$
and a relatively large $\mu_{e0}=10^{-1}$, we observe that there
are two curves in the $\kappa_{e}\times\kappa_{i}$ plane where the
condition $\chi_{e}=0$ is satisfied. Grains with positive charge
(\emph{i.e.}, with $\chi_{e}<0$) occur in the region within the curves.
Keeping $\mu_{e0}$ fixed and reducing $\mu_{i0}$, we observe that
both curves converge to the point $\kappa_{e}\approx\kappa_{i}\approx1$,
which eventually becomes an \textsf{X} point, where both curves coalesce
and split, leaving a region (for $\mu_{i0}<2.5\times10^{-3}$) with
no solution. As $\mu_{i0}$ keeps on diminishing, this region enlarges.
Hence, the null-charge condition becomes harder to be satisfied as
$\mu_{i0}$ approaches $\mu_{e0}$ from above.

Reducing the fixed value to $\mu_{e0}=10^{-14}$ and repeating the
steps with $\mu_{i0}=10^{-1}$ downwards, we observe the same process,
with the difference that the distance between the initial curves widened.
Conversely, if we increase the value of $\mu_{e0}$, the distance
shrinks, until, eventually, for $\mu_{e0}\gtrsim5\times10^{-12}$
the condition $\chi_{e}=0$ is no longer satisfied for any $\mu_{i0}$.

Therefore, the conditions for positively-charged dust when the plasma
is described by model-2 regularized distributions, are very different
from the conditions for model 1\@. Now, in order for the grains to
become positive, the kappa-indices of both electrons and ions can
be of the same order, but the regularization parameters must be very
different, with $\mu_{e0}\ll\mu_{i0}$.

\begin{figure}
\includegraphics[width=1\columnwidth]{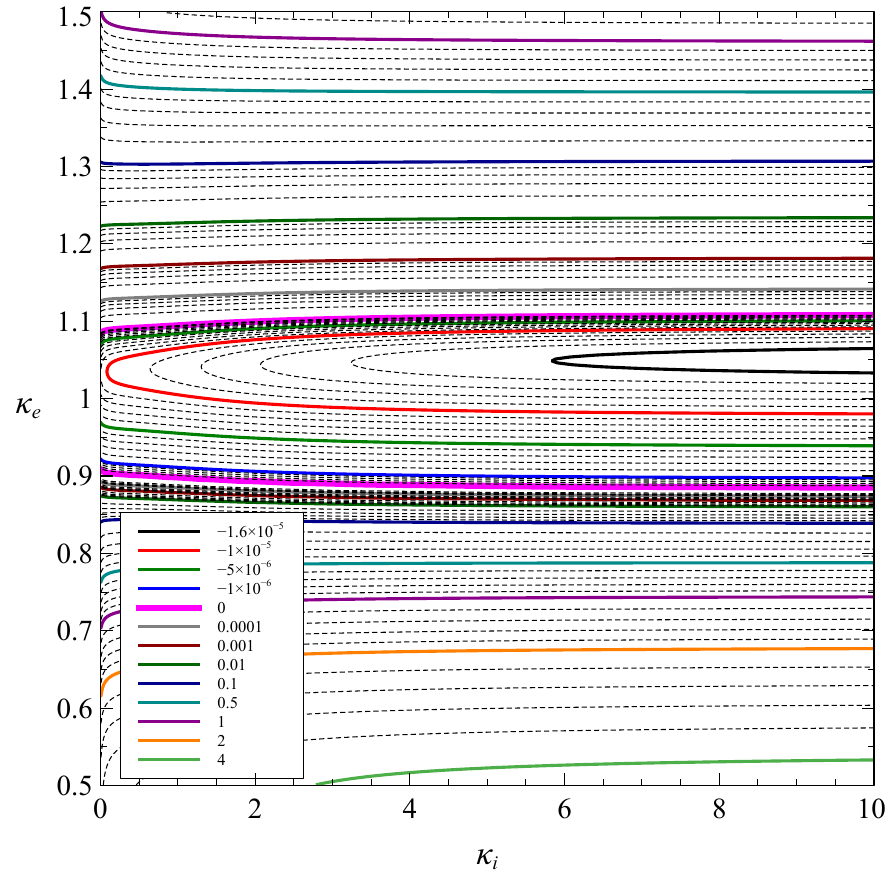}

\caption{Contour plots of the normalized potential $\chi_{e}=Z_{d}/\tau_{e}\tilde{a}$
as a function of $\kappa_{e}\times\kappa_{i}$ for model 2, with fixed
$\mu_{e0}=10^{-12}$ and $\mu_{i0}=10^{-1}$.\label{fig:DRKAP1:Chie_M1}}
\end{figure}

As a final display of results for model 2, figure \ref{fig:DRKAP1:Chie_M1}
shows contour plots of the normalized potential $\chi_{e}$ as a function
of $\kappa_{e}\times\kappa_{i}$ for the first set of parameters of
Fig. \ref{fig:DRKAP1:kappa_eic-M1}, $\mu_{e0}=10^{-12}$ and $\mu_{i0}=10^{-1}$\@.
One can clearly observe the disjoint contours marking the null-charge
condition $\left(\chi_{e}=0\right)$, with positively-charged grains
occurring withing the delimited region and negatively-charged grains
without. The largest negative potential that occurs within the black
contour $\left(\chi_{e,\mathrm{min}}=-1.67\times10^{-5}\right)$ is
very small compared with the nominal Maxwellian value.

As a summary, results from model 2 also show that in the majority
of situations the dust charge is negative, but it can be positive
under very restrictive conditions, which are quite different from
the conditions imposed by model 1\@. Now, the electronic and ionic
kappa-indices can be of the same order of magnitude, but the regularization
parameters must be vastly different, with $\mu_{e0}\ll\mu_{i0}$\@.
This particular scenario happens due to the specific characteristic
attributed to model 2, namely, that the kinetic temperature must remain
invariant to the kappa-index.

As we have concluded with model 1, such disparity of suprathermal
states between the plasma species makes it highly unlikely that such
situation will ever be found in naturally-occurring environments.

\section{Final remarks\label{sec:DRKAP1:Final_remarks}}

In this work we have discussed the charging of spherical dust grains
immersed in an infinite, homogeneous, and steady-state plasma, composed
by electrons and one positive ion species, when the velocity distribution
functions of both species are the $\kappa$-regularized distribution.
Differently from the Kappa distributions considered in Paper I, the
RKD can describe states of extreme suprathermality, which are achieved
when $\kappa+\alpha\leqslant\nicefrac{3}{2}$ (with any $\kappa>0$),
thanks to the presence of a Gaussian factor, determined by the regularization
parameter $\mu\ll1$\@. The RKD reduces to the standard Kappa distribution
when $\mu=0$ (in which case it is only valid when $\kappa+\alpha>\nicefrac{3}{2}$)
and to the Maxwellian when $\kappa\to\infty$.

From a generalized expression for the RKD, we have considered two
common models, frequently employed in the literature. Model 1 assumes
a free parameter $\theta=\sqrt{2T/m}$, for which the quantity $T$
is not the physical temperature of the system (\emph{i.e.}, is not
the kinetic temperature, nor is derived from an entropic principle),
and Model 2, for which the kinetic temperature is $\kappa$-invariant
and can be derived from an entropy formulation. 

The equation that determines the equilibrium charge of the dust grains
due to inelastic collisions with the plasma particles was solved for
both models with different combinations of velocity distribution functions:
one case, where the electrons are suprathermal and the ions are Maxwellian.
Another case, where the ions are suprathermal while the electrons
are thermal, and final cases where both species are suprathermal.
The solutions were mostly sought in the small-kappa range, which is
where the distinction between the effects from regularized Kappa distributions,
as compared to standard Kappa distributions, is more pronounced.

When solving the equation employing model 1 for the first case, it
was observed that the grains still acquire a net negative charge,
but with higher absolute value, as it was already observed in Paper
I, where standard Kappa distributions were employed. This happens
due to the excess of high-energy electrons when the species is in
a suprathermal state. For small $\kappa_{e}\lesssim1$, the negative
charge eventually saturates.

Contrastingly, in the second case (suprathermal ions, thermal electrons)
the dust charge reduces with $\kappa_{i}$, reaching a minimum value
at $\kappa_{i}=0$ that depends on the regularization parameter $\mu_{i}$\@.
For sufficiently small values of $\left\{ \mu_{i0},\kappa_{i}\right\} $,
the dust can become positively-charged due to the extreme suprathermality
of the ions as compared to electrons. This result can be obtained
even when the electrons are suprathermal. After investigating in further
details this possibility, it was concluded that indeed the dust can
become positively-charged due to collisions alone, a result that is
impossible in a thermal plasma, but only when the ions are much more
suprathermal than the electrons, meaning that $\kappa_{i}\ll\kappa_{e}$
is a necessary condition.

When investigating the possibility of positive charge of grains immersed
in plasmas described by model 2, it was observed that this can happen
when $\kappa_{i}$ and $\kappa_{e}$ are of the same order of magnitude,
but only when the regularization parameters are vastly different,
with $\mu_{e0}\ll\mu_{i0}$\@. This happens due to the requirement
of the model 2 that the kinetic temperature is independent on the
kappa index\@. As $\kappa$ reduces, the core region of the VDF becomes
more populated whereas the high-energy tail becomes depleted. Consequently,
the grain can become positively-charged because there are fewer high-energy
electrons to overcome a negative potential and more low-energy ions
with higher cross section. In this case, a balance can be achieved
where the number of positive ions attaching to the grain's surface
overcomes the number of negative electrons. This is a different situation
where the ions are more suprathermal than electrons, with the suprathermality
states determined mostly by the regularization parameters, instead
of the kappa indices.

This work considered only the collisional charging of the dust grains.
More complete treatments, to appear in future publications, will include
other relevant charging mechanisms such as photoelectric and secondary
emissions. Future work will also investigate the effect of charged
dust in suprathermal plasma in wave propagation and wave-particle
interactions.

In this Paper, we applied the formalism to a dusty plasma environment
typical of carbon-rich stars. Future applications will consider different
plasma conditions found in other space and astrophysical systems.

This line of investigation can also be extended to other plasma phenomena,
such as magnetic reconnection. There is evidence that the disruption
of the current sheet equilibrium of opposing magnetic flux tubes is
closely related to magnetic reconnection and the level of turbulence
inferred at the top of coronal mass ejections.\citep{Chian+16/12}
On one hand, theoretical studies of the effect of electronic depletion
on the Harris current sheet and other nonlinear structures can lead
to slower reconnection rates on dusty plasmas.\citep{Lazerson11/02,Yang+23/07}
On the other hand, the presence of extreme suprathermal particles
described by the regularized distributions allows the formation of
modified Harris sheets characterized by higher values of plasma parameters
such as density, current, magnetic field and temperature, which can,
contrastingly, result in higher reconnection rates.\citep{Hau+23/10}
Hence, further studies that combine both extensions to the traditional
treatment of magnetic reconnection, the presence of dust grains and
extreme suprathermal particles, could provide interesting contributions
to this important phenomenon observed in various astrophysical systems.
\begin{acknowledgments}
RG acknowledges support provided by Conselho Nacional de Desenvolvimento
Científico e Tecnológico (CNPq, Brazil), Grant No. 313330/2021-2\@.
LFZ acknowledges support from CNPq, grant No. 303189/2022-3\@. This
study was financed in part by Coordenação de Aperfeiçoamento de Pessoal
de Nível Superior (CAPES, Brazil) - Finance Code 001.
\end{acknowledgments}

\section*{Data availability}

The data that support the findings of this study are available from
the corresponding author upon reasonable request.

\appendix

\section{Properties and limiting cases of the function $\boldsymbol{\protect\Uratio mn\left(\eta,\zeta,\mu\right)}$\label{sec:DRKAP1:Limiting_cases}}

The function $\Uratio mn\left(\eta,\zeta,\mu\right)$ is defined by
(\ref{eq:DRKAP1:U_scr^mn-1})\@. The expression employed in this
work differs from the original definition from Ref. \onlinecite{Scherer+20/07}
because we are considering cases where $\zeta\geqslant0$ always.

One trivial limiting case occurs when $n=m$, 
\[
\Uratio mm\left(\eta,\zeta,\mu\right)=1.
\]

Another obvious property is the permutation of indices $m\leftrightarrow n$,
which leads to 
\begin{equation}
\Uratio nm\left(\eta,\zeta,\mu\right)=\left[\Uratio mn\left(\eta,\zeta,\mu\right)\right]^{-1}.\label{eq:DRKAP1:U_src-permutation}
\end{equation}

Yet another obvious property is 
\[
\Uratio m{\ell}\left(\eta,\zeta,\mu\right)\Uratio{\ell}n\left(\eta,\zeta,\mu\right)=\Uratio mn\left(\eta,\zeta,\mu\right).
\]

Other important identities involving the Tricomi function are given
below.\citep{Daalhuis-Full-NIST10a} Firstly, the integral representation
\begin{equation}
U\left(a,b,z\right)=\frac{1}{\Gamma\left(a\right)}\int_{0}^{\infty}dt\,t^{a-1}\left(1+t\right)^{b-a-1}\mathrm{e}^{-zt},\label{eq:DRKAP1:Tricomi-Integ_rep-1}
\end{equation}
which if valid for $\Re a>0$ and $\left|\arg\left(z\right)\right|<\nicefrac{\pi}{2}$\@.
And the Kummer transformation 
\begin{equation}
U\left(a,b,z\right)=z^{1-b}U\left(a-b+1,2-b,z\right).\label{eq:DRKAP1:Kummer_transformation}
\end{equation}

Important cases in the analysis performed in this paper involve the
possible limits of vanishing argument of the Tricomi functions in
(\ref{eq:DRKAP1:U_scr^mn-1})\@. The limit $\kappa\mu\left(\kappa\right)\longrightarrow0$
involves both limits $\kappa\to0$ and $\kappa\to\infty$\@. These
limits are obtained from the general limiting cases given by Ref.
\onlinecite{Daalhuis-Full-NIST10a}, sec. 13.2(iii)\@. Given the
Tricomi function $U\left(a,b,z\right)$, the limiting cases for $z\to0$
in the aforementioned reference consider a total of 7 different possible
ranges of values of the parameter $b$\@. Since the function $\Uratio mn\left(\eta,\zeta,\mu\right)$
depends on the double set of indices $\left\{ m,n\right\} $, it means
that a total of 49 cases for $\zeta$ must be considered. 

A straightforward but rather tedious analysis of all possible cases
leads to the following limits. If $\zeta=\frac{3+m}{2}$ and $\zeta>\frac{3+n}{2}$,
or $\zeta>\frac{3+m}{2}$ and $\zeta=\frac{3+n}{2}$, 
\[
\lim_{\mu\eta\to0}\Uratio mn\left(\eta,\zeta,\mu\right)=\frac{\Gamma\left(\zeta-\frac{3+m}{2}\right)}{\Gamma\left(\zeta-\frac{3+n}{2}\right)}.
\]

The remaining cases, which occur when either $\zeta<\frac{3+m}{2}$
or $\zeta<\frac{3+n}{2}$, are not reproducible by the formula above
and are 
\[
\lim_{\mu\eta\to0}\Uratio mn\left(\eta,\zeta,\mu\right)\longrightarrow\begin{cases}
0, & n>m\\
\infty, & n<m.
\end{cases}
\]

\bibliographystyle{aipnum4-1}
\bibliography{GaelzerZiebell1124-R1}

\begin{thebibliography}{101}%
\makeatletter
\providecommand \@ifxundefined [1]{%
 \@ifx{#1\undefined}
}%
\providecommand \@ifnum [1]{%
 \ifnum #1\expandafter \@firstoftwo
 \else \expandafter \@secondoftwo
 \fi
}%
\providecommand \@ifx [1]{%
 \ifx #1\expandafter \@firstoftwo
 \else \expandafter \@secondoftwo
 \fi
}%
\providecommand \natexlab [1]{#1}%
\providecommand \enquote  [1]{``#1''}%
\providecommand \bibnamefont  [1]{#1}%
\providecommand \bibfnamefont [1]{#1}%
\providecommand \citenamefont [1]{#1}%
\providecommand \href@noop [0]{\@secondoftwo}%
\providecommand \href [0]{\begingroup \@sanitize@url \@href}%
\providecommand \@href[1]{\@@startlink{#1}\@@href}%
\providecommand \@@href[1]{\endgroup#1\@@endlink}%
\providecommand \@sanitize@url [0]{\catcode `\\12\catcode `\$12\catcode
  `\&12\catcode `\#12\catcode `\^12\catcode `\_12\catcode `\%12\relax}%
\providecommand \@@startlink[1]{}%
\providecommand \@@endlink[0]{}%
\providecommand \url  [0]{\begingroup\@sanitize@url \@url }%
\providecommand \@url [1]{\endgroup\@href {#1}{\urlprefix }}%
\providecommand \urlprefix  [0]{URL }%
\providecommand \Eprint [0]{\href }%
\providecommand \doibase [0]{http://dx.doi.org/}%
\providecommand \selectlanguage [0]{\@gobble}%
\providecommand \bibinfo  [0]{\@secondoftwo}%
\providecommand \bibfield  [0]{\@secondoftwo}%
\providecommand \translation [1]{[#1]}%
\providecommand \BibitemOpen [0]{}%
\providecommand \bibitemStop [0]{}%
\providecommand \bibitemNoStop [0]{.\EOS\space}%
\providecommand \EOS [0]{\spacefactor3000\relax}%
\providecommand \BibitemShut  [1]{\csname bibitem#1\endcsname}%
\let\auto@bib@innerbib\@empty
\bibitem [{\citenamefont {Goertz}(1989)}]{Goertz89}%
  \BibitemOpen
  \bibfield  {author} {\bibinfo {author} {\bibfnamefont {C.~K.}\ \bibnamefont
  {Goertz}},\ }\href {http://www.agu.org/journals/rg/v027/i002/RG027i002p00271}
  {\bibfield  {journal} {\bibinfo  {journal} {Rev. Geophys.}\ }\textbf
  {\bibinfo {volume} {27}},\ \bibinfo {pages} {271–292} (\bibinfo {year}
  {1989})}\BibitemShut {NoStop}%
\bibitem [{\citenamefont {Northrop}(1992)}]{Northrop92}%
  \BibitemOpen
  \bibfield  {author} {\bibinfo {author} {\bibfnamefont {T.~G.}\ \bibnamefont
  {Northrop}},\ }\href {\doibase 10.1088/0031-8949} {\bibfield  {journal}
  {\bibinfo  {journal} {Phys. Scripta}\ }\textbf {\bibinfo {volume} {45}},\
  \bibinfo {pages} {475} (\bibinfo {year} {1992})}\BibitemShut {NoStop}%
\bibitem [{\citenamefont {Mendis}\ and\ \citenamefont
  {Rosenberg}(1994)}]{MendisRosenberg94/09}%
  \BibitemOpen
  \bibfield  {author} {\bibinfo {author} {\bibfnamefont {D.~A.}\ \bibnamefont
  {Mendis}}\ and\ \bibinfo {author} {\bibfnamefont {M.}~\bibnamefont
  {Rosenberg}},\ }\href {\doibase 10.1146/annurev.aa.32.090194.002223}
  {\bibfield  {journal} {\bibinfo  {journal} {Annu. Rev. Astron. Astrophys.}\
  }\textbf {\bibinfo {volume} {32}},\ \bibinfo {pages} {419–463} (\bibinfo
  {year} {1994})}\BibitemShut {NoStop}%
\bibitem [{\citenamefont {Shukla}\ and\ \citenamefont
  {Mamun}(2002)}]{ShuklaMamun02}%
  \BibitemOpen
  \bibfield  {author} {\bibinfo {author} {\bibfnamefont {P.~K.}\ \bibnamefont
  {Shukla}}\ and\ \bibinfo {author} {\bibfnamefont {A.~A.}\ \bibnamefont
  {Mamun}},\ }\href@noop {} {\emph {\bibinfo {title} {{Introduction to Dusty
  Plasma Physics}}}},\ edited by\ \bibinfo {editor} {\bibfnamefont {A.~A.}\
  \bibnamefont {Mamun}},\ {Series in Plasma Physics}\ (\bibinfo  {publisher}
  {Taylor \& Francis, Inc.},\ \bibinfo {address} {New York},\ \bibinfo {year}
  {2002})\ p.\ \bibinfo {pages} {450}\BibitemShut {NoStop}%
\bibitem [{\citenamefont {Tsytovich}, \citenamefont {Morfill},\ and\
  \citenamefont {Thomas}(2004)}]{Tsytovich+04/10}%
  \BibitemOpen
  \bibfield  {author} {\bibinfo {author} {\bibfnamefont {V.~N.}\ \bibnamefont
  {Tsytovich}}, \bibinfo {author} {\bibfnamefont {G.~E.}\ \bibnamefont
  {Morfill}}, \ and\ \bibinfo {author} {\bibfnamefont {H.}~\bibnamefont
  {Thomas}},\ }\href {\doibase 10.1134/1.1809401} {\bibfield  {journal}
  {\bibinfo  {journal} {Plasma Phys. Rep.}\ }\textbf {\bibinfo {volume} {30}},\
  \bibinfo {pages} {816–864} (\bibinfo {year} {2004})}\BibitemShut {NoStop}%
\bibitem [{\citenamefont {Fortov}\ \emph {et~al.}(2005)\citenamefont {Fortov},
  \citenamefont {Ivlev}, \citenamefont {Khrapak}, \citenamefont {Khrapak},\
  and\ \citenamefont {Morfill}}]{Fortov+05/12}%
  \BibitemOpen
  \bibfield  {author} {\bibinfo {author} {\bibfnamefont {V.~E.}\ \bibnamefont
  {Fortov}}, \bibinfo {author} {\bibfnamefont {A.~V.}\ \bibnamefont {Ivlev}},
  \bibinfo {author} {\bibfnamefont {S.~A.}\ \bibnamefont {Khrapak}}, \bibinfo
  {author} {\bibfnamefont {A.~G.}\ \bibnamefont {Khrapak}}, \ and\ \bibinfo
  {author} {\bibfnamefont {G.~E.}\ \bibnamefont {Morfill}},\ }\href {\doibase
  10.1016/j.physrep.2005.08.007} {\bibfield  {journal} {\bibinfo  {journal}
  {Physics Reports}\ }\textbf {\bibinfo {volume} {421}},\ \bibinfo {pages}
  {1–103} (\bibinfo {year} {2005})}\BibitemShut {NoStop}%
\bibitem [{\citenamefont {Mann}\ \emph {et~al.}(2010)\citenamefont {Mann},
  \citenamefont {Czechowski}, \citenamefont {Meyer-Vernet}, \citenamefont
  {Zaslavsky},\ and\ \citenamefont {Lamy}}]{Mann+10/12}%
  \BibitemOpen
  \bibfield  {author} {\bibinfo {author} {\bibfnamefont {I.}~\bibnamefont
  {Mann}}, \bibinfo {author} {\bibfnamefont {A.}~\bibnamefont {Czechowski}},
  \bibinfo {author} {\bibfnamefont {N.}~\bibnamefont {Meyer-Vernet}}, \bibinfo
  {author} {\bibfnamefont {A.}~\bibnamefont {Zaslavsky}}, \ and\ \bibinfo
  {author} {\bibfnamefont {H.}~\bibnamefont {Lamy}},\ }\href {\doibase
  10.1088/0741-3335} {\bibfield  {journal} {\bibinfo  {journal} {Plasma Phys.
  Cont. Fusion}\ }\textbf {\bibinfo {volume} {52}},\ \bibinfo {pages} {124012}
  (\bibinfo {year} {2010})}\BibitemShut {NoStop}%
\bibitem [{\citenamefont {Mann}, \citenamefont {Meyer-Vernet},\ and\
  \citenamefont {Czechowski}(2014)}]{Mann+14/03}%
  \BibitemOpen
  \bibfield  {author} {\bibinfo {author} {\bibfnamefont {I.}~\bibnamefont
  {Mann}}, \bibinfo {author} {\bibfnamefont {N.}~\bibnamefont {Meyer-Vernet}},
  \ and\ \bibinfo {author} {\bibfnamefont {A.}~\bibnamefont {Czechowski}},\
  }\href {\doibase 10.1016/j.physrep.2013.11.001} {\bibfield  {journal}
  {\bibinfo  {journal} {Phys. Rep.}\ }\textbf {\bibinfo {volume} {536}},\
  \bibinfo {pages} {1–39} (\bibinfo {year} {2014})}\BibitemShut {NoStop}%
\bibitem [{\citenamefont {Spahn}\ \emph {et~al.}(2019)\citenamefont {Spahn},
  \citenamefont {Sachse}, \citenamefont {Seiß}, \citenamefont {Hsu},
  \citenamefont {Kempf},\ and\ \citenamefont {Horányi}}]{Spahn+19/02}%
  \BibitemOpen
  \bibfield  {author} {\bibinfo {author} {\bibfnamefont {F.}~\bibnamefont
  {Spahn}}, \bibinfo {author} {\bibfnamefont {M.}~\bibnamefont {Sachse}},
  \bibinfo {author} {\bibfnamefont {M.}~\bibnamefont {Seiß}}, \bibinfo
  {author} {\bibfnamefont {H.-W.}\ \bibnamefont {Hsu}}, \bibinfo {author}
  {\bibfnamefont {S.}~\bibnamefont {Kempf}}, \ and\ \bibinfo {author}
  {\bibfnamefont {M.}~\bibnamefont {Horányi}},\ }\href {\doibase
  10.1007/s11214-018-0577-3} {\bibfield  {journal} {\bibinfo  {journal} {Space
  Sci. Rev.}\ }\textbf {\bibinfo {volume} {215}},\ \bibinfo {pages} {11}
  (\bibinfo {year} {2019})}\BibitemShut {NoStop}%
\bibitem [{\citenamefont {Melzer}\ \emph {et~al.}(2021)\citenamefont {Melzer},
  \citenamefont {Krüger}, \citenamefont {Maier},\ and\ \citenamefont
  {Schütt}}]{Melzer+21/11}%
  \BibitemOpen
  \bibfield  {author} {\bibinfo {author} {\bibfnamefont {A.}~\bibnamefont
  {Melzer}}, \bibinfo {author} {\bibfnamefont {H.}~\bibnamefont {Krüger}},
  \bibinfo {author} {\bibfnamefont {D.}~\bibnamefont {Maier}}, \ and\ \bibinfo
  {author} {\bibfnamefont {S.}~\bibnamefont {Schütt}},\ }\href {\doibase
  10.1007/s41614-021-00060-2} {\bibfield  {journal} {\bibinfo  {journal} {Rev.
  Mod. Plasma Phys.}\ }\textbf {\bibinfo {volume} {5}},\ \bibinfo {pages} {11}
  (\bibinfo {year} {2021})}\BibitemShut {NoStop}%
\bibitem [{\citenamefont {Sterken}\ \emph {et~al.}(2022)\citenamefont
  {Sterken}, \citenamefont {Baalmann}, \citenamefont {Draine}, \citenamefont
  {Godenko}, \citenamefont {Herbst}, \citenamefont {Hsu}, \citenamefont
  {Hunziker}, \citenamefont {Izmodenov}, \citenamefont {Lallement},\ and\
  \citenamefont {Slavin}}]{Sterken+22/12}%
  \BibitemOpen
  \bibfield  {author} {\bibinfo {author} {\bibfnamefont {V.~J.}\ \bibnamefont
  {Sterken}}, \bibinfo {author} {\bibfnamefont {L.~R.}\ \bibnamefont
  {Baalmann}}, \bibinfo {author} {\bibfnamefont {B.~T.}\ \bibnamefont
  {Draine}}, \bibinfo {author} {\bibfnamefont {E.}~\bibnamefont {Godenko}},
  \bibinfo {author} {\bibfnamefont {K.}~\bibnamefont {Herbst}}, \bibinfo
  {author} {\bibfnamefont {H.-W.}\ \bibnamefont {Hsu}}, \bibinfo {author}
  {\bibfnamefont {S.}~\bibnamefont {Hunziker}}, \bibinfo {author}
  {\bibfnamefont {V.}~\bibnamefont {Izmodenov}}, \bibinfo {author}
  {\bibfnamefont {R.}~\bibnamefont {Lallement}}, \ and\ \bibinfo {author}
  {\bibfnamefont {J.~D.}\ \bibnamefont {Slavin}},\ }\href {\doibase
  10.1007/s11214-022-00939-7} {\bibfield  {journal} {\bibinfo  {journal} {Space
  Sci. Rev.}\ }\textbf {\bibinfo {volume} {218}},\ \bibinfo {pages} {71}
  (\bibinfo {year} {2022})}\BibitemShut {NoStop}%
\bibitem [{\citenamefont {Godenko}\ and\ \citenamefont
  {Izmodenov}(2023)}]{GodenkoIzmodenov23/12}%
  \BibitemOpen
  \bibfield  {author} {\bibinfo {author} {\bibfnamefont {E.}~\bibnamefont
  {Godenko}}\ and\ \bibinfo {author} {\bibfnamefont {V.}~\bibnamefont
  {Izmodenov}},\ }\href {\doibase 10.1016/j.asr.2023.09.016} {\bibfield
  {journal} {\bibinfo  {journal} {Adv. Space Res.}\ }\textbf {\bibinfo {volume}
  {72}},\ \bibinfo {pages} {5142–5158} (\bibinfo {year} {2023})}\BibitemShut
  {NoStop}%
\bibitem [{\citenamefont {Lieb}\ \emph {et~al.}(2025)\citenamefont {Lieb},
  \citenamefont {Lau}, \citenamefont {Hoffman}, \citenamefont {Corcoran},
  \citenamefont {{Garcia Marin}}, \citenamefont {Gull}, \citenamefont
  {Hamaguchi}, \citenamefont {Han}, \citenamefont {Hankins}, \citenamefont
  {Jones}, \citenamefont {Madura}, \citenamefont {Marchenko}, \citenamefont
  {Matsuhara}, \citenamefont {Millour}, \citenamefont {Moffat}, \citenamefont
  {Morris}, \citenamefont {Morris}, \citenamefont {Onaka}, \citenamefont
  {Perrin}, \citenamefont {Rest}, \citenamefont {Richardson}, \citenamefont
  {Russell}, \citenamefont {Sanchez-Bermudez}, \citenamefont {Soulain},
  \citenamefont {Tuthill}, \citenamefont {Weigelt},\ and\ \citenamefont
  {Williams}}]{Lieb+25/01}%
  \BibitemOpen
  \bibfield  {author} {\bibinfo {author} {\bibfnamefont {E.~P.}\ \bibnamefont
  {Lieb}}, \bibinfo {author} {\bibfnamefont {R.~M.}\ \bibnamefont {Lau}},
  \bibinfo {author} {\bibfnamefont {J.~L.}\ \bibnamefont {Hoffman}}, \bibinfo
  {author} {\bibfnamefont {M.~F.}\ \bibnamefont {Corcoran}}, \bibinfo {author}
  {\bibfnamefont {M.}~\bibnamefont {{Garcia Marin}}}, \bibinfo {author}
  {\bibfnamefont {T.~R.}\ \bibnamefont {Gull}}, \bibinfo {author}
  {\bibfnamefont {K.}~\bibnamefont {Hamaguchi}}, \bibinfo {author}
  {\bibfnamefont {Y.}~\bibnamefont {Han}}, \bibinfo {author} {\bibfnamefont
  {M.~J.}\ \bibnamefont {Hankins}}, \bibinfo {author} {\bibfnamefont {O.~C.}\
  \bibnamefont {Jones}}, \bibinfo {author} {\bibfnamefont {T.~I.}\ \bibnamefont
  {Madura}}, \bibinfo {author} {\bibfnamefont {S.~V.}\ \bibnamefont
  {Marchenko}}, \bibinfo {author} {\bibfnamefont {H.}~\bibnamefont
  {Matsuhara}}, \bibinfo {author} {\bibfnamefont {F.}~\bibnamefont {Millour}},
  \bibinfo {author} {\bibfnamefont {A.~F.~J.}\ \bibnamefont {Moffat}}, \bibinfo
  {author} {\bibfnamefont {M.~R.}\ \bibnamefont {Morris}}, \bibinfo {author}
  {\bibfnamefont {P.~W.}\ \bibnamefont {Morris}}, \bibinfo {author}
  {\bibfnamefont {T.}~\bibnamefont {Onaka}}, \bibinfo {author} {\bibfnamefont
  {M.~D.}\ \bibnamefont {Perrin}}, \bibinfo {author} {\bibfnamefont
  {A.}~\bibnamefont {Rest}}, \bibinfo {author} {\bibfnamefont {N.}~\bibnamefont
  {Richardson}}, \bibinfo {author} {\bibfnamefont {C.~M.~P.}\ \bibnamefont
  {Russell}}, \bibinfo {author} {\bibfnamefont {J.}~\bibnamefont
  {Sanchez-Bermudez}}, \bibinfo {author} {\bibfnamefont {A.}~\bibnamefont
  {Soulain}}, \bibinfo {author} {\bibfnamefont {P.}~\bibnamefont {Tuthill}},
  \bibinfo {author} {\bibfnamefont {G.}~\bibnamefont {Weigelt}}, \ and\
  \bibinfo {author} {\bibfnamefont {P.~M.}\ \bibnamefont {Williams}},\ }\href
  {\doibase 10.3847/2041-8213/ad9aa9} {\bibfield  {journal} {\bibinfo
  {journal} {Astrophys. J. Lett.}\ }\textbf {\bibinfo {volume} {979}},\
  \bibinfo {pages} {L3} (\bibinfo {year} {2025})}\BibitemShut {NoStop}%
\bibitem [{\citenamefont {Kimura}\ and\ \citenamefont
  {Mann}(1998)}]{KimuraMann98/05}%
  \BibitemOpen
  \bibfield  {author} {\bibinfo {author} {\bibfnamefont {H.}~\bibnamefont
  {Kimura}}\ and\ \bibinfo {author} {\bibfnamefont {I.}~\bibnamefont {Mann}},\
  }\href {\doibase 10.1086/305613} {\bibfield  {journal} {\bibinfo  {journal}
  {Astrophys. J.}\ }\textbf {\bibinfo {volume} {499}},\ \bibinfo {pages}
  {454–462} (\bibinfo {year} {1998})}\BibitemShut {NoStop}%
\bibitem [{\citenamefont {Ignatov}(2009)}]{Ignatov09/09}%
  \BibitemOpen
  \bibfield  {author} {\bibinfo {author} {\bibfnamefont {A.~M.}\ \bibnamefont
  {Ignatov}},\ }\href {\doibase 10.1134/S1063780X09080042} {\bibfield
  {journal} {\bibinfo  {journal} {Plasma Phys. Rep.}\ }\textbf {\bibinfo
  {volume} {35}},\ \bibinfo {pages} {647–650} (\bibinfo {year}
  {2009})}\BibitemShut {NoStop}%
\bibitem [{\citenamefont {Galvão}\ and\ \citenamefont
  {Ziebell}(2012)}]{GalvaoZiebell12/09}%
  \BibitemOpen
  \bibfield  {author} {\bibinfo {author} {\bibfnamefont {R.~A.}\ \bibnamefont
  {Galvão}}\ and\ \bibinfo {author} {\bibfnamefont {L.~F.}\ \bibnamefont
  {Ziebell}},\ }\href {\doibase 10.1063/1.4748932} {\bibfield  {journal}
  {\bibinfo  {journal} {Phys. Plasmas}\ }\textbf {\bibinfo {volume} {19}},\
  \bibinfo {eid} {093702} (\bibinfo {year} {2012})}\BibitemShut {NoStop}%
\bibitem [{\citenamefont {Ibáñez-Mejía}\ \emph {et~al.}(2019)\citenamefont
  {Ibáñez-Mejía}, \citenamefont {Walch}, \citenamefont {Ivlev},
  \citenamefont {Clarke}, \citenamefont {Caselli},\ and\ \citenamefont
  {Joshi}}]{Ibanez-Mejia+19/05}%
  \BibitemOpen
  \bibfield  {author} {\bibinfo {author} {\bibfnamefont {J.~C.}\ \bibnamefont
  {Ibáñez-Mejía}}, \bibinfo {author} {\bibfnamefont {S.}~\bibnamefont
  {Walch}}, \bibinfo {author} {\bibfnamefont {A.~V.}\ \bibnamefont {Ivlev}},
  \bibinfo {author} {\bibfnamefont {S.}~\bibnamefont {Clarke}}, \bibinfo
  {author} {\bibfnamefont {P.}~\bibnamefont {Caselli}}, \ and\ \bibinfo
  {author} {\bibfnamefont {P.~R.}\ \bibnamefont {Joshi}},\ }\href {\doibase
  10.1093/mnras/stz207} {\bibfield  {journal} {\bibinfo  {journal} {Mont. Not.
  R. Astron. Soc.}\ }\textbf {\bibinfo {volume} {485}},\ \bibinfo {pages}
  {1220–1247} (\bibinfo {year} {2019})}\BibitemShut {NoStop}%
\bibitem [{\citenamefont {Kodanova}\ \emph {et~al.}(2019)\citenamefont
  {Kodanova}, \citenamefont {Bastykova}, \citenamefont {Ramazanov},
  \citenamefont {Nigmetova}, \citenamefont {Maiorov},\ and\ \citenamefont
  {Moldabekov}}]{Kodanova+19/07}%
  \BibitemOpen
  \bibfield  {author} {\bibinfo {author} {\bibfnamefont {S.~K.}\ \bibnamefont
  {Kodanova}}, \bibinfo {author} {\bibfnamefont {N.~K.}\ \bibnamefont
  {Bastykova}}, \bibinfo {author} {\bibfnamefont {T.~S.}\ \bibnamefont
  {Ramazanov}}, \bibinfo {author} {\bibfnamefont {G.~N.}\ \bibnamefont
  {Nigmetova}}, \bibinfo {author} {\bibfnamefont {S.~A.}\ \bibnamefont
  {Maiorov}}, \ and\ \bibinfo {author} {\bibfnamefont {Z.~A.}\ \bibnamefont
  {Moldabekov}},\ }\href {\doibase 10.1109/TPS.2019.2916303} {\bibfield
  {journal} {\bibinfo  {journal} {IEEE Trans. Plasma Sci.}\ }\textbf {\bibinfo
  {volume} {47}},\ \bibinfo {pages} {3052–3056} (\bibinfo {year}
  {2019})}\BibitemShut {NoStop}%
\bibitem [{\citenamefont {Masheyeva}, \citenamefont {Dzhumagulova},\ and\
  \citenamefont {Myrzaly}(2022)}]{Masheyeva+22/11}%
  \BibitemOpen
  \bibfield  {author} {\bibinfo {author} {\bibfnamefont {R.~U.}\ \bibnamefont
  {Masheyeva}}, \bibinfo {author} {\bibfnamefont {K.~N.}\ \bibnamefont
  {Dzhumagulova}}, \ and\ \bibinfo {author} {\bibfnamefont {M.}~\bibnamefont
  {Myrzaly}},\ }\href {\doibase 10.1134/S1063780X22600888} {\bibfield
  {journal} {\bibinfo  {journal} {Plasma Phys. Rep.}\ }\textbf {\bibinfo
  {volume} {48}},\ \bibinfo {pages} {1203–1210} (\bibinfo {year}
  {2022})}\BibitemShut {NoStop}%
\bibitem [{\citenamefont {D'Angelo}(1990)}]{DAngelo90/09}%
  \BibitemOpen
  \bibfield  {author} {\bibinfo {author} {\bibfnamefont {N.}~\bibnamefont
  {D'Angelo}},\ }\href {\doibase 10.1016/0032-0633(90)90022-I} {\bibfield
  {journal} {\bibinfo  {journal} {Planet. Space Sci.}\ }\textbf {\bibinfo
  {volume} {38}},\ \bibinfo {pages} {1143–1146} (\bibinfo {year}
  {1990})}\BibitemShut {NoStop}%
\bibitem [{\citenamefont {Amin}(1996)}]{Amin96/09}%
  \BibitemOpen
  \bibfield  {author} {\bibinfo {author} {\bibfnamefont {M.~R.}\ \bibnamefont
  {Amin}},\ }\href {\doibase 10.1103/PhysRevE.54.R2232} {\bibfield  {journal}
  {\bibinfo  {journal} {Phys. Rev. E}\ }\textbf {\bibinfo {volume} {54}},\
  \bibinfo {pages} {R2232–R2235} (\bibinfo {year} {1996})}\BibitemShut
  {NoStop}%
\bibitem [{\citenamefont {Baluku}\ and\ \citenamefont
  {Hellberg}(2015)}]{BalukuHellberg15/08}%
  \BibitemOpen
  \bibfield  {author} {\bibinfo {author} {\bibfnamefont {T.~K.}\ \bibnamefont
  {Baluku}}\ and\ \bibinfo {author} {\bibfnamefont {M.~A.}\ \bibnamefont
  {Hellberg}},\ }\href {\doibase 10.1063/1.4927581} {\bibfield  {journal}
  {\bibinfo  {journal} {Phys. Plasmas}\ }\textbf {\bibinfo {volume} {22}},\
  \bibinfo {eid} {083701} (\bibinfo {year} {2015}),\
  10.1063/1.4927581}\BibitemShut {NoStop}%
\bibitem [{\citenamefont {Bhakta}, \citenamefont {Ghosh},\ and\ \citenamefont
  {Sarkar}(2017)}]{Bhakta+17/02}%
  \BibitemOpen
  \bibfield  {author} {\bibinfo {author} {\bibfnamefont {S.}~\bibnamefont
  {Bhakta}}, \bibinfo {author} {\bibfnamefont {U.}~\bibnamefont {Ghosh}}, \
  and\ \bibinfo {author} {\bibfnamefont {S.}~\bibnamefont {Sarkar}},\ }\href
  {\doibase 10.1063/1.4976711} {\bibfield  {journal} {\bibinfo  {journal}
  {Phys. Plasmas}\ }\textbf {\bibinfo {volume} {24}},\ \bibinfo {pages}
  {023704} (\bibinfo {year} {2017})}\BibitemShut {NoStop}%
\bibitem [{\citenamefont {Momot}, \citenamefont {Zagorodny},\ and\
  \citenamefont {Momot}(2018)}]{Momot+18/07}%
  \BibitemOpen
  \bibfield  {author} {\bibinfo {author} {\bibfnamefont {A.~I.}\ \bibnamefont
  {Momot}}, \bibinfo {author} {\bibfnamefont {A.~G.}\ \bibnamefont
  {Zagorodny}}, \ and\ \bibinfo {author} {\bibfnamefont {O.~V.}\ \bibnamefont
  {Momot}},\ }\href {\doibase 10.1063/1.5042161} {\bibfield  {journal}
  {\bibinfo  {journal} {Phys. Plasmas}\ }\textbf {\bibinfo {volume} {25}},\
  \bibinfo {pages} {073706} (\bibinfo {year} {2018})}\BibitemShut {NoStop}%
\bibitem [{\citenamefont {de~Juli}\ and\ \citenamefont
  {Schneider}(1998)}]{deJuliSchneider98/09}%
  \BibitemOpen
  \bibfield  {author} {\bibinfo {author} {\bibfnamefont {M.~C.}\ \bibnamefont
  {de~Juli}}\ and\ \bibinfo {author} {\bibfnamefont {R.~S.}\ \bibnamefont
  {Schneider}},\ }\href {\doibase 10.1017/S0022377898006849} {\bibfield
  {journal} {\bibinfo  {journal} {J. Plasma Phys.}\ }\textbf {\bibinfo {volume}
  {60}},\ \bibinfo {pages} {243–263} (\bibinfo {year} {1998})}\BibitemShut
  {NoStop}%
\bibitem [{\citenamefont {Falceta-Gonçalves}\ and\ \citenamefont
  {Jatenco-Pereira}(2002)}]{Falceta-GoncalvesJatenco-Pereira02/09}%
  \BibitemOpen
  \bibfield  {author} {\bibinfo {author} {\bibfnamefont {D.}~\bibnamefont
  {Falceta-Gonçalves}}\ and\ \bibinfo {author} {\bibfnamefont
  {V.}~\bibnamefont {Jatenco-Pereira}},\ }\href {\doibase 10.1086/341794}
  {\bibfield  {journal} {\bibinfo  {journal} {Astrophys. J.}\ }\textbf
  {\bibinfo {volume} {576}},\ \bibinfo {pages} {976–981} (\bibinfo {year}
  {2002})}\BibitemShut {NoStop}%
\bibitem [{\citenamefont {de~Juli}\ \emph {et~al.}(2005)\citenamefont
  {de~Juli}, \citenamefont {Schneider}, \citenamefont {Ziebell},\ and\
  \citenamefont {Jatenco-Pereira}}]{deJuli+05/05}%
  \BibitemOpen
  \bibfield  {author} {\bibinfo {author} {\bibfnamefont {M.~C.}\ \bibnamefont
  {de~Juli}}, \bibinfo {author} {\bibfnamefont {R.~S.}\ \bibnamefont
  {Schneider}}, \bibinfo {author} {\bibfnamefont {L.~F.}\ \bibnamefont
  {Ziebell}}, \ and\ \bibinfo {author} {\bibfnamefont {V.}~\bibnamefont
  {Jatenco-Pereira}},\ }\href {\doibase 10.1063/1.1899647} {\bibfield
  {journal} {\bibinfo  {journal} {Phys. Plasmas}\ }\textbf {\bibinfo {volume}
  {12}},\ \bibinfo {pages} {052109} (\bibinfo {year} {2005})}\BibitemShut
  {NoStop}%
\bibitem [{\citenamefont {Vidotto}\ and\ \citenamefont
  {Jatenco-Pereira}(2006)}]{VidottoJatenco-Pereira06/03}%
  \BibitemOpen
  \bibfield  {author} {\bibinfo {author} {\bibfnamefont {A.~A.}\ \bibnamefont
  {Vidotto}}\ and\ \bibinfo {author} {\bibfnamefont {V.}~\bibnamefont
  {Jatenco-Pereira}},\ }\href {\doibase 10.1086/499329} {\bibfield  {journal}
  {\bibinfo  {journal} {Astrophys. J.}\ }\textbf {\bibinfo {volume} {639}},\
  \bibinfo {pages} {416–422} (\bibinfo {year} {2006})}\BibitemShut {NoStop}%
\bibitem [{\citenamefont {{de Juli}}\ \emph {et~al.}(2007)\citenamefont {{de
  Juli}}, \citenamefont {Schneider}, \citenamefont {Ziebell},\ and\
  \citenamefont {Gaelzer}}]{deJuli+07/02}%
  \BibitemOpen
  \bibfield  {author} {\bibinfo {author} {\bibfnamefont {M.~C.}\ \bibnamefont
  {{de Juli}}}, \bibinfo {author} {\bibfnamefont {R.~S.}\ \bibnamefont
  {Schneider}}, \bibinfo {author} {\bibfnamefont {L.~F.}\ \bibnamefont
  {Ziebell}}, \ and\ \bibinfo {author} {\bibfnamefont {R.}~\bibnamefont
  {Gaelzer}},\ }\href {\doibase 10.1063/1.2435704} {\bibfield  {journal}
  {\bibinfo  {journal} {Phys. Plasmas}\ }\textbf {\bibinfo {volume} {14}},\
  \bibinfo {eid} {022104} (\bibinfo {year} {2007})}\BibitemShut {NoStop}%
\bibitem [{\citenamefont {de~Juli}\ \emph {et~al.}(2007)\citenamefont
  {de~Juli}, \citenamefont {Schneider}, \citenamefont {Ziebell},\ and\
  \citenamefont {Gaelzer}}]{deJuli+07/10}%
  \BibitemOpen
  \bibfield  {author} {\bibinfo {author} {\bibfnamefont {M.~C.}\ \bibnamefont
  {de~Juli}}, \bibinfo {author} {\bibfnamefont {R.~S.}\ \bibnamefont
  {Schneider}}, \bibinfo {author} {\bibfnamefont {L.~F.}\ \bibnamefont
  {Ziebell}}, \ and\ \bibinfo {author} {\bibfnamefont {R.}~\bibnamefont
  {Gaelzer}},\ }\href {\doibase 10.1029/2007JA012434} {\bibfield  {journal}
  {\bibinfo  {journal} {J. Geophys. Res.}\ }\textbf {\bibinfo {volume} {112}},\
  \bibinfo {pages} {A10105} (\bibinfo {year} {2007})}\BibitemShut {NoStop}%
\bibitem [{\citenamefont {Gaelzer}\ \emph {et~al.}(2009)\citenamefont
  {Gaelzer}, \citenamefont {de~Juli}, \citenamefont {Schneider},\ and\
  \citenamefont {Ziebell}}]{Gaelzer+09/01}%
  \BibitemOpen
  \bibfield  {author} {\bibinfo {author} {\bibfnamefont {R.}~\bibnamefont
  {Gaelzer}}, \bibinfo {author} {\bibfnamefont {M.~C.}\ \bibnamefont
  {de~Juli}}, \bibinfo {author} {\bibfnamefont {R.~S.}\ \bibnamefont
  {Schneider}}, \ and\ \bibinfo {author} {\bibfnamefont {L.~F.}\ \bibnamefont
  {Ziebell}},\ }\href {\doibase 10.1088/0741-3335} {\bibfield  {journal}
  {\bibinfo  {journal} {Plasma Phys. Cont. Fusion}\ }\textbf {\bibinfo {volume}
  {51}},\ \bibinfo {pages} {015011 (17pp)} (\bibinfo {year}
  {2009})}\BibitemShut {NoStop}%
\bibitem [{\citenamefont {de~Juli}\ \emph {et~al.}(2009)\citenamefont
  {de~Juli}, \citenamefont {Schneider}, \citenamefont {Ziebell},\ and\
  \citenamefont {Gaelzer}}]{deJuli+09/03}%
  \BibitemOpen
  \bibfield  {author} {\bibinfo {author} {\bibfnamefont {M.~C.}\ \bibnamefont
  {de~Juli}}, \bibinfo {author} {\bibfnamefont {R.~S.}\ \bibnamefont
  {Schneider}}, \bibinfo {author} {\bibfnamefont {L.~F.}\ \bibnamefont
  {Ziebell}}, \ and\ \bibinfo {author} {\bibfnamefont {R.}~\bibnamefont
  {Gaelzer}},\ }\href {\doibase 10.1590/S0103-97332009000100019} {\bibfield
  {journal} {\bibinfo  {journal} {Braz. J. Phys.}\ }\textbf {\bibinfo {volume}
  {39}},\ \bibinfo {pages} {111–132} (\bibinfo {year} {2009})}\BibitemShut
  {NoStop}%
\bibitem [{\citenamefont {Jatenco-Pereira}, \citenamefont {Chian},\ and\
  \citenamefont {Rubab}(2014)}]{Jatenco-Pereira+14/03}%
  \BibitemOpen
  \bibfield  {author} {\bibinfo {author} {\bibfnamefont {V.}~\bibnamefont
  {Jatenco-Pereira}}, \bibinfo {author} {\bibfnamefont {A.~C.-L.}\ \bibnamefont
  {Chian}}, \ and\ \bibinfo {author} {\bibfnamefont {N.}~\bibnamefont
  {Rubab}},\ }\href {\doibase 10.5194/npg-21-405-2014} {\bibfield  {journal}
  {\bibinfo  {journal} {Nonl. Proc. Geophys.}\ }\textbf {\bibinfo {volume}
  {21}},\ \bibinfo {pages} {405–416} (\bibinfo {year} {2014})}\BibitemShut
  {NoStop}%
\bibitem [{\citenamefont {{De Toni}}\ and\ \citenamefont
  {Gaelzer}(2021)}]{deToniGaelzer21/11}%
  \BibitemOpen
  \bibfield  {author} {\bibinfo {author} {\bibfnamefont {L.~B.}\ \bibnamefont
  {{De Toni}}}\ and\ \bibinfo {author} {\bibfnamefont {R.}~\bibnamefont
  {Gaelzer}},\ }\href {\doibase 10.1093/mnras/stab2603} {\bibfield  {journal}
  {\bibinfo  {journal} {Mont. Not. R. Astron. Soc.}\ }\textbf {\bibinfo
  {volume} {508}},\ \bibinfo {pages} {340–351} (\bibinfo {year}
  {2021})}\BibitemShut {NoStop}%
\bibitem [{\citenamefont {{De Toni}}, \citenamefont {Gaelzer},\ and\
  \citenamefont {Ziebell}(2022{\natexlab{a}})}]{deToni+22/11}%
  \BibitemOpen
  \bibfield  {author} {\bibinfo {author} {\bibfnamefont {L.~B.}\ \bibnamefont
  {{De Toni}}}, \bibinfo {author} {\bibfnamefont {R.}~\bibnamefont {Gaelzer}},
  \ and\ \bibinfo {author} {\bibfnamefont {L.~F.}\ \bibnamefont {Ziebell}},\
  }\href {\doibase 10.1093/mnras/stac2574} {\bibfield  {journal} {\bibinfo
  {journal} {Mont. Not. R. Astron. Soc.}\ }\textbf {\bibinfo {volume} {516}},\
  \bibinfo {pages} {4650–4659} (\bibinfo {year}
  {2022}{\natexlab{a}})}\BibitemShut {NoStop}%
\bibitem [{\citenamefont {{De Toni}}, \citenamefont {Gaelzer},\ and\
  \citenamefont {Ziebell}(2022{\natexlab{b}})}]{deToni+22/05}%
  \BibitemOpen
  \bibfield  {author} {\bibinfo {author} {\bibfnamefont {L.~B.}\ \bibnamefont
  {{De Toni}}}, \bibinfo {author} {\bibfnamefont {R.}~\bibnamefont {Gaelzer}},
  \ and\ \bibinfo {author} {\bibfnamefont {L.~F.}\ \bibnamefont {Ziebell}},\
  }\href {\doibase 10.1093/mnras/stac547} {\bibfield  {journal} {\bibinfo
  {journal} {Mont. Not. R. Astron. Soc.}\ }\textbf {\bibinfo {volume} {512}},\
  \bibinfo {pages} {1795–1804} (\bibinfo {year}
  {2022}{\natexlab{b}})}\BibitemShut {NoStop}%
\bibitem [{\citenamefont {{De Toni}}, \citenamefont {Gaelzer},\ and\
  \citenamefont {Ziebell}(2024)}]{deToni+24/02}%
  \BibitemOpen
  \bibfield  {author} {\bibinfo {author} {\bibfnamefont {L.~B.}\ \bibnamefont
  {{De Toni}}}, \bibinfo {author} {\bibfnamefont {R.}~\bibnamefont {Gaelzer}},
  \ and\ \bibinfo {author} {\bibfnamefont {L.~F.}\ \bibnamefont {Ziebell}},\
  }\href {\doibase 10.1093/mnras/stae532} {\bibfield  {journal} {\bibinfo
  {journal} {Mont. Not. R. Astron. Soc.}\ }\textbf {\bibinfo {volume} {529}},\
  \bibinfo {pages} {3003–3012} (\bibinfo {year} {2024})}\BibitemShut
  {NoStop}%
\bibitem [{\citenamefont {Treumann}, \citenamefont {Jaroschek},\ and\
  \citenamefont {Scholer}(2004)}]{Treumann+04/04}%
  \BibitemOpen
  \bibfield  {author} {\bibinfo {author} {\bibfnamefont {R.~A.}\ \bibnamefont
  {Treumann}}, \bibinfo {author} {\bibfnamefont {C.~H.}\ \bibnamefont
  {Jaroschek}}, \ and\ \bibinfo {author} {\bibfnamefont {M.}~\bibnamefont
  {Scholer}},\ }\href {\doibase 10.1063/1.1667498} {\bibfield  {journal}
  {\bibinfo  {journal} {Phys. Plasmas}\ }\textbf {\bibinfo {volume} {11}},\
  \bibinfo {pages} {1317–1325} (\bibinfo {year} {2004})}\BibitemShut
  {NoStop}%
\bibitem [{\citenamefont {Goldstein}\ \emph {et~al.}(2015)\citenamefont
  {Goldstein}, \citenamefont {Wicks}, \citenamefont {Perri},\ and\
  \citenamefont {Sahraoui}}]{Goldstein+15/04}%
  \BibitemOpen
  \bibfield  {author} {\bibinfo {author} {\bibfnamefont {M.~L.}\ \bibnamefont
  {Goldstein}}, \bibinfo {author} {\bibfnamefont {R.~T.}\ \bibnamefont
  {Wicks}}, \bibinfo {author} {\bibfnamefont {S.}~\bibnamefont {Perri}}, \ and\
  \bibinfo {author} {\bibfnamefont {F.}~\bibnamefont {Sahraoui}},\ }\href
  {\doibase 10.1098/rsta.2014.0147} {\bibfield  {journal} {\bibinfo  {journal}
  {Phil. Trans. R. Soc. A}\ }\textbf {\bibinfo {volume} {373}} (\bibinfo {year}
  {2015}),\ 10.1098/rsta.2014.0147}\BibitemShut {NoStop}%
\bibitem [{\citenamefont {{Howes G. G.}}(2015)}]{Howes15/05}%
  \BibitemOpen
  \bibfield  {author} {\bibinfo {author} {\bibnamefont {{Howes G. G.}}},\
  }\href {\doibase 10.1098/rsta.2014.0145} {\bibfield  {journal} {\bibinfo
  {journal} {Phil. Trans. R. Soc. A}\ }\textbf {\bibinfo {volume} {373}}
  (\bibinfo {year} {2015}),\ 10.1098/rsta.2014.0145}\BibitemShut {NoStop}%
\bibitem [{\citenamefont {Ferrière}(2020)}]{Ferriere20/01}%
  \BibitemOpen
  \bibfield  {author} {\bibinfo {author} {\bibfnamefont {K.}~\bibnamefont
  {Ferrière}},\ }\href {\doibase 10.1088/1361-6587/ab49eb} {\bibfield
  {journal} {\bibinfo  {journal} {Plasma Phys. Cont. Fusion}\ }\textbf
  {\bibinfo {volume} {62}},\ \bibinfo {pages} {014014} (\bibinfo {year}
  {2020})}\BibitemShut {NoStop}%
\bibitem [{\citenamefont {Sahraoui}, \citenamefont {Hadid},\ and\ \citenamefont
  {Huang}(2020)}]{Sahraoui+20/12}%
  \BibitemOpen
  \bibfield  {author} {\bibinfo {author} {\bibfnamefont {F.}~\bibnamefont
  {Sahraoui}}, \bibinfo {author} {\bibfnamefont {L.}~\bibnamefont {Hadid}}, \
  and\ \bibinfo {author} {\bibfnamefont {S.}~\bibnamefont {Huang}},\ }\href
  {\doibase 10.1007/s41614-020-0040-2} {\bibfield  {journal} {\bibinfo
  {journal} {Rev. Mod. Plasma Phys.}\ }\textbf {\bibinfo {volume} {4}},\
  \bibinfo {pages} {4} (\bibinfo {year} {2020})}\BibitemShut {NoStop}%
\bibitem [{\citenamefont {Livadiotis}(2017)}]{Livadiotis17}%
  \BibitemOpen
  \bibfield  {author} {\bibinfo {author} {\bibfnamefont {G.}~\bibnamefont
  {Livadiotis}},\ }\href {https://www.elsevier.com/books/kappa-distributions/
  livadiotis/978-0-12-804638-8} {\emph {\bibinfo {title} {{Kappa Distributions:
  Theory and Applications in Plasmas}}}}\ (\bibinfo  {publisher} {Elsevier
  Science \& Technology Books},\ \bibinfo {year} {2017})\BibitemShut {NoStop}%
\bibitem [{\citenamefont {Lazar}\ and\ \citenamefont
  {Fichtner}(2021)}]{LazarFichtner21}%
  \BibitemOpen
  \bibinfo {editor} {\bibfnamefont {M.}~\bibnamefont {Lazar}}\ and\ \bibinfo
  {editor} {\bibfnamefont {H.}~\bibnamefont {Fichtner}},\ eds.,\ \href
  {\doibase 10.1007/978-3-030-82623-9} {\emph {\bibinfo {title} {{Kappa
  Distributions: From Observational Evidences via Controversial Predictions to
  a Consistent Theory of Nonequilibrium Plasmas}}}},\ \bibinfo {series}
  {{Astrophysics and Space Science Library}}, Vol.\ \bibinfo {volume} {464}\
  (\bibinfo  {publisher} {Springer},\ \bibinfo {address} {Cham},\ \bibinfo
  {year} {2021})\BibitemShut {NoStop}%
\bibitem [{\citenamefont {Gaelzer}, \citenamefont {de~Juli},\ and\
  \citenamefont {Ziebell}(2010)}]{Gaelzer+10/09}%
  \BibitemOpen
  \bibfield  {author} {\bibinfo {author} {\bibfnamefont {R.}~\bibnamefont
  {Gaelzer}}, \bibinfo {author} {\bibfnamefont {M.~C.}\ \bibnamefont
  {de~Juli}}, \ and\ \bibinfo {author} {\bibfnamefont {L.~F.}\ \bibnamefont
  {Ziebell}},\ }\href {\doibase 10.1029/2009JA015217} {\bibfield  {journal}
  {\bibinfo  {journal} {J. Geophys. Res.}\ }\textbf {\bibinfo {volume} {115}}
  (\bibinfo {year} {2010}),\ 10.1029/2009JA015217}\BibitemShut {NoStop}%
\bibitem [{\citenamefont {Rubab}\ \emph {et~al.}(2010)\citenamefont {Rubab},
  \citenamefont {Erkaev}, \citenamefont {Langmayr},\ and\ \citenamefont
  {Biernat}}]{Rubab+10/10}%
  \BibitemOpen
  \bibfield  {author} {\bibinfo {author} {\bibfnamefont {N.}~\bibnamefont
  {Rubab}}, \bibinfo {author} {\bibfnamefont {N.~V.}\ \bibnamefont {Erkaev}},
  \bibinfo {author} {\bibfnamefont {D.}~\bibnamefont {Langmayr}}, \ and\
  \bibinfo {author} {\bibfnamefont {H.~K.}\ \bibnamefont {Biernat}},\ }\href
  {\doibase 10.1063/1.3491336} {\bibfield  {journal} {\bibinfo  {journal}
  {Phys. Plasmas}\ }\textbf {\bibinfo {volume} {17}},\ \bibinfo {eid} {103704}
  (\bibinfo {year} {2010})}\BibitemShut {NoStop}%
\bibitem [{\citenamefont {Deeba}, \citenamefont {Ahmad},\ and\ \citenamefont
  {Murtaza}(2011)}]{Deeba+11/07}%
  \BibitemOpen
  \bibfield  {author} {\bibinfo {author} {\bibfnamefont {F.}~\bibnamefont
  {Deeba}}, \bibinfo {author} {\bibfnamefont {Z.}~\bibnamefont {Ahmad}}, \ and\
  \bibinfo {author} {\bibfnamefont {G.}~\bibnamefont {Murtaza}},\ }\href
  {\doibase 10.1063/1.3601763} {\bibfield  {journal} {\bibinfo  {journal}
  {Phys. Plasmas}\ }\textbf {\bibinfo {volume} {18}},\ \bibinfo {pages}
  {072104} (\bibinfo {year} {2011})}\BibitemShut {NoStop}%
\bibitem [{\citenamefont {Galvão}\ \emph {et~al.}(2011)\citenamefont
  {Galvão}, \citenamefont {Ziebell}, \citenamefont {Gaelzer},\ and\
  \citenamefont {de~Juli}}]{Galvao+11/12}%
  \BibitemOpen
  \bibfield  {author} {\bibinfo {author} {\bibfnamefont {R.}~\bibnamefont
  {Galvão}}, \bibinfo {author} {\bibfnamefont {L.~F.}\ \bibnamefont
  {Ziebell}}, \bibinfo {author} {\bibfnamefont {R.}~\bibnamefont {Gaelzer}}, \
  and\ \bibinfo {author} {\bibfnamefont {M.~C.}\ \bibnamefont {de~Juli}},\
  }\href {\doibase 10.1007/s13538-011-0041-2} {\bibfield  {journal} {\bibinfo
  {journal} {Braz. J. Phys.}\ }\textbf {\bibinfo {volume} {41}},\ \bibinfo
  {pages} {258–274} (\bibinfo {year} {2011})}\BibitemShut {NoStop}%
\bibitem [{\citenamefont {Galvão}\ \emph {et~al.}(2012)\citenamefont
  {Galvão}, \citenamefont {Ziebell}, \citenamefont {Gaelzer},\ and\
  \citenamefont {de~Juli}}]{Galvao+12/12}%
  \BibitemOpen
  \bibfield  {author} {\bibinfo {author} {\bibfnamefont {R.~A.}\ \bibnamefont
  {Galvão}}, \bibinfo {author} {\bibfnamefont {L.~F.}\ \bibnamefont
  {Ziebell}}, \bibinfo {author} {\bibfnamefont {R.}~\bibnamefont {Gaelzer}}, \
  and\ \bibinfo {author} {\bibfnamefont {M.~C.}\ \bibnamefont {de~Juli}},\
  }\href {\doibase 10.1063/1.4772771} {\bibfield  {journal} {\bibinfo
  {journal} {Phys. Plasmas}\ }\textbf {\bibinfo {volume} {19}},\ \bibinfo {eid}
  {123705} (\bibinfo {year} {2012}),\ 10.1063/1.4772771}\BibitemShut {NoStop}%
\bibitem [{\citenamefont {Shahmansouri}\ and\ \citenamefont
  {Tribeche}(2012)}]{ShahmansouriTribeche12/11}%
  \BibitemOpen
  \bibfield  {author} {\bibinfo {author} {\bibfnamefont {M.}~\bibnamefont
  {Shahmansouri}}\ and\ \bibinfo {author} {\bibfnamefont {M.}~\bibnamefont
  {Tribeche}},\ }\href {\doibase 10.1007/s10509-012-1149-8} {\bibfield
  {journal} {\bibinfo  {journal} {Astrophys. Space Sci.}\ }\textbf {\bibinfo
  {volume} {342}},\ \bibinfo {pages} {87–92} (\bibinfo {year}
  {2012})}\BibitemShut {NoStop}%
\bibitem [{\citenamefont {dos Santos}, \citenamefont {Ziebell},\ and\
  \citenamefont {Gaelzer}(2016)}]{dosSantos+16/01}%
  \BibitemOpen
  \bibfield  {author} {\bibinfo {author} {\bibfnamefont {M.~S.}\ \bibnamefont
  {dos Santos}}, \bibinfo {author} {\bibfnamefont {L.~F.}\ \bibnamefont
  {Ziebell}}, \ and\ \bibinfo {author} {\bibfnamefont {R.}~\bibnamefont
  {Gaelzer}},\ }\href {\doibase 10.1063/1.4939885} {\bibfield  {journal}
  {\bibinfo  {journal} {Phys. Plasmas}\ }\textbf {\bibinfo {volume} {23}},\
  \bibinfo {eid} {013705} (\bibinfo {year} {2016}),\
  10.1063/1.4939885}\BibitemShut {NoStop}%
\bibitem [{\citenamefont {dos Santos}, \citenamefont {Ziebell},\ and\
  \citenamefont {Gaelzer}(2017)}]{dosSantos+17/01}%
  \BibitemOpen
  \bibfield  {author} {\bibinfo {author} {\bibfnamefont {M.~S.}\ \bibnamefont
  {dos Santos}}, \bibinfo {author} {\bibfnamefont {L.~F.}\ \bibnamefont
  {Ziebell}}, \ and\ \bibinfo {author} {\bibfnamefont {R.}~\bibnamefont
  {Gaelzer}},\ }\href {\doibase 10.1007/s10509-016-2997-4} {\bibfield
  {journal} {\bibinfo  {journal} {Astrophys. Space Sci.}\ }\textbf {\bibinfo
  {volume} {362}},\ \bibinfo {pages} {18} (\bibinfo {year} {2017})}\BibitemShut
  {NoStop}%
\bibitem [{\citenamefont {Ziebell}, \citenamefont {Gaelzer},\ and\
  \citenamefont {Simões}(2017)}]{Ziebell+17/10}%
  \BibitemOpen
  \bibfield  {author} {\bibinfo {author} {\bibfnamefont {L.~F.}\ \bibnamefont
  {Ziebell}}, \bibinfo {author} {\bibfnamefont {R.}~\bibnamefont {Gaelzer}}, \
  and\ \bibinfo {author} {\bibfnamefont {F.~J.~R.}\ \bibnamefont {Simões}},\
  }\href {\doibase 10.1017/S0022377817000733} {\bibfield  {journal} {\bibinfo
  {journal} {J. Plasma Phys.}\ }\textbf {\bibinfo {volume} {83}} (\bibinfo
  {year} {2017}),\ 10.1017/S0022377817000733}\BibitemShut {NoStop}%
\bibitem [{\citenamefont {Ziebell}\ and\ \citenamefont
  {Gaelzer}(2024)}]{ZiebellGaelzer24/11-arxiv}%
  \BibitemOpen
  \bibfield  {author} {\bibinfo {author} {\bibfnamefont {L.~F.}\ \bibnamefont
  {Ziebell}}\ and\ \bibinfo {author} {\bibfnamefont {R.}~\bibnamefont
  {Gaelzer}},\ }\href {\doibase 10.48550/arXiv.2411.03536} {\enquote {\bibinfo
  {title} {{Collisional charging of dust particles by suprathermal particles.
  {I} – Standard anisotropic Kappa distributions}},}\ } (\bibinfo {year}
  {2024}),\ \bibinfo {note} {submitted for publication}\BibitemShut {NoStop}%
\bibitem [{\citenamefont {Mann}\ \emph {et~al.}(2011)\citenamefont {Mann},
  \citenamefont {Pellinen-Wannberg}, \citenamefont {Murad}, \citenamefont
  {Popova}, \citenamefont {Meyer-Vernet}, \citenamefont {Rosenberg},
  \citenamefont {Mukai}, \citenamefont {Czechowski}, \citenamefont {Mukai},
  \citenamefont {Safrankova},\ and\ \citenamefont {Nemecek}}]{Mann+11/11}%
  \BibitemOpen
  \bibfield  {author} {\bibinfo {author} {\bibfnamefont {I.}~\bibnamefont
  {Mann}}, \bibinfo {author} {\bibfnamefont {A.}~\bibnamefont
  {Pellinen-Wannberg}}, \bibinfo {author} {\bibfnamefont {E.}~\bibnamefont
  {Murad}}, \bibinfo {author} {\bibfnamefont {O.}~\bibnamefont {Popova}},
  \bibinfo {author} {\bibfnamefont {N.}~\bibnamefont {Meyer-Vernet}}, \bibinfo
  {author} {\bibfnamefont {M.}~\bibnamefont {Rosenberg}}, \bibinfo {author}
  {\bibfnamefont {T.}~\bibnamefont {Mukai}}, \bibinfo {author} {\bibfnamefont
  {A.}~\bibnamefont {Czechowski}}, \bibinfo {author} {\bibfnamefont
  {S.}~\bibnamefont {Mukai}}, \bibinfo {author} {\bibfnamefont
  {J.}~\bibnamefont {Safrankova}}, \ and\ \bibinfo {author} {\bibfnamefont
  {Z.}~\bibnamefont {Nemecek}},\ }\href {\doibase 10.1007/s11214-011-9762-3}
  {\bibfield  {journal} {\bibinfo  {journal} {Space Sci. Rev.}\ }\textbf
  {\bibinfo {volume} {161}},\ \bibinfo {pages} {1–47} (\bibinfo {year}
  {2011})}\BibitemShut {NoStop}%
\bibitem [{\citenamefont {Misra}\ and\ \citenamefont
  {Mishra}(2013)}]{MisraMishra13/07}%
  \BibitemOpen
  \bibfield  {author} {\bibinfo {author} {\bibfnamefont {S.}~\bibnamefont
  {Misra}}\ and\ \bibinfo {author} {\bibfnamefont {S.~K.}\ \bibnamefont
  {Mishra}},\ }\href {\doibase 10.1093/mnras/stt661} {\bibfield  {journal}
  {\bibinfo  {journal} {Mont. Not. R. Astron. Soc.}\ }\textbf {\bibinfo
  {volume} {432}},\ \bibinfo {pages} {2985–2993} (\bibinfo {year}
  {2013})}\BibitemShut {NoStop}%
\bibitem [{\citenamefont {Ma}\ \emph {et~al.}(2013)\citenamefont {Ma},
  \citenamefont {Matthews}, \citenamefont {Land},\ and\ \citenamefont
  {Hyde}}]{Ma+13/02}%
  \BibitemOpen
  \bibfield  {author} {\bibinfo {author} {\bibfnamefont {Q.}~\bibnamefont
  {Ma}}, \bibinfo {author} {\bibfnamefont {L.~S.}\ \bibnamefont {Matthews}},
  \bibinfo {author} {\bibfnamefont {V.}~\bibnamefont {Land}}, \ and\ \bibinfo
  {author} {\bibfnamefont {T.~W.}\ \bibnamefont {Hyde}},\ }\href {\doibase
  10.1088/0004-637X/763/2/77} {\bibfield  {journal} {\bibinfo  {journal}
  {Astrophys. J.}\ }\textbf {\bibinfo {volume} {763}},\ \bibinfo {pages} {77}
  (\bibinfo {year} {2013})}\BibitemShut {NoStop}%
\bibitem [{\citenamefont {Allen}(1992)}]{Allen92/05}%
  \BibitemOpen
  \bibfield  {author} {\bibinfo {author} {\bibfnamefont {J.~E.}\ \bibnamefont
  {Allen}},\ }\href {\doibase 10.1088/0031-8949/45/5/013} {\bibfield  {journal}
  {\bibinfo  {journal} {Phys. Scripta}\ }\textbf {\bibinfo {volume} {45}},\
  \bibinfo {pages} {497} (\bibinfo {year} {1992})}\BibitemShut {NoStop}%
\bibitem [{\citenamefont {Melzer}(2019)}]{Melzer19}%
  \BibitemOpen
  \bibfield  {author} {\bibinfo {author} {\bibfnamefont {A.}~\bibnamefont
  {Melzer}},\ }\href {\doibase 10.1007/978-3-030-20260-6} {\emph {\bibinfo
  {title} {{Physics of Dusty Plasmas: An Introduction}}}},\ {Lecture Notes in
  Physics}\ (\bibinfo  {publisher} {Springer Cham},\ \bibinfo {year} {2019})\
  \bibinfo {note} {235 + x pp.}\BibitemShut {Stop}%
\bibitem [{\citenamefont {Salimullah}, \citenamefont {Sandberg},\ and\
  \citenamefont {Shukla}(2003)}]{SalimullahSandbergShukla03/08}%
  \BibitemOpen
  \bibfield  {author} {\bibinfo {author} {\bibfnamefont {M.}~\bibnamefont
  {Salimullah}}, \bibinfo {author} {\bibfnamefont {I.}~\bibnamefont
  {Sandberg}}, \ and\ \bibinfo {author} {\bibfnamefont {P.~K.}\ \bibnamefont
  {Shukla}},\ }\href {\doibase 10.1103/PhysRevE.68.027403} {\bibfield
  {journal} {\bibinfo  {journal} {Phys. Rev. E}\ }\textbf {\bibinfo {volume}
  {68}},\ \bibinfo {eid} {027403} (\bibinfo {year} {2003})}\BibitemShut
  {NoStop}%
\bibitem [{\citenamefont {Chang}\ and\ \citenamefont
  {Spariosu}(1993)}]{ChangSpariosu93/01}%
  \BibitemOpen
  \bibfield  {author} {\bibinfo {author} {\bibfnamefont {J.~S.}\ \bibnamefont
  {Chang}}\ and\ \bibinfo {author} {\bibfnamefont {K.}~\bibnamefont
  {Spariosu}},\ }\href {\doibase 10.1143/JPSJ.62.97} {\bibfield  {journal}
  {\bibinfo  {journal} {J. Phys. Soc. Jpn.}\ }\textbf {\bibinfo {volume}
  {62}},\ \bibinfo {pages} {97–104} (\bibinfo {year} {1993})}\BibitemShut
  {NoStop}%
\bibitem [{\citenamefont {Davari}, \citenamefont {Farokhi},\ and\ \citenamefont
  {{Ali Asgarian}}(2023)}]{Davari+23/01}%
  \BibitemOpen
  \bibfield  {author} {\bibinfo {author} {\bibfnamefont {H.}~\bibnamefont
  {Davari}}, \bibinfo {author} {\bibfnamefont {B.}~\bibnamefont {Farokhi}}, \
  and\ \bibinfo {author} {\bibfnamefont {M.}~\bibnamefont {{Ali Asgarian}}},\
  }\href {\doibase 10.1038/s41598-023-28310-y} {\bibfield  {journal} {\bibinfo
  {journal} {Sci. Rep.}\ }\textbf {\bibinfo {volume} {13}},\ \bibinfo {pages}
  {1111} (\bibinfo {year} {2023})}\BibitemShut {NoStop}%
\bibitem [{\citenamefont {Vasyliunas}(1968)}]{Vasyliunas68/05}%
  \BibitemOpen
  \bibfield  {author} {\bibinfo {author} {\bibfnamefont {V.~M.}\ \bibnamefont
  {Vasyliunas}},\ }\href {\doibase 10.1029/JA073i009p02839} {\bibfield
  {journal} {\bibinfo  {journal} {J. Geophys. Res.}\ }\textbf {\bibinfo
  {volume} {73}},\ \bibinfo {pages} {2839–2884} (\bibinfo {year}
  {1968})}\BibitemShut {NoStop}%
\bibitem [{\citenamefont {Hellberg}\ \emph {et~al.}(2009)\citenamefont
  {Hellberg}, \citenamefont {Mace}, \citenamefont {Baluku}, \citenamefont
  {Kourakis},\ and\ \citenamefont {Saini}}]{Hellberg+09/09}%
  \BibitemOpen
  \bibfield  {author} {\bibinfo {author} {\bibfnamefont {M.~A.}\ \bibnamefont
  {Hellberg}}, \bibinfo {author} {\bibfnamefont {R.~L.}\ \bibnamefont {Mace}},
  \bibinfo {author} {\bibfnamefont {T.~K.}\ \bibnamefont {Baluku}}, \bibinfo
  {author} {\bibfnamefont {I.}~\bibnamefont {Kourakis}}, \ and\ \bibinfo
  {author} {\bibfnamefont {N.~S.}\ \bibnamefont {Saini}},\ }\href {\doibase
  10.1063/1.3213388} {\bibfield  {journal} {\bibinfo  {journal} {Phys.
  Plasmas}\ }\textbf {\bibinfo {volume} {16}},\ \bibinfo {eid} {094701}
  (\bibinfo {year} {2009})}\BibitemShut {NoStop}%
\bibitem [{\citenamefont {Hau}, \citenamefont {Fu},\ and\ \citenamefont
  {Chuang}(2009)}]{Hau+09/09}%
  \BibitemOpen
  \bibfield  {author} {\bibinfo {author} {\bibfnamefont {L.-N.}\ \bibnamefont
  {Hau}}, \bibinfo {author} {\bibfnamefont {W.-Z.}\ \bibnamefont {Fu}}, \ and\
  \bibinfo {author} {\bibfnamefont {S.-H.}\ \bibnamefont {Chuang}},\ }\href
  {\doibase 10.1063/1.3213389} {\bibfield  {journal} {\bibinfo  {journal}
  {Phys. Plasmas}\ }\textbf {\bibinfo {volume} {16}},\ \bibinfo {eid} {094702}
  (\bibinfo {year} {2009})}\BibitemShut {NoStop}%
\bibitem [{\citenamefont {Scherer}, \citenamefont {Fichtner},\ and\
  \citenamefont {Lazar}(2017)}]{Scherer+17/12}%
  \BibitemOpen
  \bibfield  {author} {\bibinfo {author} {\bibfnamefont {K.}~\bibnamefont
  {Scherer}}, \bibinfo {author} {\bibfnamefont {H.}~\bibnamefont {Fichtner}}, \
  and\ \bibinfo {author} {\bibfnamefont {M.}~\bibnamefont {Lazar}},\ }\href
  {\doibase 10.1209/0295-5075/120/50002} {\bibfield  {journal} {\bibinfo
  {journal} {Europhys. Lett.}\ }\textbf {\bibinfo {volume} {120}},\ \bibinfo
  {pages} {50002} (\bibinfo {year} {2017})}\BibitemShut {NoStop}%
\bibitem [{\citenamefont {Fichtner}\ \emph {et~al.}(2018)\citenamefont
  {Fichtner}, \citenamefont {Scherer}, \citenamefont {Lazar}, \citenamefont
  {Fahr},\ and\ \citenamefont {Vörös}}]{Fichtner+18/11}%
  \BibitemOpen
  \bibfield  {author} {\bibinfo {author} {\bibfnamefont {H.}~\bibnamefont
  {Fichtner}}, \bibinfo {author} {\bibfnamefont {K.}~\bibnamefont {Scherer}},
  \bibinfo {author} {\bibfnamefont {M.}~\bibnamefont {Lazar}}, \bibinfo
  {author} {\bibfnamefont {H.~J.}\ \bibnamefont {Fahr}}, \ and\ \bibinfo
  {author} {\bibfnamefont {Z.}~\bibnamefont {Vörös}},\ }\href {\doibase
  10.1103/PhysRevE.98.053205} {\bibfield  {journal} {\bibinfo  {journal} {Phys.
  Rev. E}\ }\textbf {\bibinfo {volume} {98}},\ \bibinfo {pages} {053205}
  (\bibinfo {year} {2018})}\BibitemShut {NoStop}%
\bibitem [{\citenamefont {Gaelzer}\ and\ \citenamefont
  {Ziebell}(2014)}]{GaelzerZiebell14/12}%
  \BibitemOpen
  \bibfield  {author} {\bibinfo {author} {\bibfnamefont {R.}~\bibnamefont
  {Gaelzer}}\ and\ \bibinfo {author} {\bibfnamefont {L.~F.}\ \bibnamefont
  {Ziebell}},\ }\href {\doibase 10.1002/2014JA020667} {\bibfield  {journal}
  {\bibinfo  {journal} {J. Geophys. Res.}\ }\textbf {\bibinfo {volume} {119}},\
  \bibinfo {pages} {9334–9356} (\bibinfo {year} {2014})}\BibitemShut
  {NoStop}%
\bibitem [{\citenamefont {Gaelzer}\ and\ \citenamefont
  {Ziebell}(2016)}]{GaelzerZiebell16/02}%
  \BibitemOpen
  \bibfield  {author} {\bibinfo {author} {\bibfnamefont {R.}~\bibnamefont
  {Gaelzer}}\ and\ \bibinfo {author} {\bibfnamefont {L.~F.}\ \bibnamefont
  {Ziebell}},\ }\href {\doibase 10.1063/1.4941260} {\bibfield  {journal}
  {\bibinfo  {journal} {Phys. Plasmas}\ }\textbf {\bibinfo {volume} {23}},\
  \bibinfo {eid} {022110} (\bibinfo {year} {2016})}\BibitemShut {NoStop}%
\bibitem [{\citenamefont {Askey}\ and\ \citenamefont
  {Roy}(2010)}]{AskeyRoy-Full-NIST10}%
  \BibitemOpen
  \bibfield  {author} {\bibinfo {author} {\bibfnamefont {R.~A.}\ \bibnamefont
  {Askey}}\ and\ \bibinfo {author} {\bibfnamefont {R.}~\bibnamefont {Roy}},\
  }in\ \href {http://dlmf.nist.gov/5} {\emph {\bibinfo {booktitle} {{NIST
  Handbook of Mathematical Functions}}}},\ \bibinfo {editor} {edited by\
  \bibinfo {editor} {\bibfnamefont {F.~W.~J.}\ \bibnamefont {Olver}}, \bibinfo
  {editor} {\bibfnamefont {D.~W.}\ \bibnamefont {Lozier}}, \bibinfo {editor}
  {\bibfnamefont {R.~F.}\ \bibnamefont {Boisvert}}, \ and\ \bibinfo {editor}
  {\bibfnamefont {C.~W.}\ \bibnamefont {Clark}}}\ (\bibinfo  {publisher}
  {Cambridge},\ \bibinfo {address} {New York},\ \bibinfo {year} {2010})\
  Chap.~\bibinfo {chapter} {5}, p.\ \bibinfo {pages} {135–147}\BibitemShut
  {NoStop}%
\bibitem [{\citenamefont {Gaelzer}, \citenamefont {Ziebell},\ and\
  \citenamefont {Meneses}(2016)}]{Gaelzer+16/06}%
  \BibitemOpen
  \bibfield  {author} {\bibinfo {author} {\bibfnamefont {R.}~\bibnamefont
  {Gaelzer}}, \bibinfo {author} {\bibfnamefont {L.~F.}\ \bibnamefont
  {Ziebell}}, \ and\ \bibinfo {author} {\bibfnamefont {A.~R.}\ \bibnamefont
  {Meneses}},\ }\href {\doibase 10.1063/1.4953430} {\bibfield  {journal}
  {\bibinfo  {journal} {Phys. Plasmas}\ }\textbf {\bibinfo {volume} {23}},\
  \bibinfo {eid} {062108} (\bibinfo {year} {2016})}\BibitemShut {NoStop}%
\bibitem [{\citenamefont {Maksimovic}\ \emph {et~al.}(2005)\citenamefont
  {Maksimovic}, \citenamefont {Zouganelis}, \citenamefont {Chaufray},
  \citenamefont {Issautier}, \citenamefont {Scime}, \citenamefont {Littleton},
  \citenamefont {Marsch}, \citenamefont {McComas}, \citenamefont {Salem},
  \citenamefont {Lin},\ and\ \citenamefont {Elliot}}]{Maksimovic+05/09}%
  \BibitemOpen
  \bibfield  {author} {\bibinfo {author} {\bibfnamefont {M.}~\bibnamefont
  {Maksimovic}}, \bibinfo {author} {\bibfnamefont {I.}~\bibnamefont
  {Zouganelis}}, \bibinfo {author} {\bibfnamefont {J.~Y.}\ \bibnamefont
  {Chaufray}}, \bibinfo {author} {\bibfnamefont {K.}~\bibnamefont {Issautier}},
  \bibinfo {author} {\bibfnamefont {E.~E.}\ \bibnamefont {Scime}}, \bibinfo
  {author} {\bibfnamefont {J.~E.}\ \bibnamefont {Littleton}}, \bibinfo {author}
  {\bibfnamefont {E.}~\bibnamefont {Marsch}}, \bibinfo {author} {\bibfnamefont
  {D.~J.}\ \bibnamefont {McComas}}, \bibinfo {author} {\bibfnamefont
  {C.}~\bibnamefont {Salem}}, \bibinfo {author} {\bibfnamefont {R.~P.}\
  \bibnamefont {Lin}}, \ and\ \bibinfo {author} {\bibfnamefont
  {H.}~\bibnamefont {Elliot}},\ }\href {\doibase 10.1029/2005JA011119}
  {\bibfield  {journal} {\bibinfo  {journal} {J. Geophys. Res.}\ }\textbf
  {\bibinfo {volume} {110}},\ \bibinfo {pages} {A09104} (\bibinfo {year}
  {2005})}\BibitemShut {NoStop}%
\bibitem [{\citenamefont {Štverák}\ \emph {et~al.}(2008)\citenamefont
  {Štverák}, \citenamefont {Trávníček}, \citenamefont {Maksimovic},
  \citenamefont {Marsch}, \citenamefont {Fazakerley},\ and\ \citenamefont
  {Scime}}]{Stverak+08/03}%
  \BibitemOpen
  \bibfield  {author} {\bibinfo {author} {\bibfnamefont {S.}~\bibnamefont
  {Štverák}}, \bibinfo {author} {\bibfnamefont {P.}~\bibnamefont
  {Trávníček}}, \bibinfo {author} {\bibfnamefont {M.}~\bibnamefont
  {Maksimovic}}, \bibinfo {author} {\bibfnamefont {E.}~\bibnamefont {Marsch}},
  \bibinfo {author} {\bibfnamefont {A.~N.}\ \bibnamefont {Fazakerley}}, \ and\
  \bibinfo {author} {\bibfnamefont {E.~E.}\ \bibnamefont {Scime}},\ }\href
  {\doibase 10.1029/2007JA012733} {\bibfield  {journal} {\bibinfo  {journal}
  {J. Geophys. Res.}\ }\textbf {\bibinfo {volume} {113}},\ \bibinfo {pages}
  {A03103} (\bibinfo {year} {2008})}\BibitemShut {NoStop}%
\bibitem [{\citenamefont {Štverák}\ \emph {et~al.}(2009)\citenamefont
  {Štverák}, \citenamefont {Maksimovic}, \citenamefont {Trávníček},
  \citenamefont {Marsch}, \citenamefont {Fazakerley},\ and\ \citenamefont
  {Scime}}]{Stverak+09/05}%
  \BibitemOpen
  \bibfield  {author} {\bibinfo {author} {\bibfnamefont {S.}~\bibnamefont
  {Štverák}}, \bibinfo {author} {\bibfnamefont {M.}~\bibnamefont
  {Maksimovic}}, \bibinfo {author} {\bibfnamefont {P.~M.}\ \bibnamefont
  {Trávníček}}, \bibinfo {author} {\bibfnamefont {E.}~\bibnamefont
  {Marsch}}, \bibinfo {author} {\bibfnamefont {A.~N.}\ \bibnamefont
  {Fazakerley}}, \ and\ \bibinfo {author} {\bibfnamefont {E.~E.}\ \bibnamefont
  {Scime}},\ }\href {\doibase 10.1029/2008JA013883} {\bibfield  {journal}
  {\bibinfo  {journal} {J. Geophys. Res.}\ }\textbf {\bibinfo {volume} {114}},\
  \bibinfo {pages} {A05104} (\bibinfo {year} {2009})}\BibitemShut {NoStop}%
\bibitem [{\citenamefont {{Wilson III}}\ \emph {et~al.}(2019)\citenamefont
  {{Wilson III}}, \citenamefont {Chen}, \citenamefont {Wang}, \citenamefont
  {Schwartz}, \citenamefont {Turner}, \citenamefont {Stevens}, \citenamefont
  {Kasper}, \citenamefont {Osmane}, \citenamefont {Caprioli}, \citenamefont
  {Bale}, \citenamefont {Pulupa}, \citenamefont {Salem},\ and\ \citenamefont
  {Goodrich}}]{WilsonIII+19/07}%
  \BibitemOpen
  \bibfield  {author} {\bibinfo {author} {\bibfnamefont {L.~B.}\ \bibnamefont
  {{Wilson III}}}, \bibinfo {author} {\bibfnamefont {L.-J.}\ \bibnamefont
  {Chen}}, \bibinfo {author} {\bibfnamefont {S.}~\bibnamefont {Wang}}, \bibinfo
  {author} {\bibfnamefont {S.~J.}\ \bibnamefont {Schwartz}}, \bibinfo {author}
  {\bibfnamefont {D.~L.}\ \bibnamefont {Turner}}, \bibinfo {author}
  {\bibfnamefont {M.~L.}\ \bibnamefont {Stevens}}, \bibinfo {author}
  {\bibfnamefont {J.~C.}\ \bibnamefont {Kasper}}, \bibinfo {author}
  {\bibfnamefont {A.}~\bibnamefont {Osmane}}, \bibinfo {author} {\bibfnamefont
  {D.}~\bibnamefont {Caprioli}}, \bibinfo {author} {\bibfnamefont {S.~D.}\
  \bibnamefont {Bale}}, \bibinfo {author} {\bibfnamefont {M.~P.}\ \bibnamefont
  {Pulupa}}, \bibinfo {author} {\bibfnamefont {C.~S.}\ \bibnamefont {Salem}}, \
  and\ \bibinfo {author} {\bibfnamefont {K.~A.}\ \bibnamefont {Goodrich}},\
  }\href {\doibase 10.3847/1538-4365/ab22bd} {\bibfield  {journal} {\bibinfo
  {journal} {apjss}\ }\textbf {\bibinfo {volume} {243}},\ \bibinfo {pages} {8}
  (\bibinfo {year} {2019})}\BibitemShut {NoStop}%
\bibitem [{\citenamefont {{Wilson III}}\ \emph {et~al.}(2022)\citenamefont
  {{Wilson III}}, \citenamefont {Goodrich}, \citenamefont {Turner},
  \citenamefont {Cohen}, \citenamefont {Whittlesey},\ and\ \citenamefont
  {Schwartz}}]{WilsonIII+22/11}%
  \BibitemOpen
  \bibfield  {author} {\bibinfo {author} {\bibfnamefont {L.~B.}\ \bibnamefont
  {{Wilson III}}}, \bibinfo {author} {\bibfnamefont {K.~A.}\ \bibnamefont
  {Goodrich}}, \bibinfo {author} {\bibfnamefont {D.~L.}\ \bibnamefont
  {Turner}}, \bibinfo {author} {\bibfnamefont {I.~J.}\ \bibnamefont {Cohen}},
  \bibinfo {author} {\bibfnamefont {P.~L.}\ \bibnamefont {Whittlesey}}, \ and\
  \bibinfo {author} {\bibfnamefont {S.~J.}\ \bibnamefont {Schwartz}},\ }\href
  {\doibase 10.3389/fspas.2022.1063841} {\bibfield  {journal} {\bibinfo
  {journal} {Front. Astron. Space Sci.}\ }\textbf {\bibinfo {volume} {9}}
  (\bibinfo {year} {2022}),\ 10.3389/fspas.2022.1063841}\BibitemShut {NoStop}%
\bibitem [{\citenamefont {Livadiotis}\ \emph {et~al.}(2022)\citenamefont
  {Livadiotis}, \citenamefont {McComas}, \citenamefont {Funsten}, \citenamefont
  {Schwadron}, \citenamefont {Szalay},\ and\ \citenamefont
  {Zirnstein}}]{Livadiotis+22/10}%
  \BibitemOpen
  \bibfield  {author} {\bibinfo {author} {\bibfnamefont {G.}~\bibnamefont
  {Livadiotis}}, \bibinfo {author} {\bibfnamefont {D.~J.}\ \bibnamefont
  {McComas}}, \bibinfo {author} {\bibfnamefont {H.~O.}\ \bibnamefont
  {Funsten}}, \bibinfo {author} {\bibfnamefont {N.~A.}\ \bibnamefont
  {Schwadron}}, \bibinfo {author} {\bibfnamefont {J.~R.}\ \bibnamefont
  {Szalay}}, \ and\ \bibinfo {author} {\bibfnamefont {E.}~\bibnamefont
  {Zirnstein}},\ }\href {\doibase 10.3847/1538-4365/ac8b88} {\bibfield
  {journal} {\bibinfo  {journal} {Astrophys. J. Suppl. Ser.}\ }\textbf
  {\bibinfo {volume} {262}},\ \bibinfo {pages} {53} (\bibinfo {year}
  {2022})}\BibitemShut {NoStop}%
\bibitem [{\citenamefont {Scherer}\ \emph {et~al.}(2020)\citenamefont
  {Scherer}, \citenamefont {Husidic}, \citenamefont {Lazar},\ and\
  \citenamefont {Fichtner}}]{Scherer+20/07}%
  \BibitemOpen
  \bibfield  {author} {\bibinfo {author} {\bibfnamefont {K.}~\bibnamefont
  {Scherer}}, \bibinfo {author} {\bibfnamefont {E.}~\bibnamefont {Husidic}},
  \bibinfo {author} {\bibfnamefont {M.}~\bibnamefont {Lazar}}, \ and\ \bibinfo
  {author} {\bibfnamefont {H.}~\bibnamefont {Fichtner}},\ }\href {\doibase
  10.1093/mnras/staa1969} {\bibfield  {journal} {\bibinfo  {journal} {Mont.
  Not. R. Astron. Soc.}\ }\textbf {\bibinfo {volume} {497}},\ \bibinfo {pages}
  {1738–1756} (\bibinfo {year} {2020})}\BibitemShut {NoStop}%
\bibitem [{\citenamefont {Gaelzer}, \citenamefont {Fichtner},\ and\
  \citenamefont {Scherer}(2024)}]{Gaelzer+24/07}%
  \BibitemOpen
  \bibfield  {author} {\bibinfo {author} {\bibfnamefont {R.}~\bibnamefont
  {Gaelzer}}, \bibinfo {author} {\bibfnamefont {H.}~\bibnamefont {Fichtner}}, \
  and\ \bibinfo {author} {\bibfnamefont {K.}~\bibnamefont {Scherer}},\ }\href
  {\doibase 10.1063/5.0212434} {\bibfield  {journal} {\bibinfo  {journal}
  {Phys. Plasmas}\ }\textbf {\bibinfo {volume} {31}},\ \bibinfo {pages}
  {072112} (\bibinfo {year} {2024})}\BibitemShut {NoStop}%
\bibitem [{\citenamefont {Daalhuis}(2010)}]{Daalhuis-Full-NIST10a}%
  \BibitemOpen
  \bibfield  {author} {\bibinfo {author} {\bibfnamefont {A.~B.~O.}\
  \bibnamefont {Daalhuis}},\ }in\ \href {http://dlmf.nist.gov/13} {\emph
  {\bibinfo {booktitle} {{NIST Handbook of Mathematical Functions}}}},\
  \bibinfo {editor} {edited by\ \bibinfo {editor} {\bibfnamefont {F.~W.~J.}\
  \bibnamefont {Olver}}, \bibinfo {editor} {\bibfnamefont {D.~W.}\ \bibnamefont
  {Lozier}}, \bibinfo {editor} {\bibfnamefont {R.~F.}\ \bibnamefont
  {Boisvert}}, \ and\ \bibinfo {editor} {\bibfnamefont {C.~W.}\ \bibnamefont
  {Clark}}}\ (\bibinfo  {publisher} {Cambridge},\ \bibinfo {address} {New
  York},\ \bibinfo {year} {2010})\ Chap.~\bibinfo {chapter} {13}, p.\ \bibinfo
  {pages} {321–349}\BibitemShut {NoStop}%
\bibitem [{\citenamefont {Husidic}\ \emph {et~al.}(2022)\citenamefont
  {Husidic}, \citenamefont {Scherer}, \citenamefont {Lazar}, \citenamefont
  {Fichtner},\ and\ \citenamefont {Poedts}}]{Husidic+22/03}%
  \BibitemOpen
  \bibfield  {author} {\bibinfo {author} {\bibfnamefont {E.}~\bibnamefont
  {Husidic}}, \bibinfo {author} {\bibfnamefont {K.}~\bibnamefont {Scherer}},
  \bibinfo {author} {\bibfnamefont {M.}~\bibnamefont {Lazar}}, \bibinfo
  {author} {\bibfnamefont {H.}~\bibnamefont {Fichtner}}, \ and\ \bibinfo
  {author} {\bibfnamefont {S.}~\bibnamefont {Poedts}},\ }\href {\doibase
  10.3847/1538-4357/ac4af4} {\bibfield  {journal} {\bibinfo  {journal}
  {Astrophys. J.}\ }\textbf {\bibinfo {volume} {927}},\ \bibinfo {pages} {159}
  (\bibinfo {year} {2022})}\BibitemShut {NoStop}%
\bibitem [{\citenamefont {{Scherer}}\ \emph {et~al.}(2022)\citenamefont
  {{Scherer}}, \citenamefont {{Husidic}}, \citenamefont {{Lazar}},\ and\
  \citenamefont {{Fichtner}}}]{Scherer+22/07}%
  \BibitemOpen
  \bibfield  {author} {\bibinfo {author} {\bibfnamefont {K.}~\bibnamefont
  {{Scherer}}}, \bibinfo {author} {\bibfnamefont {E.}~\bibnamefont
  {{Husidic}}}, \bibinfo {author} {\bibfnamefont {M.}~\bibnamefont {{Lazar}}},
  \ and\ \bibinfo {author} {\bibfnamefont {H.}~\bibnamefont {{Fichtner}}},\
  }\href {\doibase 10.1051/0004-6361/202243477} {\bibfield  {journal} {\bibinfo
   {journal} {Astron. Astrophys.}\ }\textbf {\bibinfo {volume} {663}},\
  \bibinfo {eid} {A67} (\bibinfo {year} {2022})}\BibitemShut {NoStop}%
\bibitem [{\citenamefont {Pierrard}\ \emph {et~al.}(2023)\citenamefont
  {Pierrard}, \citenamefont {{Péters de Bonhome}}, \citenamefont {Halekas},
  \citenamefont {Audoor}, \citenamefont {Whittlesey},\ and\ \citenamefont
  {Livi}}]{Pierrard+23/09}%
  \BibitemOpen
  \bibfield  {author} {\bibinfo {author} {\bibfnamefont {V.}~\bibnamefont
  {Pierrard}}, \bibinfo {author} {\bibfnamefont {M.}~\bibnamefont {{Péters de
  Bonhome}}}, \bibinfo {author} {\bibfnamefont {J.}~\bibnamefont {Halekas}},
  \bibinfo {author} {\bibfnamefont {C.}~\bibnamefont {Audoor}}, \bibinfo
  {author} {\bibfnamefont {P.}~\bibnamefont {Whittlesey}}, \ and\ \bibinfo
  {author} {\bibfnamefont {R.}~\bibnamefont {Livi}},\ }\href {\doibase
  10.3390/plasma6030036} {\bibfield  {journal} {\bibinfo  {journal} {Plasma}\
  }\textbf {\bibinfo {volume} {6}},\ \bibinfo {pages} {518–540} (\bibinfo
  {year} {2023})}\BibitemShut {NoStop}%
\bibitem [{\citenamefont {Hau}, \citenamefont {Chang},\ and\ \citenamefont
  {Lazar}(2023)}]{Hau+23/10}%
  \BibitemOpen
  \bibfield  {author} {\bibinfo {author} {\bibfnamefont {L.-N.}\ \bibnamefont
  {Hau}}, \bibinfo {author} {\bibfnamefont {C.-K.}\ \bibnamefont {Chang}}, \
  and\ \bibinfo {author} {\bibfnamefont {M.}~\bibnamefont {Lazar}},\ }\href
  {\doibase 10.3847/1538-4357/acf851} {\bibfield  {journal} {\bibinfo
  {journal} {Astrophys. J.}\ }\textbf {\bibinfo {volume} {956}},\ \bibinfo
  {pages} {144} (\bibinfo {year} {2023})}\BibitemShut {NoStop}%
\bibitem [{\citenamefont {Liu}(2024)}]{Liu24/02}%
  \BibitemOpen
  \bibfield  {author} {\bibinfo {author} {\bibfnamefont {Y.}~\bibnamefont
  {Liu}},\ }\href {\doibase 10.1007/s13538-023-01385-8} {\bibfield  {journal}
  {\bibinfo  {journal} {Braz. J. Phys.}\ }\textbf {\bibinfo {volume} {54}},\
  \bibinfo {pages} {5} (\bibinfo {year} {2024})}\BibitemShut {NoStop}%
\bibitem [{\citenamefont {Yoon}\ \emph {et~al.}(2018)\citenamefont {Yoon},
  \citenamefont {Lazar}, \citenamefont {Scherer}, \citenamefont {Fichtner},\
  and\ \citenamefont {Schlickeiser}}]{Yoon+18/12}%
  \BibitemOpen
  \bibfield  {author} {\bibinfo {author} {\bibfnamefont {P.~H.}\ \bibnamefont
  {Yoon}}, \bibinfo {author} {\bibfnamefont {M.}~\bibnamefont {Lazar}},
  \bibinfo {author} {\bibfnamefont {K.}~\bibnamefont {Scherer}}, \bibinfo
  {author} {\bibfnamefont {H.}~\bibnamefont {Fichtner}}, \ and\ \bibinfo
  {author} {\bibfnamefont {R.}~\bibnamefont {Schlickeiser}},\ }\href {\doibase
  10.3847/1538-4357/aaeb94} {\bibfield  {journal} {\bibinfo  {journal}
  {Astrophys. J.}\ }\textbf {\bibinfo {volume} {868}},\ \bibinfo {pages} {131}
  (\bibinfo {year} {2018})}\BibitemShut {NoStop}%
\bibitem [{\citenamefont {Yoon}(2021)}]{Yoon21}%
  \BibitemOpen
  \bibfield  {author} {\bibinfo {author} {\bibfnamefont {P.~H.}\ \bibnamefont
  {Yoon}},\ }\enquote {\bibinfo {title} {{Non-equilibrium Statistical Mechanics
  of Electron Kappa Distribution}},}\ p.\ \bibinfo {pages} {235–277},\ vol.\
  \bibinfo {volume} {464}\ of\  \cite{LazarFichtner21} (\bibinfo {year}
  {2021})\BibitemShut {NoStop}%
\bibitem [{\citenamefont {Livadiotis}\ and\ \citenamefont
  {McComas}(2009)}]{LivadiotisMcComas09/11}%
  \BibitemOpen
  \bibfield  {author} {\bibinfo {author} {\bibfnamefont {G.}~\bibnamefont
  {Livadiotis}}\ and\ \bibinfo {author} {\bibfnamefont {D.~J.}\ \bibnamefont
  {McComas}},\ }\href {\doibase 10.1029/2009JA014352} {\bibfield  {journal}
  {\bibinfo  {journal} {J. Geophys. Res.}\ }\textbf {\bibinfo {volume} {114}},\
  \bibinfo {pages} {A11105} (\bibinfo {year} {2009})}\BibitemShut {NoStop}%
\bibitem [{\citenamefont {Livadiotis}\ and\ \citenamefont
  {McComas}(2021)}]{LivadiotisMcComas21/12}%
  \BibitemOpen
  \bibfield  {author} {\bibinfo {author} {\bibfnamefont {G.}~\bibnamefont
  {Livadiotis}}\ and\ \bibinfo {author} {\bibfnamefont {D.~J.}\ \bibnamefont
  {McComas}},\ }\href {\doibase 10.3390/e23121683} {\bibfield  {journal}
  {\bibinfo  {journal} {Entropy}\ }\textbf {\bibinfo {volume} {23}},\ \bibinfo
  {pages} {1683} (\bibinfo {year} {2021})}\BibitemShut {NoStop}%
\bibitem [{\citenamefont {Livadiotis}\ and\ \citenamefont
  {McComas}(2022)}]{LivadiotisMcComas22/11}%
  \BibitemOpen
  \bibfield  {author} {\bibinfo {author} {\bibfnamefont {G.}~\bibnamefont
  {Livadiotis}}\ and\ \bibinfo {author} {\bibfnamefont {D.~J.}\ \bibnamefont
  {McComas}},\ }\href {\doibase 10.3847/1538-4357/ac99df} {\bibfield  {journal}
  {\bibinfo  {journal} {Astrophys. J.}\ }\textbf {\bibinfo {volume} {940}},\
  \bibinfo {pages} {83} (\bibinfo {year} {2022})}\BibitemShut {NoStop}%
\bibitem [{\citenamefont {Livadiotis}\ and\ \citenamefont
  {McComas}(2024)}]{LivadiotisMcComas24/05}%
  \BibitemOpen
  \bibfield  {author} {\bibinfo {author} {\bibfnamefont {G.}~\bibnamefont
  {Livadiotis}}\ and\ \bibinfo {author} {\bibfnamefont {D.~J.}\ \bibnamefont
  {McComas}},\ }\href {\doibase 10.1209/0295-5075/ad4415} {\bibfield  {journal}
  {\bibinfo  {journal} {Europhys. Lett.}\ }\textbf {\bibinfo {volume} {146}},\
  \bibinfo {pages} {41003} (\bibinfo {year} {2024})}\BibitemShut {NoStop}%
\bibitem [{\citenamefont {Leubner}(2002)}]{Leubner02/11}%
  \BibitemOpen
  \bibfield  {author} {\bibinfo {author} {\bibfnamefont {M.~P.}\ \bibnamefont
  {Leubner}},\ }\href {\doibase 10.1023/A:1020990413487} {\bibfield  {journal}
  {\bibinfo  {journal} {Astrophys. Space Sci.}\ }\textbf {\bibinfo {volume}
  {282}},\ \bibinfo {pages} {573–579} (\bibinfo {year} {2002})}\BibitemShut
  {NoStop}%
\bibitem [{\citenamefont {{Lazar, M.}}\ \emph {et~al.}(2017)\citenamefont
  {{Lazar, M.}}, \citenamefont {{Pierrard, V.}}, \citenamefont {{Shaaban, S.
  M.}}, \citenamefont {{Fichtner, H.}},\ and\ \citenamefont {{Poedts,
  S.}}}]{Lazar+17/06}%
  \BibitemOpen
  \bibfield  {author} {\bibinfo {author} {\bibnamefont {{Lazar, M.}}}, \bibinfo
  {author} {\bibnamefont {{Pierrard, V.}}}, \bibinfo {author} {\bibnamefont
  {{Shaaban, S. M.}}}, \bibinfo {author} {\bibnamefont {{Fichtner, H.}}}, \
  and\ \bibinfo {author} {\bibnamefont {{Poedts, S.}}},\ }\href {\doibase
  10.1051/0004-6361/201630194} {\bibfield  {journal} {\bibinfo  {journal}
  {Astron. Astrophys.}\ }\textbf {\bibinfo {volume} {602}},\ \bibinfo {pages}
  {A44} (\bibinfo {year} {2017})}\BibitemShut {NoStop}%
\bibitem [{\citenamefont {Eyelade}\ \emph {et~al.}(2021)\citenamefont
  {Eyelade}, \citenamefont {Stepanova}, \citenamefont {Espinoza},\ and\
  \citenamefont {Moya}}]{Eyelade+21/04}%
  \BibitemOpen
  \bibfield  {author} {\bibinfo {author} {\bibfnamefont {A.~V.}\ \bibnamefont
  {Eyelade}}, \bibinfo {author} {\bibfnamefont {M.}~\bibnamefont {Stepanova}},
  \bibinfo {author} {\bibfnamefont {C.~M.}\ \bibnamefont {Espinoza}}, \ and\
  \bibinfo {author} {\bibfnamefont {P.~S.}\ \bibnamefont {Moya}},\ }\href
  {\doibase 10.3847/1538-4365/abdec9} {\bibfield  {journal} {\bibinfo
  {journal} {Astrophys. J. Suppl. Ser.}\ }\textbf {\bibinfo {volume} {253}},\
  \bibinfo {pages} {34} (\bibinfo {year} {2021})}\BibitemShut {NoStop}%
\bibitem [{\citenamefont {Kirpichev}\ \emph {et~al.}(2021)\citenamefont
  {Kirpichev}, \citenamefont {Antonova}, \citenamefont {Stepanova},
  \citenamefont {Eyelade}, \citenamefont {Espinoza}, \citenamefont
  {Ovchinnikov}, \citenamefont {Vorobjev},\ and\ \citenamefont
  {Yagodkina}}]{Kirpichev+21/10}%
  \BibitemOpen
  \bibfield  {author} {\bibinfo {author} {\bibfnamefont {I.~P.}\ \bibnamefont
  {Kirpichev}}, \bibinfo {author} {\bibfnamefont {E.~E.}\ \bibnamefont
  {Antonova}}, \bibinfo {author} {\bibfnamefont {M.}~\bibnamefont {Stepanova}},
  \bibinfo {author} {\bibfnamefont {A.~V.}\ \bibnamefont {Eyelade}}, \bibinfo
  {author} {\bibfnamefont {C.~M.}\ \bibnamefont {Espinoza}}, \bibinfo {author}
  {\bibfnamefont {I.~L.}\ \bibnamefont {Ovchinnikov}}, \bibinfo {author}
  {\bibfnamefont {V.~G.}\ \bibnamefont {Vorobjev}}, \ and\ \bibinfo {author}
  {\bibfnamefont {O.~I.}\ \bibnamefont {Yagodkina}},\ }\href {\doibase
  10.1029/2021JA029409} {\bibfield  {journal} {\bibinfo  {journal} {J. Geophys.
  Res.}\ }\textbf {\bibinfo {volume} {126}},\ \bibinfo {pages} {e2021JA029409}
  (\bibinfo {year} {2021})}\BibitemShut {NoStop}%
\bibitem [{\citenamefont {Zheng}\ \emph {et~al.}(2024)\citenamefont {Zheng},
  \citenamefont {Martinović}, \citenamefont {Pierrard}, \citenamefont {Klein},
  \citenamefont {Liu}, \citenamefont {Abraham}, \citenamefont {Liu},
  \citenamefont {Luo}, \citenamefont {Lin}, \citenamefont {Liu},\ and\
  \citenamefont {Li}}]{Zheng+24/12}%
  \BibitemOpen
  \bibfield  {author} {\bibinfo {author} {\bibfnamefont {X.}~\bibnamefont
  {Zheng}}, \bibinfo {author} {\bibfnamefont {M.~M.}\ \bibnamefont
  {Martinović}}, \bibinfo {author} {\bibfnamefont {V.}~\bibnamefont
  {Pierrard}}, \bibinfo {author} {\bibfnamefont {K.~G.}\ \bibnamefont {Klein}},
  \bibinfo {author} {\bibfnamefont {M.}~\bibnamefont {Liu}}, \bibinfo {author}
  {\bibfnamefont {J.~B.}\ \bibnamefont {Abraham}}, \bibinfo {author}
  {\bibfnamefont {Y.}~\bibnamefont {Liu}}, \bibinfo {author} {\bibfnamefont
  {J.}~\bibnamefont {Luo}}, \bibinfo {author} {\bibfnamefont {X.}~\bibnamefont
  {Lin}}, \bibinfo {author} {\bibfnamefont {G.}~\bibnamefont {Liu}}, \ and\
  \bibinfo {author} {\bibfnamefont {J.}~\bibnamefont {Li}},\ }\href {\doibase
  10.3847/1538-4357/ad7d05} {\bibfield  {journal} {\bibinfo  {journal}
  {Astrophys. J.}\ }\textbf {\bibinfo {volume} {977}},\ \bibinfo {pages} {39}
  (\bibinfo {year} {2024})}\BibitemShut {NoStop}%
\bibitem [{\citenamefont {Livadiotis}, \citenamefont {Desai},\ and\
  \citenamefont {Wilson}(2018)}]{Livadiotis+18/02}%
  \BibitemOpen
  \bibfield  {author} {\bibinfo {author} {\bibfnamefont {G.}~\bibnamefont
  {Livadiotis}}, \bibinfo {author} {\bibfnamefont {M.~I.}\ \bibnamefont
  {Desai}}, \ and\ \bibinfo {author} {\bibfnamefont {L.~B.}\ \bibnamefont
  {Wilson}, \bibfnamefont {III}},\ }\href {\doibase 10.3847/1538-4357/aaa713}
  {\bibfield  {journal} {\bibinfo  {journal} {Astrophys. J.}\ }\textbf
  {\bibinfo {volume} {853}},\ \bibinfo {pages} {142} (\bibinfo {year}
  {2018})}\BibitemShut {NoStop}%
\bibitem [{\citenamefont {Cuesta}\ \emph {et~al.}(2024)\citenamefont {Cuesta},
  \citenamefont {Cummings}, \citenamefont {Livadiotis}, \citenamefont
  {McComas}, \citenamefont {Cohen}, \citenamefont {Khoo}, \citenamefont
  {Sharma}, \citenamefont {Shen}, \citenamefont {Bandyopadhyay}, \citenamefont
  {Rankin}, \citenamefont {Szalay}, \citenamefont {Farooki}, \citenamefont
  {Xu}, \citenamefont {Muro}, \citenamefont {Stevens},\ and\ \citenamefont
  {Bale}}]{Cuesta+24/10}%
  \BibitemOpen
  \bibfield  {author} {\bibinfo {author} {\bibfnamefont {M.~E.}\ \bibnamefont
  {Cuesta}}, \bibinfo {author} {\bibfnamefont {A.~T.}\ \bibnamefont
  {Cummings}}, \bibinfo {author} {\bibfnamefont {G.}~\bibnamefont
  {Livadiotis}}, \bibinfo {author} {\bibfnamefont {D.~J.}\ \bibnamefont
  {McComas}}, \bibinfo {author} {\bibfnamefont {C.~M.~S.}\ \bibnamefont
  {Cohen}}, \bibinfo {author} {\bibfnamefont {L.~Y.}\ \bibnamefont {Khoo}},
  \bibinfo {author} {\bibfnamefont {T.}~\bibnamefont {Sharma}}, \bibinfo
  {author} {\bibfnamefont {M.~M.}\ \bibnamefont {Shen}}, \bibinfo {author}
  {\bibfnamefont {R.}~\bibnamefont {Bandyopadhyay}}, \bibinfo {author}
  {\bibfnamefont {J.~S.}\ \bibnamefont {Rankin}}, \bibinfo {author}
  {\bibfnamefont {J.~R.}\ \bibnamefont {Szalay}}, \bibinfo {author}
  {\bibfnamefont {H.~A.}\ \bibnamefont {Farooki}}, \bibinfo {author}
  {\bibfnamefont {Z.}~\bibnamefont {Xu}}, \bibinfo {author} {\bibfnamefont
  {G.~D.}\ \bibnamefont {Muro}}, \bibinfo {author} {\bibfnamefont {M.~L.}\
  \bibnamefont {Stevens}}, \ and\ \bibinfo {author} {\bibfnamefont {S.~D.}\
  \bibnamefont {Bale}},\ }\href {\doibase 10.3847/1538-4357/ad68fd} {\bibfield
  {journal} {\bibinfo  {journal} {Astrophys. J.}\ }\textbf {\bibinfo {volume}
  {973}},\ \bibinfo {pages} {76} (\bibinfo {year} {2024})}\BibitemShut
  {NoStop}%
\bibitem [{\citenamefont {Chian}\ \emph {et~al.}(2016)\citenamefont {Chian},
  \citenamefont {Feng}, \citenamefont {Hu}, \citenamefont {Loew}, \citenamefont
  {Miranda}, \citenamefont {Muñoz}, \citenamefont {Sibeck},\ and\
  \citenamefont {Wu}}]{Chian+16/12}%
  \BibitemOpen
  \bibfield  {author} {\bibinfo {author} {\bibfnamefont {A.~C.-L.}\
  \bibnamefont {Chian}}, \bibinfo {author} {\bibfnamefont {H.~Q.}\ \bibnamefont
  {Feng}}, \bibinfo {author} {\bibfnamefont {Q.}~\bibnamefont {Hu}}, \bibinfo
  {author} {\bibfnamefont {M.~H.}\ \bibnamefont {Loew}}, \bibinfo {author}
  {\bibfnamefont {R.~A.}\ \bibnamefont {Miranda}}, \bibinfo {author}
  {\bibfnamefont {P.~R.}\ \bibnamefont {Muñoz}}, \bibinfo {author}
  {\bibfnamefont {D.~G.}\ \bibnamefont {Sibeck}}, \ and\ \bibinfo {author}
  {\bibfnamefont {D.~J.}\ \bibnamefont {Wu}},\ }\href {\doibase
  10.3847/0004-637X/832/2/179} {\bibfield  {journal} {\bibinfo  {journal}
  {Astrophys. J.}\ }\textbf {\bibinfo {volume} {832}},\ \bibinfo {pages} {179}
  (\bibinfo {year} {2016})}\BibitemShut {NoStop}%
\bibitem [{\citenamefont {Lazerson}(2011)}]{Lazerson11/02}%
  \BibitemOpen
  \bibfield  {author} {\bibinfo {author} {\bibfnamefont {S.~A.}\ \bibnamefont
  {Lazerson}},\ }\href {\doibase 10.1017/S0022377810000218} {\bibfield
  {journal} {\bibinfo  {journal} {J. Plasma Phys.}\ }\textbf {\bibinfo {volume}
  {77}},\ \bibinfo {pages} {31–37} (\bibinfo {year} {2011})}\BibitemShut
  {NoStop}%
\bibitem [{\citenamefont {Yang}, \citenamefont {Wang},\ and\ \citenamefont
  {Dong}(2023)}]{Yang+23/07}%
  \BibitemOpen
  \bibfield  {author} {\bibinfo {author} {\bibfnamefont {S.-D.}\ \bibnamefont
  {Yang}}, \bibinfo {author} {\bibfnamefont {L.}~\bibnamefont {Wang}}, \ and\
  \bibinfo {author} {\bibfnamefont {C.}~\bibnamefont {Dong}},\ }\href {\doibase
  10.1093/mnras/stad1453} {\bibfield  {journal} {\bibinfo  {journal} {Mont.
  Not. R. Astron. Soc.}\ }\textbf {\bibinfo {volume} {523}},\ \bibinfo {pages}
  {928–933} (\bibinfo {year} {2023})}\BibitemShut {NoStop}%
\end{thebibliography}%

\end{document}